%% file: ms.tex
\documentclass[iop,twocolumn,natbib209,numberedappendix,twocolappendix]{emulateapj}    

\usepackage{natbib}
\usepackage{amsmath}
\usepackage{color}
\usepackage{longtable}
\usepackage{graphicx}
\usepackage{lscape}
\usepackage{textcomp}

\usepackage{hyperref}

\shorttitle{YSOs in L1641: Disks, Accretion, and SFH.}
\newcommand{\Teff}{$T_{\rm{eff}}$}
\newcommand{\kms}{\,km\,s$^{-1}$}

\newcommand{\ergps}{erg\,s$^{-1}$}

\newcommand{\mum}{\,$\mu$m}

\newcommand{\Msun}{$M_{\odot}$}

\newcommand{\accunit}{$_{\odot}$\,yr$^{-1}$}
\newcommand{\sek}{Sect.}
\newcommand{\tab}{table}

\newcommand{\rev}{ }
\newcommand{\newrev}{ }
\newcommand{\newnewrev}{ }
\newcommand{\lrev}{ }
\newcommand{\CaII}{Ca\,II}
\newcommand{\HeI}{He\,I}
\newcommand{\fig}{Fig.\,}
\newcommand{\sect}{Sect.\,}
\newcommand{\AV}{$A_{\rm{V}}$}
\newcommand{\LiI}{Li\,I\,$\lambda$6708\,\AA}
\newcommand{\FeII}{\mbox{Fe\,\textsc{ii}}}%

\begin{document}

\title{Young Stellar Objects in Lynds~1641: Disks, Accretion, and Star Formation History.}

\author{Min~Fang\altaffilmark{1,2,3}, Jinyoung~Serena~Kim\altaffilmark{4}, Roy~van~Boekel\altaffilmark{2}, Aurora~Sicilia-Aguilar\altaffilmark{5}, Thomas~Henning\altaffilmark{2}, Kevin~Flaherty\altaffilmark{4}}
\altaffiltext{1}{Purple Mountain Observatory and Key Laboratory of Radio Astronomy, Chinese Academy of Sciences, 2 West Beijing Road, 210008 Nanjing, China}
\altaffiltext{2}{Max-Planck Institute for Astronomy, K\"onigstuhl 17, D-69117 Heidelberg, Germany}
\altaffiltext{3}{ Key Laboratory of Modern Astronomy and Astrophysics (Nanjing University)，Ministry of Education, Nanjing 210093, China}
\altaffiltext{4}{Steward Observatory, University of Arizona, 933 North Cherry Avenue, Tucson, AZ 85721-0065}
\altaffiltext{5}{Departamento de Fisica Teorica, Facultad de Ciencias, Universidad Autonoma de Madrid, 28049 Cantoblanco, Madrid, Spain}
\email{mfang@pmo.ac.cn}

\begin{abstract} 

We investigate the young stellar objects (YSOs) in the Lynds~1641 (L1641) cloud  using multi-wavelength data including Spitzer, WISE, 2MASS, and XMM covering {\newnewrev $\sim$1390} YSOs across a range of evolutionary stages. In addition, we targeted a sub-sample of YSOs for optical spectroscopy. We use this data, along with archival photometric data, to derive spectral types, extinction values, masses, ages, as well as accretion rates. We obtain a disk fraction of $\sim${\newnewrev 50\%} in L1641. The disk frequency is almost constant as a function of stellar mass with a slight peak at log($M_*$/\Msun)$\approx$$-$0.25. The analysis of multi-epoch spectroscopic data indicates that the accretion variability of YSOs cannot explain the two orders of magnitude of scatter for YSOs with similar masses. Forty-six new transition disk (TD) objects are confirmed in this work, and we find that the fraction of accreting TDs is lower than for optically thick disks ({\newnewrev 40--45\% vs. 77--79\%} respectively). We confirm our previous result that the accreting TDs have a similar median accretion rate to normal optically thick disks. We confirm that two star formation modes (isolated vs. clustered) exist in L1641. We find that the diskless YSOs are statistically older than the YSOs with optically-thick disks and the transition disk objects have a median age which is intermediate between the two populations. We tentatively study the star formation history in L1641 based on the age distribution and find that star formation started to be active 2--3\,Myr ago.

\end{abstract}

\maketitle

\section{Introduction}

Circumstellar disks, as a byproduct of the star-formation process via angular momentum conservation, play a key role in the formation of new stars, and subsequent planetary systems. The investigation of the disk evolution is thus important to understand both star formation and  planet formation.

The disk dissipation processes have been constrained by surveys of large samples of young stars. These surveys probe the inner disk regions using excess emission above the stellar photosphere at infrared wavelengths and find that the lifetime is several Myrs \citep{1989AJ.....97.1451S,2001ApJ...553L.153H, 2002astro.ph.10520H,2007ApJ...662.1067H,2006ApJ...638..897S,2013A&A...549A..15F}. The large variety of disk morphologies observed at a given age suggests that the disk evolution is controlled by several parameters: stellar and disk mass, multiplicity, local environments, etc. \citep[][]{2006ApJ...648..484H,2006ApJ...653L..57B,2007A&A...462..245G,2009A&A...504..461F,2012A&A...539A.119F}.

Observations at infrared wavelengths suggest that there are two types of evolved disks: radially depleted disks and globally depleted disks. The radially depleted disks show very weak or no infrared excess at near infrared wavelengths, but strong excess emission at mid-infrared and longer wavelengths, suggesting dust in these disks are cleared starting from the inside and moving outward \citep{1985prpl.conf.1100H,1995ApJ...439..288O,2000LNP...548..341M,2004ApJ...612..496M,2005ApJ...631.1134A,2005AJ....129.1049C}. The globally depleted disks exhibit an approximately uniformly reduced infrared excess compared to primordial disks over all wavelengths out to 24\mum\ \citep[][]{2006AJ....131.1574L,2008ApJ...687.1145S,2009ApJ...701.1188S,2009AJ....138..703C,2009ApJ...698....1C,2010arXiv1002.1715C}, indicating there is global dust depletion in these disks. The investigation of the two types of evolved disks are important to understand the disk dissipation processes. 

The radially depleted disks,  usually named TDs, are better studied in the literature than the  globally depleted disks. In the following, without specification we only refer to the  radially depleted disks as TDs. There are several processes proposed to produce TDs, including (1) giant planet formation \citep{2003MNRAS.342...79R,2004ApJ...612L.137Q}, (2)  tidal truncation in close binaries, (3) photo-evaporation \citep{1998ApJ...499..758J,1999ApJ...515..669S,2000ApJ...539..258R,2000prpl.conf..401H,2001MNRAS.328..485C,2003MNRAS.342.1139A,2006MNRAS.369..229A}, {\newrev (4) magnetorotational instability \citep{2007NatPh...3..604C}}, and (5) dust grain growth \citep{2011ApJ...742...39S}. To distinguish these mechanisms, an important diagnostic is the comparison of the accretion rate of matter onto the central star between transition disk objects and less evolved classical T\,Tauri stars \citep{2007MNRAS.378..369N,2010ApJ...710..597S,2010ApJ...712..925C}.

 For the normal classical T~Tauri stars (CTTS),  various studies have suggested that there is an empirical correlation between the average accretion rate and the mass of the central star of $\dot{M}_{\rm acc}$\,$\propto$\,$M_*^{\alpha}$, with $\alpha$$\approx$1-3 \citep{2003ApJ...582.1109W,2003ApJ...592..266M,2004AJ....128.1294C,2005ApJ...625..906M,2005ApJ...626..498M,2006A&A...452..245N,2006A&A...459..837G,2008ApJ...681..594H,2008A&A...481..423G,2009A&A...504..461F}. Although the correlation between accretion rate and stellar mass is obvious for a large sample of young stars, individual objects with similar ages and masses commonly scatter around the average relation by up to 2 orders of magnitude. Multiple epoch observations of the same stars can show large variations in the H${\alpha}$ emission strength and line profile, pointing at substantial temporal variations in the accretion \citep{1995ApJ...449..341J,1996A&A...314..835G,2001AJ....122.3335A,2002ApJ...571..378A,2010ApJ...710..597S}, which provides one plausible potential explanation for the large scatter in the $\dot{M}_{\rm acc}$ vs. $M_*$ relation. Other explanations for such a large scatter relate to initial conditions, angular momentum, disk mass, and substantial time evolution \citep{1998ApJ...495..385H,2006ApJ...645L..69D}.

To address all of these questions, we need to study  a large sample of well characterized young stars. L1641 is one of the best sites for this investigation. L1641  is located in the Orion molecular cloud complex at a distance of 400-500\,pc \citep[e.g.][]{1982AJ.....87.1213A,2007PASJ...59..897H}, and the region probably has a ``depth'' of at least several tens of parsecs. In this work, we will assume a distance of 450\,pc. In L1641, a large population of young stars are formed in relative isolation, in addition to a population of young stars in many clusters or aggregates \citep{1993ApJ...412..233S,1995PhDT..........A,2009A&A...504..461F}. The Spitzer observations of this region have yielded excellent samples of thousands of young stars down to very low mass including both substantial numbers of TDs as well as normal T~Tauri stars \citep{2009A&A...504..461F,2012AJ....144..192M}. {\rev In \citet{2009A&A...504..461F} (hereafter Paper\,I), we performed a large spectroscopic survey of YSOs in L1641 with VLT/VIMOS. Using these data, in combination with the optical and near-infrared photometry, we derived the masses, ages, and accretion properties of a subsample of YSOs in L1641, and related these properties to their disk properties traced by the Spitzer observations. Following the work in Paper I, we perform a new optical spectroscopic survey of young stars in the L1641 cloud with the MMT/Hectospec and the MMT/Hectochelle. In this work, we report on the results from these observations. The data from Hectospec can cover a wavelength range from 3700--9000\,\AA, and are used to confirm the youth and determine the spectral types of YSO candidates in L1641, especially for the TD candidates proposed in Paper\,I. Furthermore, the spectra from Hectochelle can cover the H$\alpha$ line with high spectral resolution. In this work, a group of YSOs in L1641 have been observed at multiple epochs with Hectochelle, aimed at monitoring the accretion variations of YSOs in L1641. }

 We arrange the paper as follows: in $\S$2 we describe our observations and data reduction, in $\S$3 we delineate our data analysis, we present our results in $\S$4, followed by a discussion in $\S$5, and we summarize our efforts in $\S$6.

\begin{figure}
\begin{center}
\includegraphics[width=\columnwidth]{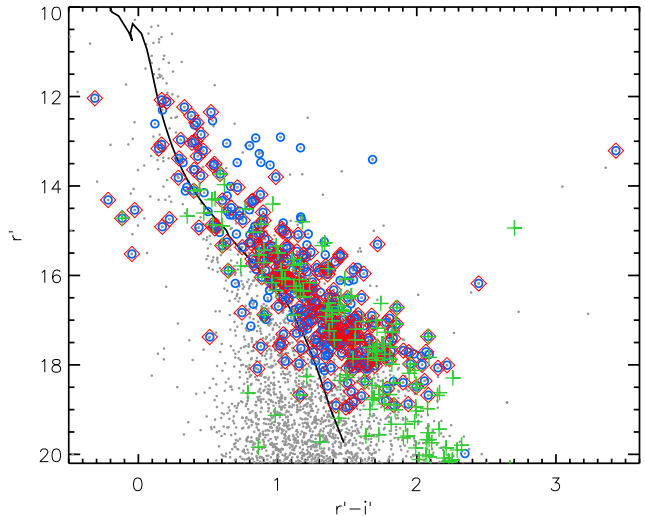}
\caption{ $r'$ vs. $r'-i'$ color-magnitude diagram. The grey filled circles show all the sources found in the L1641 field (Paper\,I), the open circles and diamonds are the targets for the spectroscopic survey with Hectospec and Hectochelle, respectively. The pluses mark the YSOs in Paper\,I. The solid line shows the 10\,Myr pre-main-sequence (PMS) isochrone from \cite{2008ApJS..178...89D}}\label{Fig:CMD}
\end{center}
\end{figure}

\section{Observations and data reduction}

We combine spectroscopic data newly obtained with Hectospec \citep{2005PASP..117.1411F} and Hectochelle \citep{2011PASP..123.1188S} with photometric data from Paper\,I. We use the spectroscopic data to do the spectral classification, and characterize the accretion properties of YSOs in L1641.

\subsection{Photometry}

Our photometric data used in this work are mainly extracted from Paper\,I  (see the description in Paper\,I). We complement them with the data from the UKIRT infrared deep sky survey \citep[UKIDSS,][]{2007MNRAS.379.1599L}, and Wide-field Infrared Survey Explorer \citep[WISE,][]{2010AJ....140.1868W}. We briefly describe  each dataset as follows.

Our optical photometry are partially taken from the Sloan Digital Sky Survey \citep[SDSS,][]{2000AJ....120.1579Y} in the $u'g'r'i'z'$ bands centered on 0.35, 0.48, 0.62, 0.76 and 0.91\mum, respectively. The 10$\sigma$ limiting magnitudes for the SDSS survey of the L1641 cloud are $\sim$20.5, 21.7, 21.4, 21.1 and 19.7 in the $u'g'r'i'z'$ bands, respectively. The SDSS survey has not covered the south-east half of the L1641 cloud, toward which we complemented the SDSS photometry of L1641 with CCD imaging in the SDSS $g'r'i'z'$ bands performed at the Calar Alto 3.5m telescope, using the Large Area Imager for Calar Alto (LAICA). Conditions were photometric, but the seeing was poor (2--3\arcsec), somewhat limiting the sensitivity for faint point sources ($\sim$19.6, 19.6, 19.9, and 18.9 mag at $g'r'i'z'$ bands, respectively).

The near-infrared photometry in the $JHK_{\rm S}$ bands was taken from the Two-Micron All Sky Survey \citep[2MASS, ][]{2006AJ....131.1163S}, with 10$\sigma$ limiting magnitudes of $\sim$16.2, 15.3, and 14.6~mag, respectively. The north-west part of the L1641 cloud (DEC (J2000)$>-$6.44) has been observed by the UKIRT infrared deep sky survey \citep[UKIDSS,][]{2007MNRAS.379.1599L}, in  $JHK$ bands, with 10$\sigma$ limiting magnitudes of $\sim$19.3, 18.2, and 17.6~mag, respectively. For a few objects which are too faint to be detected in the 2MASS survey, we extract their photometry  in  the UKIDSS catalog.

The mid-infrared photometry for the sources in the L1641 cloud are extracted from the Spitzer IRAC and MIPS imaging data (see the detail description for the data reduction in Paper\,I). The 10$\sigma$ limiting magnitudes for the IRAC and MIPS imaging survey of the L1641 cloud are $\sim$17.5, 17.2, 14.4, 12.2, and 9.0~mag in the IRAC [3.6], [4.5], [5.8], [8.0], and [24] bands, respectively. The L1641 cloud has been  covered by the WISE survey at wavelengths of 3.4, 4.6, 12, and 22\mum\ with spatial resolutions of  6\farcs1, 6\farcs4, 6\farcs5, and 12\farcs0. We also extracted the mid-infrared  photometry from the WISE survey. The 10$\sigma$ limiting magnitudes are estimated to be $\sim$16.4, 15.1, 10.7, and 7.1~mag in the WISE [3.4], [4.6], [12], and [22] bands, respectively.

\subsection{X-ray emission}
Young stars can be identified by their X-ray emission, which is 2 to 3 orders of magnitude brighter than seen in the field population \citep{1999ARA&A..37..363F}. Thus, complementary X-ray data can be  extremely useful to distinguish young stars from  field stars in star forming regions. The L1641 cloud has been almost fully covered by the XMM survey \citep{2009AIPC.1094..959W}. The sky coverage of the XMM survey is shown in Fig.~\ref{Fig:yso_dis}. All the X-ray emission sources detected in the survey have been published in the XMM-Newton Serendipitous Source Catalogue \citep{2009A&A...493..339W}. We have extracted $\sim$940 X-ray sources from the  XMM Serendipitous Source Catalogue in the field of L1641. We matched the X-ray sources to our targets using 2$''$ tolerance, and found optical and/or infrared counterparts for 66\% of the X-ray sources. We consider these X-ray emission sources to be likely YSO candidates. We must stress that a small fraction of ``YSOs'' in our catalog can be bright AGN and nearby foreground stars. Without spectroscopic data, these sources are difficult to be excluded.

\subsection{Optical spectroscopy}

\subsubsection{Target selection}
We select the targets for our spectroscopic observations if they obey any of the following  criteria:

\begin{enumerate}
\item \label{sel_crit:IR_excess} X-ray emission
\item \label{sel_crit:LiI} Infrared excess
\item \label{sel_crit:Halpha} Above the 10\,Myr PMS isochrone 
\end{enumerate}

To reach enough SNR for spectral classification, we select the sources with SDSS r$'$ magnitude  brighter than than $\sim$19\,mag. One additional target, fainter than 19\,mag in at $r'$ band but very bright at $i'$ band, is also included in our spectroscopic sample with Hectospec. {\newnewrev For the spectroscopic survey with Hectochelle, we select the YSOs with SDSS r$'\sim$12--19\,mag}. Figure~\ref{Fig:CMD} shows the $r'$ vs. $r'-i'$ color-magnitude diagram of our targets for both spectroscopic surveys.

\scriptsize
\renewcommand{\tabcolsep}{0.08cm}
\input{tab1.tex}

\renewcommand{\tabcolsep}{0.2cm}
\normalsize

\subsubsection{Spectroscopic observations and data reduction}

\begin{figure*}[p]
\begin{center}
\includegraphics[width=1.8\columnwidth]{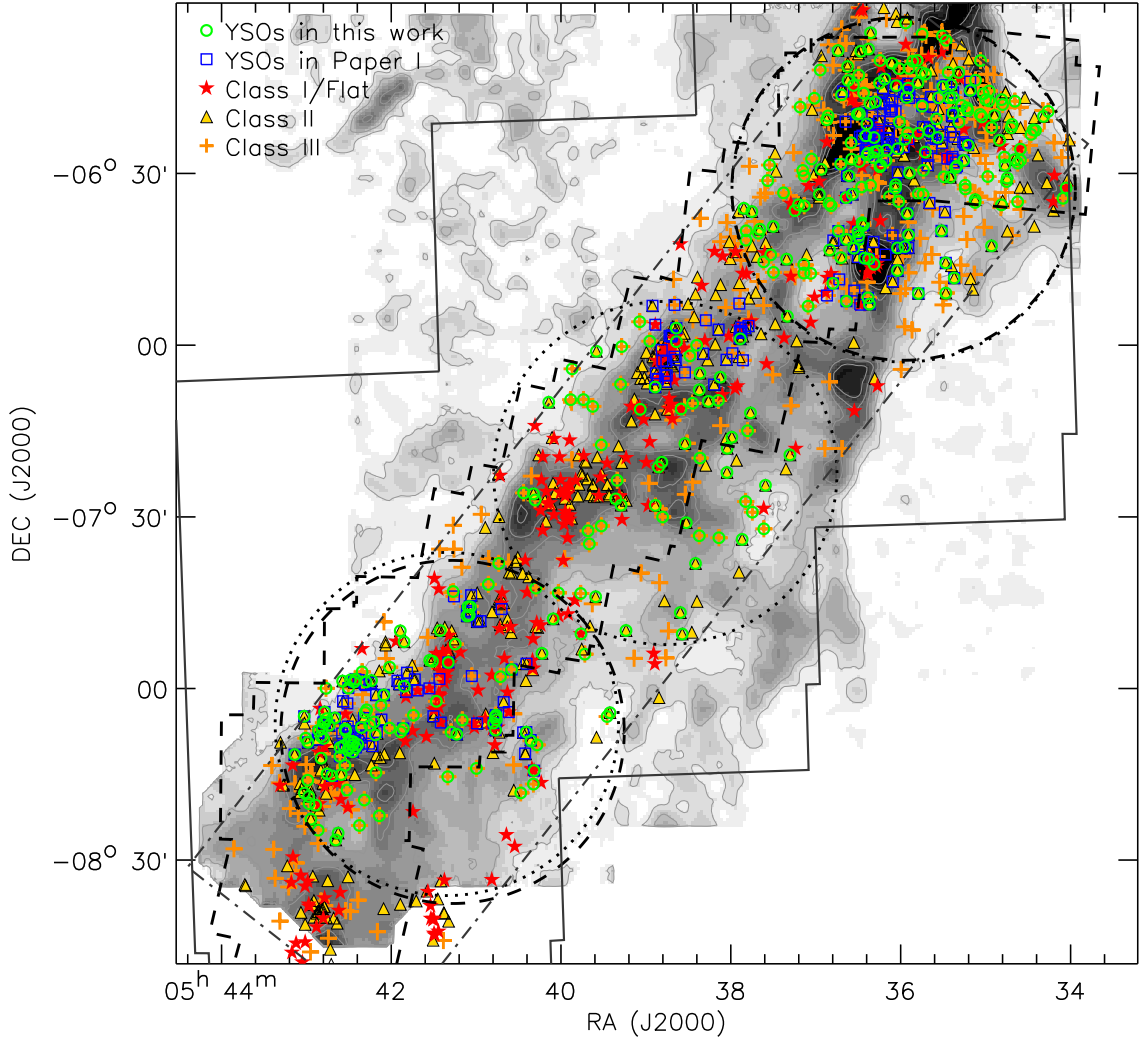}
\caption{The distribution of YSOs in L1641 overplotted on a  $^{13}$CO integrated intensity map which is a combination of our observations with the Delingha 14\,m telescope and data from  \citet{1987ApJ...312L..45B}.The green open circles mark the YSOs observed with Hectospec. The blue open squares show the YSOs studied in Paper\,I. The red filled {\lrev star symbols} and  yellow filled triangles represent the Class\,I/Flat-spectrum and Class II YSOs, respectively. The orange plus marks show the Class\,III sources. The big dotted-line circles show the fields of view (FOV) of our spectroscopic surveys with Hectospec, {\newnewrev and the big dashed-line circles display the FOVs of our Hectochelle survey}. The dashed lines enclose the fields observed with XMM observation. The dash-dotted lines encircle the regions observed with Spitzer IRAC observations, and the solid lines enclose the fields observed with Spitzer MIPS 24\mum\ observations.}\label{Fig:yso_dis}
\end{center}
\end{figure*}

The intermediate resolution spectra of sources are taken with the Hectospec multi-object spectrograph which can take a maximum of 300 spectra simultaneously. We used  the 270 groove mm$^{-1}$ grating and obtained spectra in the range 3700--9000\,\AA\ with a resolution of $\sim$5\,\AA.  We have $\sim$450 targets, which are distributed at 3 pointings (see Fig.~\ref{Fig:yso_dis}). The observational data were taken  at the nights on 2011 Jan 23, 24, and 25. Table~\ref{Tab:obs_log} lists the observational logs.

The high-resolution H$\alpha$ spectroscopy of sources are obtained with the Hectochelle multi-object spectrograph which can take a maximum of 240 spectra simultaneously  {\newnewrev with a spectral resolution of 34000}. We used  the OB 26 filter which simultaneously cover H$\alpha$ emission line and \LiI\ absorption line. The  observational data were taken at the night on 2010 Feb 5 with two pointings. On the nights on 2010 Mar 3, 2010 Nov 29, and 2011 Oct 19, a major fraction of our targets were observed again. The observational logs are listed in \tab~\ref{Tab:obs_log}.

Toward each pointing observed with Hectochelle or Hectospec, we have taken at least one extra set of sky spectra by offsetting the telescope by $\sim$5--10~arcsec between the science exposures, which can provide us with a sky spectrum  close to each object. Together with the rest of sky fibers at each science exposure, this enables us to construct the appropriate nebular and sky spectrum for subtraction, taking into account the different transmission of every fiber and the variability of the nebular and sky emission throughout the field of view. 

We use the IRAF routines to reduce the Hectospec and Hectochelle data according to a standard procedure. We flat and extract the spectra using dome flats with the IRAF task \textit{\textbf{dofibers}} under the package \textit{\textbf{specred}}. The wavelength solution is achieved with HeNeAr and ThAr comparison spectra for Hectospec and Hectochelle, respectively, using the IRAF task \textit{\textbf{identify}} and \textit{\textbf{reidentify}} under the package \textit{\textbf{specred}}.  We calibrate the spectra with a wavelength solution constructed using the IRAF task \textit{\textbf{dispcor}} under the package \textit{\textbf{specred}}. For each pointing, we have done the observations at several exposures. We extract the spectra for each exposure. Finally, we obtain the spectra for each target and the corresponding sky spectra close to this target. We subtract the sky from the spectra of each target, and combine the sky-subtracted spectra into one final spectrum.

\subsubsection{Complementary literature spectroscopy}
An additional sample of 12 stars with  spectral types, of which 10 are from \citet{1995PhDT..........A} and  2 from our VIMOS spectroscopic survey, are also included in this work. These stars show X-ray emission, but are not in Paper\,I. Thus, we include them in this work.

\section{Analysis}
\subsection{YSO selection criteria}
\label{sec:analysis:YSO_selection_cirteria}
A star in our sample is classified as a young star if it obeys any of the following criteria:

\begin{enumerate}
\item \label{crit:LiI} X-ray emission
\item \label{crit:IR_excess} Infrared excess
\item \label{crit:Halpha} H$\alpha$ emission
\end{enumerate}

\noindent

When  good SNR Hectochelle spectra are available, we also use the \LiI\ absorption line as an indicator of youth. We find that sources with strong infrared excess emission always show H$\alpha$ emission, but do not always show X-ray emission. We also find that some stars with X-ray emission show  H$\alpha$ absorption. These stars typically have spectral types from late A to late G. We note that there may be a small contamination of our sample from dMe stars, which are old, M-type stars that show H$\alpha$ emission due to chromospheric activity, and from extragalactic sources which show infrared excesses and/or X-ray emission.

\begin{figure*}
\centering
\includegraphics[width=1.8\columnwidth]{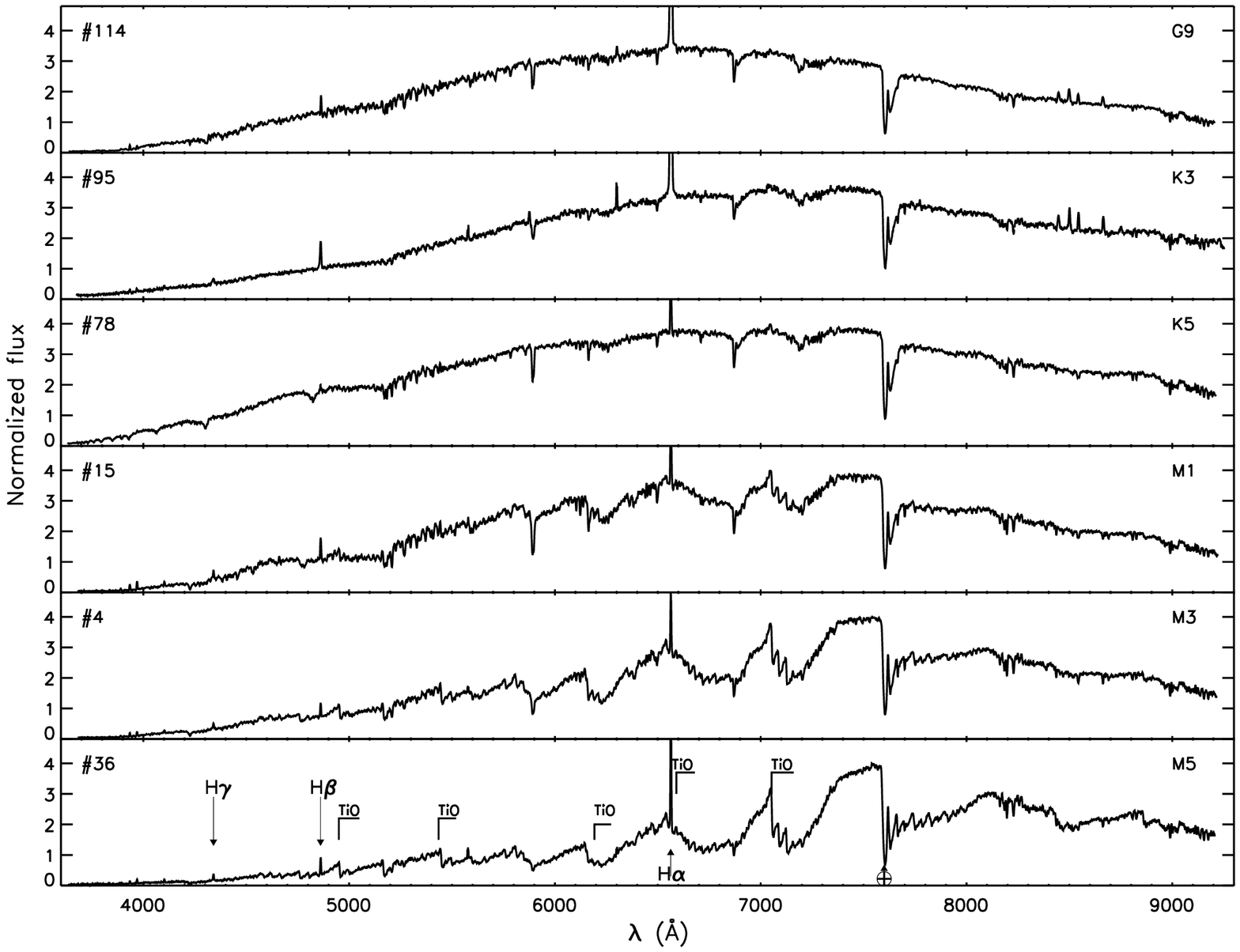}
\caption{\label{fig:spectral_sequence} Example spectra from our Hectospec observations covering a range of spectral types. The 
prominent Balmer lines (H$\alpha$, H$\beta$, H$\gamma$) and TiO absorption features are indicated.}
\end{figure*}

\subsection{Spectral classification}\label{sec:analysis_spt}
We classified the Hectospec spectra  using the scheme developed by \citet{2004AJ....127.1682H}, which employs the empirical relation between  the equivalent widths of selected atomic and molecular absorption lines and the effective temperature. The classification scheme consists of of 3 subregimes, each combines  a number of absorption features and spans a range in spectral types \citep[See the detailed description of the method in][]{2004AJ....127.1682H}.

\begin{figure}
\begin{center}
\includegraphics[width=\columnwidth]{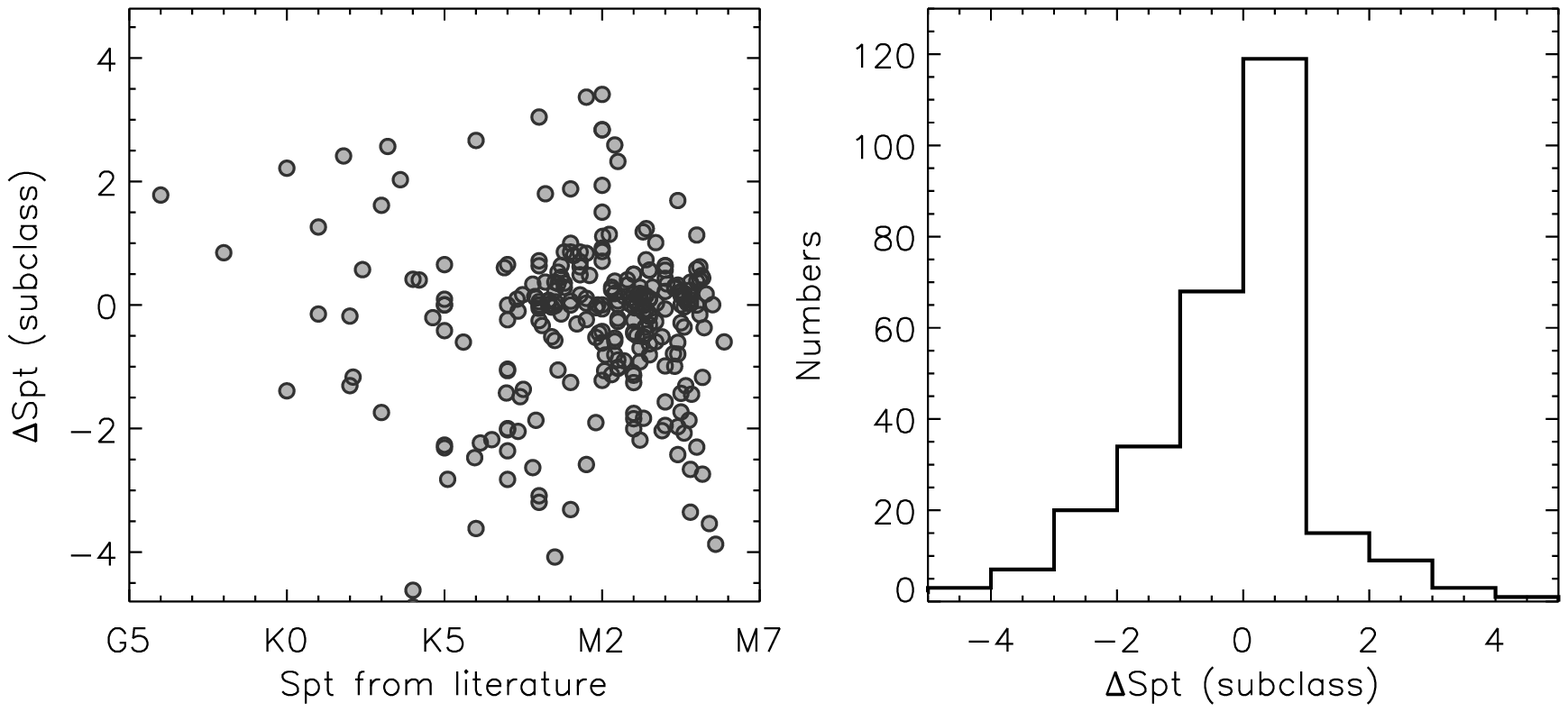}
\caption{ Left panel: the differences between the spectral types in this work and in the literature vs. their spectral types in literature. Right panel: the distribution of spectral-type differences in left panel.}\label{Fig:spt_dif}
\end{center}
\end{figure}

In Fig.~\ref{fig:spectral_sequence}, we show sample Hectospec spectra of stars with spectral types from late-G to late-M from our survey. There is a clear change in the spectral shape with spectral type,  especially the TiO strength, which is a prominent diagnostic for late K to M type stars.

\begin{figure}
\begin{center}
\includegraphics[width=1\columnwidth]{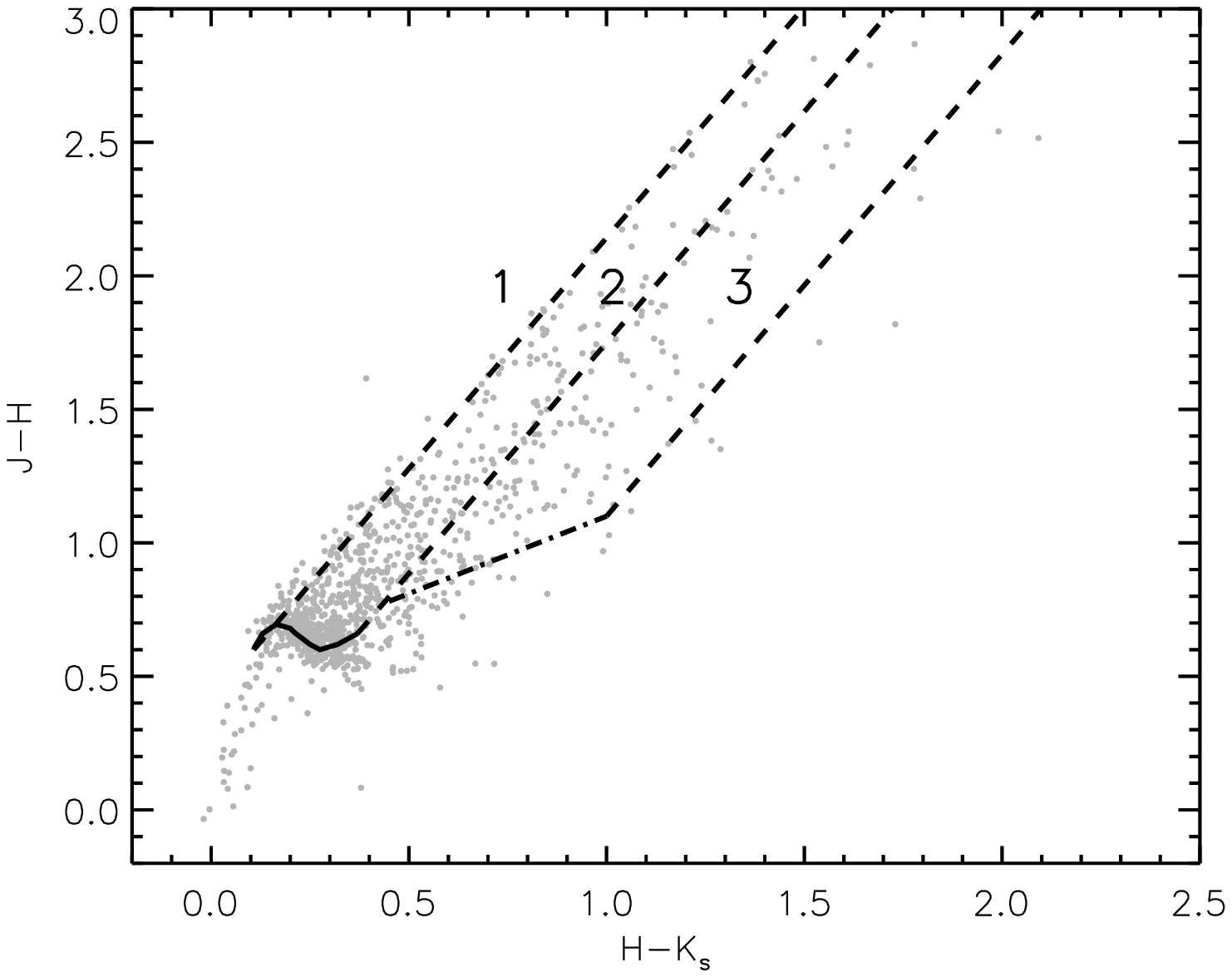}
\caption{\label{fig:twomass_ccmap} The $H$$-$$K_{\rm s}$ vs. $J$$-$$H$ color-color diagram for the young stars in L1641. The solid line show the intrinsic colors for the  main-sequence stars of $\sim$K5--M6 type \citep{1988PASP..100.1134B}, and the dash-dotted line are the locus of T~Tauri stars from \citet{1997AJ....114..288M}. The dashed lines show the reddening vectors from \citet{2007ApJ...663.1069F}. The dashed lines separate the diagram into three regions marked as numbers  1, 2, and 3, in the figure. In each region, we use different ways to estimate the extinction of the sources without spectral types.}
\end{center}
\end{figure}

Of the $\sim$430 targets with spectral-type estimates in our spectroscopic sample, $\sim${\newnewrev 290} sources have been classified in the literature  \citep[Paper\,I;][]{1995PhDT..........A,2008A&A...489.1409G,2012ApJ...752...59H}. Figure~\ref{Fig:spt_dif} shows the comparison of the spectral types in this work and those quoted  in the literature. About 90\% of this sample agrees within 1~subclass, {\newnewrev 95}\% within 2~subclasses.  

\subsection{Determining the stellar properties}\label{Sec:ana_stellar_properties}

We converted the spectral types to effective temperatures using the relation from \citet{1995ApJS..101..117K} for stars earlier than M0, and the one from  \citet{2003ApJ...593.1093L} for stars later than M0. We determined the extinction and bolometric luminosity of the central stars in the YSOs by fitting their optical and near-infrared photometry with a (reddened) model atmosphere with the effective temperature derived from spectral classification. {\lrev We adopted models by \cite{1979ApJS...40....1K,1994KurCD..19.....K} at temperatures above 4500\,K, and the MARCS models at lower temperatures  \citep{2008A&A...486..951G}.} The fitting employs two free parameters: the angular diameter ($\theta$) and the extinction ($A_{\rm V}$) in $V$ band. In general, we use the photometry in $g'$, $r'$, $i'$, $z'$ from SDSS, and in $J$ bands for our SED fit, but for stars without near-infrared excess emission we also included the $H$ and $K_{\rm s}$ bands. We calculated the synthetic photometry  by integrating the intensity of the (reddened) model atmospheres over the spectral response curve of the system for each filter, and compared them with the observations. By minimizing the $\chi^{2}$ in  an automated iterative procedure, we obtained the optimum values for $\theta$ and $A_{\rm V}$.  We  used the extinction law of \cite{1989ApJ...345..245C} to redden model atmospheres, and adopted a total to selective extinction value typical of ISM dust ($R_{\rm{V}}$=3.1).

We calculated  the bolometric luminosity of stars from the effective temperatures and angular diameters assuming a distance of 450\,pc  \citep{1981ApJ...244..884G,1982AJ.....87.1213A,1986ApJ...303..375M,2005A&A...430..523W}, using the following formulae:
 
\begin{equation}
\label{eq:Lbol}
\ \ \ \ L_{\rm{bol}} = \pi \theta^2 d^2 \sigma T_{\rm{eff}}^4
\end{equation}  

\noindent where $\theta$ is the angular diameter, $d$ is the distance, $\sigma$ is the Stefan-Boltzmann constant, and \Teff\ is the effective temperature.

With effective temperatures and bolometric luminosities, we can place the YSOs in  the HR diagram from which stellar masses and ages can be estimated by comparison to theoretical pre-main sequence (PMS) evolutionary tracks. However, several sets of such tracks exist, by various authors, which yield significantly different results on stellar ages \cite[see e.g.][for a discussion]{2008ASPC..384..200H}. We estimated masses and ages using five different sets of publicly available PMS evolutionary tracks by \cite{1997MmSAI..68..807D} (DM97), \cite{1998A&A...337..403B} (B98), \cite{2000A&A...358..593S} (S00), \cite{2008ApJS..178...89D} (D08), and \cite{2011A&A...533A.109T} (Pisa11).  In the remainder of the discussion, if not specifically mentioned, we will adopt the values obtained from the tracks of \citet{2008ApJS..178...89D}, as these have the best resolution in both mass and age. We stress, however, that there are substantial systematic differences between the different sets of tracks  \citep[see ][for a detailed discussion of the various sets of PMS evolutionary tracks available in the literature]{2004ApJ...604..741H,2008ASPC..384..200H}, and our motives for choosing those by \citet{2008ApJS..178...89D} are pragmatic.

We employ the same method as in Paper\,I to estimate the uncertainties in mass and age of the individual stars. We use a simple Monte-Carlo method to created a large number of synthetic [\Teff, $L_*$] points for each star, assuming the errors in both quantities to be normally distributed. For each point in the HR diagram, we derived the mass and age, and use the standard deviations in the resulting mass and age distribution as their uncertainties. This procedure can account for the observational errors, but the systematic uncertainties from different PMS evolution tracks remain. In addition, for highly extincted sources, the shape of the adopted extinction law can affect the resulting mass and age by affecting the stellar luminosity estimate. Higher values for the total to selective extinction $R_{\rm V}$ can lead to higher stellar luminosities, which yield younger ages for YSOs, and higher masses for YSOs with the earlier spectral types.

\begin{figure*}
\begin{center}
\includegraphics[width=2\columnwidth]{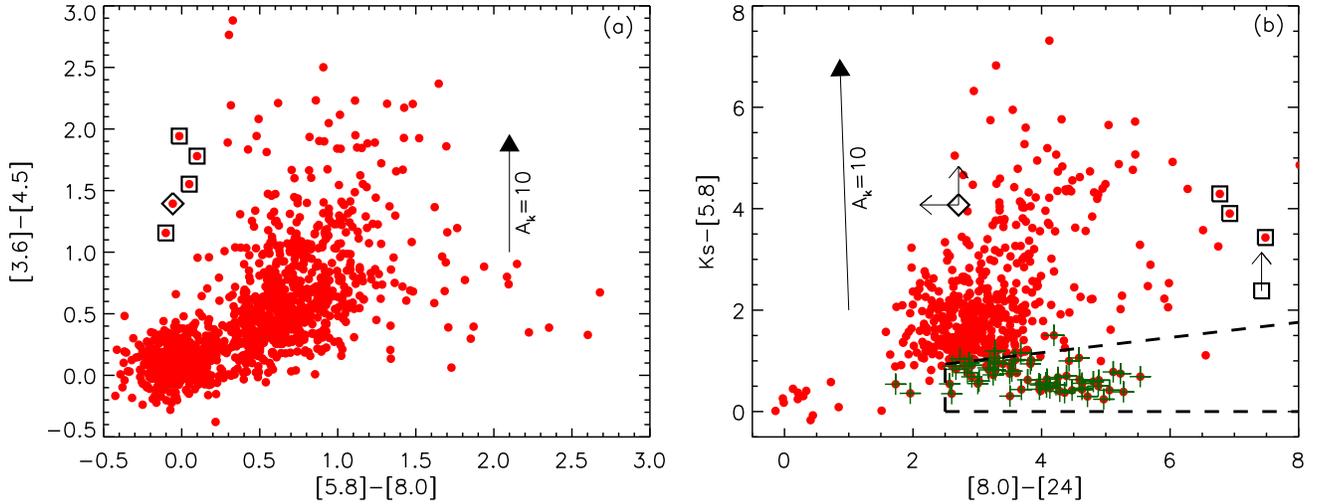}
\caption{\label{fig:allccmap} (a)[3.6]$-$[4.5] vs. [5.8]$-$[8.0] color-color diagram.  The open diamond marks the source ID\#1074, and the open boxes show the sources ID\#1218, 1220, 1233, and 1244. These five sources show [5.8]-[8.0] colors close to zero, but [3.6]$-$[4.5]$>$1.0. (b)[8.0]$-$[24] vs. $K_{\rm s}$$-$[5.8] color-color diagram. The dashed lines enclose the region for selecting TDs. The pluses mark the TDs in Paper\,I and this work. The large arrow in each panel show the reddening vector with $K_{\rm s}$-band extinction of 10\,mag. The open diamond and boxes are similar to panel(a). The small arrows close to the open symbols mark the upper or lower limits for the colors of the sources.}\end{center}
\end{figure*}

\subsection{Determining the disk properties}\label{Sec:ana_dis_properties}

\subsubsection{Infrared spectral slopes and  Classifications}\label{Sec:ana_SED_classification}

We calculate the infrared spectral slope for each YSO, defined as $\alpha=d log(\lambda F_{\lambda})/d log(\lambda)$, with the dereddened photometry in the infrared broad bands. We calculate four sets of infrared slopes, $\alpha_{2-8}$, $\alpha_{3.6-8}$, $\alpha_{2-24}$, and $\alpha_{3.6-24}$, corresponding to the spectral range of $K_{\rm s}$ to [8.0], [3.6] to [8.0], $K_{\rm s}$ to [24], and [3.6] to [24], respectively. The extinction of each YSO is estimated individually.  For the YSOs with spectral-type measurements, we can  accurately estimate their extinction by fitting their optical and near-infrared photometry as described in \sek~\ref{Sec:ana_stellar_properties}. However, for the YSOs without spectral types, we can only roughly determine their extinction with a method similar to that used by \citet{2008ApJ...674..336G} in which the extinction is obtained employing the  $H$$-$$K_{\rm s}$ vs. $J$$-$$H$ color-color diagram.  We describe the method  as follows.

The location of each YSO in the $H$$-$$K_{\rm s}$ vs. $J$$-$$H$ color-color diagram depends on both its intrinsic colors and its extinction. To estimate the extinction of individual YSOs, we need to have knowledge of their intrinsic colors. For the diskless YSOs, their intrinsic colors  are mainly determined by their spectral types. For YSOs with disks, the excess emissions from hot inner regions of disks can also contribute to their intrinsic colors and make them redder than diskless stars. The CTTS locus in the $H$$-$$K_{\rm s}$ vs. $J$$-$$H$ color-color diagram is observationally defined as the colors of  dereddened classical T~Tauri stars. In consideration of these different origins of intrinsic colors of YSOs,  we divide the diagram into 3 sub-regions (see Fig.~\ref{fig:twomass_ccmap}). Within each region, we use a different way to obtain the intrinsic color [$J$$-$$H$]$_{0}$ of the sources. Extinction values of individual YSOs are estimated from their observed  [$J$$-$$H$] ([$J$$-$$H$]$_{obs}$)  and [$J$$-$$H$]$_{0}$ using the  equation: $A_{\rm K}$=([$J$$-$$H$]$_{obs}$$-$[$J$$-$$H$]$_{0}$)/0.95. In Region\,1, the sources  show slightly bluer [$J$$-$$H$] colors than expected from normal reddening of diskless YSOs, probably due to the photometric uncertainties. For such YSOs, their [$J$$-$$H$]$_{0}$ values  are simply assumed to be 0.6 (the typical value for a K5-type dwarf star). In Region\,2, the color excess of each source is mainly due to reddening. Its intrinsic [$J$$-$$H$]  color is obtained from the intersection between the reddening vector and the locus of main-sequence stars\footnote{For simplicity, we force [$J$$-$$H$]$_{0}$$\ge$0.6 as done in \citet{2008ApJ...674..336G}}. In Region\,3, the sources are supposed to be the reddened classical T~Tauri stars. The intrinsic [$J$$-$$H$] color is derived from where the reddening vector and the CTTS locus  intersects. For the stars outside of these three regions, or without detection in all the three 2MASS bands, their extinction is assumed to be zero. The extinction values derived from the $H$$-$$K_{\rm s}$ vs. $J$$-$$H$ color-color diagram are less accurate than those of YSOs with spectral types. {\newnewrev For a sample of spectroscopic YSOs with substantial extinction ($A_{\rm v}$$>$1), we compare the extinctions estimated from the above rough method with the more accurate ones described in \sect\ref{Sec:ana_stellar_properties}. We find the extinctions from the  rough method are statistically slightly lower than the relatively accurate ones.} We estimate how the extinction affect the spectral slopes. For $A_{\rm K}$=1, the dereddened $\alpha$$_{2-8}$, $\alpha$$_{3.6-8}$, $\alpha$$_{2-24}$, and $\alpha$$_{3.6-24}$ are 0.36, 0.14, 0.15, and 0.04 less than the observed spectral slopes, respectively. Thus, given the typical extinction for our YSOs in L1641, extinction only slightly influences our estimates of the spectral slopes of the YSOs, except for a few extremely embedded YSOs.

We classify the YSOs into different classes according to the two spectral slopes, i.e. $\alpha$$_{3.6-8}$ and $\alpha$$_{3.6-24}$.  For the sources with MIPS 24\mum\ detection, we classify the sources as diskless stars if $\alpha$$_{3.6-24}$$\le$$-2.2$, Class\,II sources if $-2.2$$<$$\alpha$$_{3.6-24}$$<$$-0.3$, Flat-spectrum when  $-0.3$$\le$$\alpha$$_{3.6-24}$$\le$$+0.3$, and Class\,I source if $\alpha_{3.6-24}$$>$$+0.3$ \citep{2008ApJ...675.1375L}. A part of YSOs have not been detected at 24\mum. For these YSOs, we use the dereddened infrared slope ($\alpha$$_{3.6-8}$) to classify the young stars into Class\,III if $\alpha$$_{3.6-8}$$\le$$-2.56$, Class\,II if $-2.56$$<$$\alpha$$_{3.6-8}$$<$$-0.3$,  Flat-spectrum when  $-0.3$$\le$$\alpha$$_{3.6-8.0}$$\le$$+0.3$, and Class\,I sources if $\alpha$$_{3.6-8.0}$$>$$+0.3$. For 857 YSOs with reliable estimates of spectral types, their classes will be revised by comparing their reddened model photospheric fluxes with their observed fluxes at infrared wavelengths.

\subsubsection{Accretion rates}\label{sec:ana_acc}
Gas from the disk accretes onto the star along the magnetic field lines and hits the stellar surface at approximately the free-fall velocity, causing a strong accretion shock and associated hot-spots on the stellar surface. Various emission lines, such as the hydrogen Balmer series, \HeI\,5876~\AA, Br$\gamma$, etc., are formed in the infalling magnetospheric flow \citep{1994ApJ...426..669H,1998ApJ...492..323G,2001ApJ...550..944M,2010A&A...522A.104L,2011MNRAS.416.2623K}. Optical/ultraviolet excess continuum emission is produced in the accretion shocks. All these emission lines or excess emission can be used to estimate the accretion rates  with the help of models or the empirical relation between line luminosity and accretion luminosity \citep[Paper\,I;][]{1994ApJ...426..669H,1995ApJ...452..736H,1998ApJ...492..323G,2001ApJ...550..944M,2004AJ....128.1294C,2004A&A...424..603N,2008ApJ...681..594H}.

We use two methods to estimate the accretion rates: (1) the empirical relation between accretion luminosity and  H$\alpha$ or  H$\beta$  lines luminosity from Paper\,I, (2) the empirical relation between  the full width of H$\alpha$ at 10\% ($FW_{H\alpha, 10\%}$) of the peak intensity and accretion rates from \cite{2004A&A...424..603N}.

In the former method, we estimate the accretion rates from the observed H$\alpha$ and H$\beta$ line luminosity. The H$\alpha$ and H$\beta$  line luminosity  are calculated by integrating over the line profile, adopting the best fit model atmosphere spectrum (see \sek~\ref{Sec:ana_stellar_properties}) as the continuum level. The line luminosities are converted to the accretion luminosity via the empirical relation given in Paper\,I:
\begin{equation}
log(L_{\rm{acc}}/L_{\odot})=(2.27\pm0.23)+(1.25\pm0.07)\times log (L_{\rm{H}\alpha}/L_{\odot})
\end{equation}
\begin{equation}
log(L_{\rm{acc}}/L_{\odot})=(3.01\pm0.19)+(1.28\pm0.05)\times log (L_{\rm{H}\beta}/L_{\odot})
\end{equation}

The inferred accretion luminosities are then converted into mass accretion rates using the following relation:

\begin{equation}
\dot{M}_{\rm acc}=\frac{L_{{\rm acc}}R_{\star}}{{\rm G}M_{\star} (1-\frac{R_{\star}}{R_{\rm in}})},
\end{equation}

\noindent where $R_{\rm in}$ denotes the truncation radius of the disk, which is taken to be 5\,$R_{\star}$ \citep{1998ApJ...492..323G}. G is the gravitational constant, $M_{\star}$ is the stellar mass as estimated from  the location of each star in the HR~diagram, and $R_{\star}$ is the stellar radius derived using the fitting procedure described in \sect\ref{Sec:ana_stellar_properties}.   

For the  YSOs with high-resolution H$\alpha$ spectroscopy, {\newnewrev we distinguish accretors from non-accretors based on the criterion ($FW_{H\alpha, 10\%}$$>$250\kms) described in Appendix~\ref{Appen:Halpha}}, and estimate the accretion rates with the empirical relation between accretion rates and $FW_{H\alpha, 10\%}$, using the formula given by \citet{2004A&A...424..603N}: 

 \begin{equation}
  Log \dot{M}_{acc}=-12.89(\pm0.3)+9.7(\pm0.7)\times10^{-3} FW_{H\alpha, 10\%}
 \end{equation}

{\newnewrev A comparison of accretion rates derived from the above two methods is discussed in Appendix~\ref{Appen:Halpha_lum_FW}.}

\section{Results}

\subsection{YSOs in L1641}\label{Sec:census}
In this section, we will perform a census of YSOs in L1641 based on the data collected from several projects. Then, we will calculate the infrared spectral slopes of these YSOs, and classify them into different evolutionary stages.

\subsubsection{A census of YSOs}
As one of the best studied star-forming regions in the literature, L1641 has been observed by various projects, including infrared imaging surveys \citep{1993ApJ...412..233S,2012AJ....144..192M}, spectroscopic surveys \citep[Paper\,I;][]{1995PhDT..........A}, and an X-ray emission survey \citep{2009AIPC.1094..959W,2009A&A...493..339W} (see FOVs of each survey in Fig.~\ref{Fig:yso_dis}). However, there has no been no available census of YSOs in this region in the literature. In this work, we provide an inventory of YSO candidates according to the following selection criteria: (1) confirmed with spectroscopy in Paper\,I and this work, (2) showing infrared excess emission, or (3) showing X-ray emission. A source is considered as a YSO if it obeys one of the above criteria. We have identified 1247 YSOs.

{\newnewrev Recently \citet{2012ApJ...752...59H} presented a catalog of $\sim$860  young stars in L1641 confirmed with optical spectroscopy, among which $\sim$720 stars fall within the region that we study in this paper. We compare our YSO sample with that of  \citet{2012ApJ...752...59H}, and find  $\sim$580 common sources in both catalog. About 140 YSOs in the catalog of \citet{2012ApJ...752...59H} are absent in our census. Among them, $\sim$120 sources are diskless stars, and $\sim$20 sources are YSOs with the infrared excess emission. We did not identify these diskless YSOs because they do not show X-ray emissions and have not been observed with spectroscopy prior to the survey of \citet{2012ApJ...752...59H}. The 20 disked YSOs are located close to the edge of the coverage of the Spitzer survey, and lack photometry in several IRAC bands in our previous photometric catalog. Here, we include all these additional YSOs in our YSO catalog.} In table~\ref{tab:allYSO_L1641_optical} and ~\ref{tab:allYSO_L1641_infrared}, we list the YSO criteria that they abide by as well as their optical and infrared photometry.

In our YSO catalog, there are 1273 sources with photometry in all four IRAC bands. Figure~\ref{fig:allccmap}(a) shows their [3.6]$-$[4.5] vs. [5.8]$-$[8.0] color-color diagram. The sources in the color-color diagram fall into two populations. A major fraction of stars are located towards the top-right of the origin due to infrared excess emission from  hot dust in the inner disk, while other sources reside in a cluster near the origin, consistent with photospheric infrared colors. In Fig.~\ref{fig:allccmap}(a), there are five sources (ID\#1074, 1218, 1220, 1233, and 1244) showing IRAC colors distinct from the others. Those sources will be discussed in \sect\ref{sec:results:blue_source}.

Figure~\ref{fig:allccmap}(b) shows the $K_{\rm s}$$-$[5.8] vs. [8.0]$-$[24] color-color diagram. Due to the lower sensitivity of MIPS, only a few diskless stars are detected at 24\mum, and appear around the origin in Fig.~\ref{fig:allccmap}(b). A major fraction of YSOs detected at 24\mum\ are harboring optically-thick inner disks and are located within $K_{\rm s}$$-$[5.8]$>1$, and [8.0]$-$[24]$>$2. There is a minor group of YSOs  with $K_{\rm s}$$-$[5.8]$\lesssim$1, and [8.0]$-$[24]$\gtrsim$2.5, indicating these sources show very weak or no excess emission at the shorter IRAC bands, but strong excess emission at 24\mum,  characteristic of TDs (see Paper\,I). The five unusual sources in Fig.~\ref{fig:allccmap}(a)
are also marked Fig.~\ref{fig:allccmap}(b). Among them, two sources (ID\#1074 and 1233) are not detected in both the 2MASS survey and the UKIDSS survey, and an upper $K_{\rm s}$-band limiting magnitude of 15.3 is assumed for them. The source ID\#1074 is not detected at 24\mum, and its upper limiting magnitude at 24\mum\ is used for the plot.

\subsubsection{The completeness and contamination of YSO census}\label{Sec:completeness}

The datasets used in our YSO census consist of different surveys, each of which has slightly different sky coverage (see Fig.~\ref{Fig:yso_dis}), and different sensitivities to detect the YSOs. The XMM survey is more sensitive to Class\,III sources than Class\,II sources \citep{1999ARA&A..37..363F}, and insensitive to the intermediate mass stars \citep{2003ApJ...584..911F}. The Spitzer imaging survey can be used to identify YSOs with circumstellar disks. The spectroscopic surveys can be used to confirm the youth properties of stars, but are limited by extinction. Thus, it is difficult to quantify the completeness of our YSO census. Here, we only qualitatively evaluate it based on the K-band luminosity function (KLF) of our YSO sample.

\begin{figure}
\begin{center}
\includegraphics[width=1.\columnwidth]{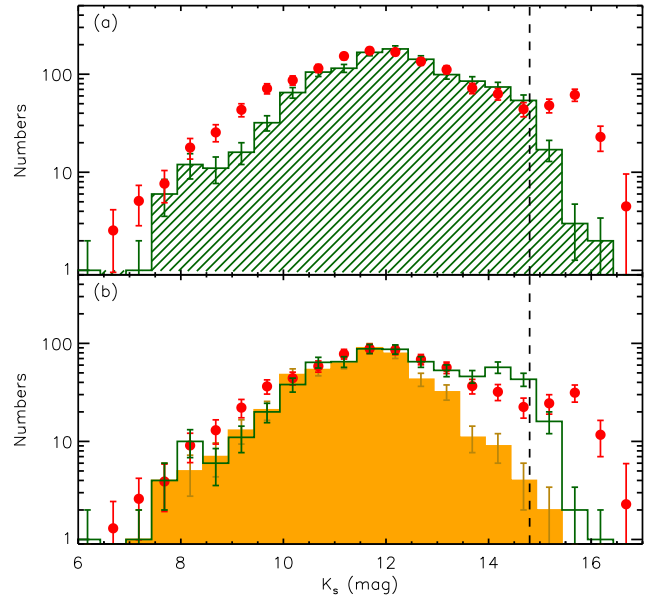}
\caption{\label{fig:KLF} (a) The K-band luminosity functions (KLF) for all YSOs in L1641 (hatched histograms) and in the Trapezium cluster (filled circles).  The Trapezium  KLF  from \citet{2002ApJ...573..366M}  has been shifted from 414\,pc \citep{2007A&A...474..515M} to 450\,pc, and scaled to match the peak of the KLF of L1641. The dashed line marks the completeness limit of the 2MASS survey. {\newnewrev (b) The K-band luminosity functions (KLF) for YSOs with infrared excess (open histograms), and for YSOs with X-ray emission (filled histograms) in L1641. The scaled KLF in the Trapezium cluster (filled circles)  is also shown as comparison.}}\end{center}
\end{figure}

The shapes of KLFs of young clusters are sensitive to their initial mass functions (IMF) and the mean ages \citep{2000ApJ...533..358M}, and thus can be used to constrain these properties \citep{1993ApJ...412..233S,1995AJ....109.1682L,2002ApJ...573..366M,2003AJ....125.2029M}. In our YSO catalog, 84\% of the YSOs have been detected in the 2MASS survey. In Fig.~\ref{fig:KLF}(a), we show the KLF of these YSOs. The KLF of L1641 peaks at $\sim$12\,mag (see Fig.~\ref{fig:KLF}(a)), consistent with the KLFs of $\sim$1\,Myr old clusters \citep{2000ApJ...533..358M}. As a comparison we also show the KLF (with the $K$-band completeness limit of $\sim$17.5\,mag) of the Trapezium cluster which is at an age ($\sim$1\,Myr) similar to L1641 \citep{2002ApJ...573..366M,1997AJ....113.1733H}. With the assumption that L1641 and Trapezium  have a similar IMF, one would expect they show a similar KLF shape. Thus, the comparison of the KLFs of L1641 and Trapezium can be used to qualitatively evaluate the completeness of our YSO census in L1641.

In general, the KLF of L1641 (see Fig.~\ref{fig:KLF}(a)) fits to that of the Trapezium cluster within the range of 10 to 15\,mag, corresponding to a mass range of 1.0-0.04\,\Msun\ for 1\,Myr PMS stars\citep{2008ApJS..178...89D,1998A&A...337..403B}. However, at the faint ($\gtrsim$15\,mag) and bright (8 to 10\,mag) ends of the KLF, L1641 appears to be deficient in the stars, compared to the Trapezium cluster. The lack of faint YSOs in L1641 is mainly attributed to incompleteness of the 2MASS survey. In our YSO catalog, there are 16\% of the YSOs which are not detected by the 2MASS survey. These objects can partially contribute to the faint part of the KLF of L1641. Within the range of 8 to 10\,mag, corresponding to intermediate mass 1\,Myr PMS stars \citep{2008ApJS..178...89D}, there is another deficiency in YSOs, which could be attributed to the fact that intermediate-mass stars usually show very weak X-ray emission, and would not be picked up by the XMM survey \citep{2003ApJ...584..911F}. These intermediate-mass YSOs can be detected in the optical and infrared imaging surveys. But if they have lost their circumstellar disks, they cannot be identified as YSOs according to their infrared colors. The comparison of the KLFs of L1641 and Trapezium suggests our YSO census may lack $\sim$60 intermediate-mass diskless stars in L1641. Recently, \citet{2012ApJ...752...59H,2013ApJ...764..114H} have proposed that L1641 and  the Trapezium cluster may have different IMFs, and L1641 is deficient in the stars at the upper mass end of IMF, compared with that of the Trapezium cluster. Thus, the actually numbers of missing intermediate-mass diskless stars  in our YSO census may be less than that we expected.

\begin{figure*}
\centering
\includegraphics[width=1.8\columnwidth]{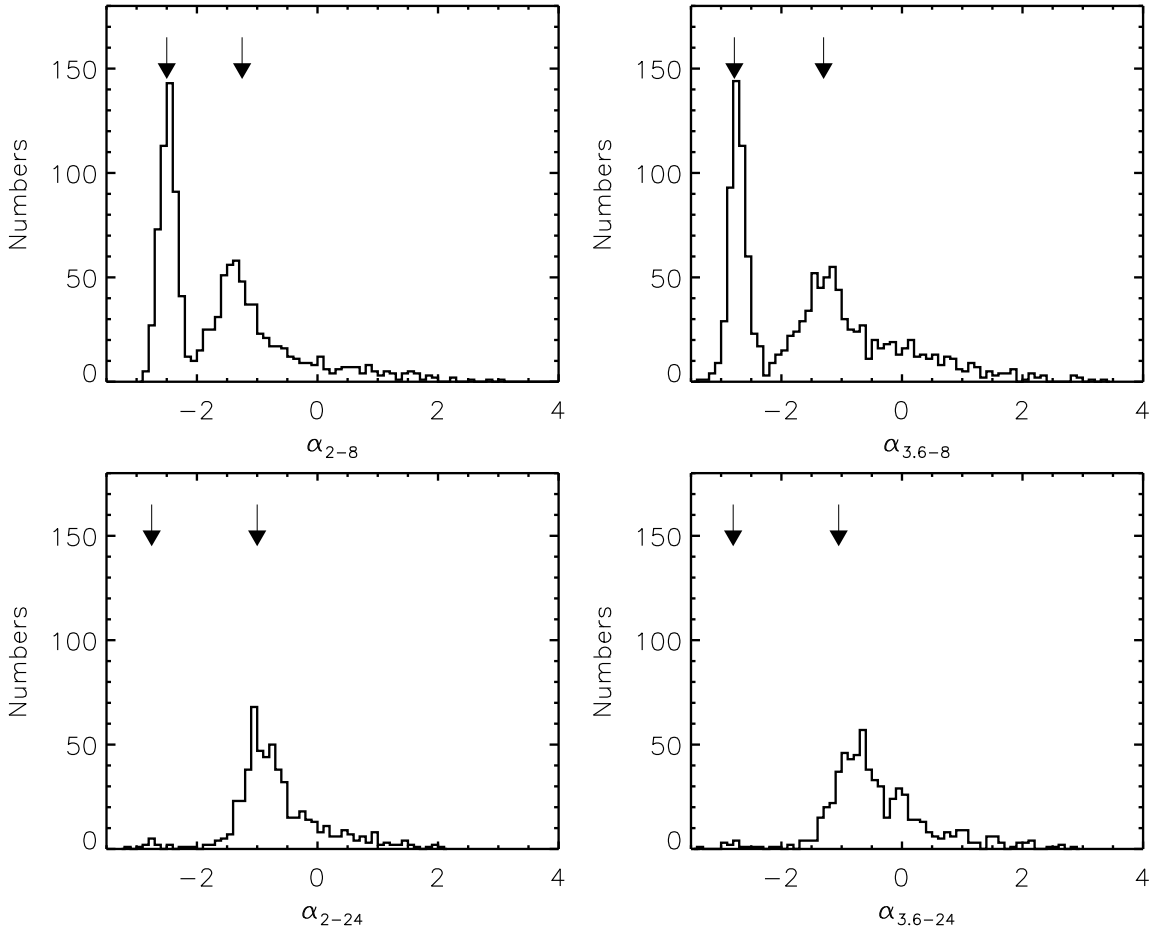}
\caption{\label{fig:alpha} Distributions of the dereddened spectral slopes for YSOs in L1641. The arrows in each panel mark the peak positions of the distributions of the spectral slopes for YSOs in Taurus \citep{2010ApJS..186..111L}.}\end{figure*}

 In Fig.~\ref{fig:KLF}(b) we show the KLFs of two populations of YSOs in our sample: one for the young stars with infrared excesses, and the other for those with X-ray emissions. It can be noted that both KLFs are similar when $K_{\rm s}$$<$12.5\,mag, corresponding to a 1\,Myr PMS stars with masses$\gtrsim$0.2\,\Msun, though both populations only share the common 159 stars with  $K_{\rm s}$$<$12.5\,mag. At the faint end of KLF ($K_{\rm s}$$>$12.5\,mag), X-ray emitting YSOs are clearly deficient in stars compared with the YSOs with infrared excesses, which could be due to the incompleteness of the XMM survey. We evaluate the completeness of the XMM observations in L1641 using the tool, Flux Limits from Images from XMM-Newton (FLIX)\footnote{FLIX is a on-line tool provided by the XMM-Newton Survey Science Center (see http://www.ledas.ac.uk/flix/flix.html). It provides robust estimates of the X-ray upper limit to a given point in the sky where there are no sources detected in the 2XMMi catalog. See \citet{2007A&A...469...27C} for a detail discussion of the upper limit algorithm.}, which can be used to roughly estimate the upper limit from the observed XMM image. We used the FLIX tool to estimate the 5-$\sigma$ upper limit of the 0.2--2\,keV X-ray luminosity in the field of L1641. We found that the 5-$\sigma$ upper limits vary from region to region with a typical value $\sim$2.4$\times$10$^{29}$\,\ergps\ at the distance ($\sim$450\,pc) of L1641. According to the relation between the X-ray luminosity and stellar mass for the young stars in Taurus ($\sim$1.5\,Myr) from \citet{2007A&A...468..353G}, the typical 5-$\sigma$ upper limit of the X-ray observations corresponds to the X-ray luminosity of 1.5\,Myr old PMS star with masses of 0.2--0.3\,\Msun, which is consistent with the estimate of incompleteness from the KLF of X-ray sources. In the FOVs of the XMM observations, we have identified 395 Class\,III sources with spectroscopy. Among them, 253 sources are brighter than 12.5\,mag at $K_{\rm s}$ band, and $\sim$78\% (197/253) of them show X-ray emissions. The fraction of X-ray emitting sources among the faint Class\,III sources ($K_{\rm s}$$>$12.5\,mag) decrease to $\sim$38\% (54/142). On one hand these statistics confirm the results on the completeness of the YSOs selected from the X-ray data  estimated using FLIX and KLF; on the other hand they indicate that the spectroscopic surveys have partially relieved the incompleteness of Class\,III sources suffered from the X-ray survey.


{\lrev  We estimate the approximate completeness of our YSO census based on the study of \citet{2012ApJ...752...59H} in which the targets for the spectroscopic survey are selected from $V$ vs. $V$$-$$I$ color-magnitude diagram. The spectroscopic survey of \citet{2012ApJ...752...59H} is complete to  $V$$\lesssim$21\,mag for disk population, and has $\sim$95\% completeness for diskless population till $V$$\lesssim$21\,mag, which corresponds to a 1.5\,Myr old, $\sim$0.2\,\Msun\ PMS star with $A_{\rm V}$$\sim$3. The faint disk population  ($V$$\gtrsim$21\,mag), most of which are heavily extincted objects, are below their detection limit, but can easily be detected in the Spizter imaging survey. In our catalog, there are 533 Class\,II sources. Among them, 330 Class\,II sources are in \citet{2012ApJ...752...59H}. The other 203 Class\,II sources are too faint to be observed. In the same field, \citet{2012ApJ...752...59H} have found 388 Class\,III sources. If we assume that both Class\,II and Class\,III sources have a similar fraction of faint sources, which are unobservable in the spectroscopic survey of \citet{2012ApJ...752...59H}, the total number of Class\,III sources is expected to be $\sim$660. In our YSO catalog, 507 Class\,III sources are inventoried. Thus, the completeness of our YSO census is estimated to be $\sim$90\% (1388/1541). In addition, we can evaluate the completeness of our YSO sample using the $r$ vs. $r$$-$$i$ color-magnitude diagram (see the detail discussion in Appendix~\ref{Appen:CMD}), and find that our census is complete to $r$$\lesssim$20 which corresponds to a 1.5\,Myr old, 0.2\,\Msun\ PMS star with an extinction of $A_{\rm V}$$\sim$3.}




  {\newnewrev Our YSO catalog may be contaminated by a small fraction of extragalactic sources and nearby foreground stars. These contaminators show similar SEDs to those of YSOs or X-ray emissions, thus are selected as YSO candidates. The main  extragalactic contaminators include star-forming galaxies and AGNs \citep{2008ApJ...674..336G}. According to the criteria of \cite{2008ApJ...674..336G}, which are used to identify potential extragalactic contaminators based on the IRAC photometry and colors, 92 among 1273 sources with detection in four IRAC bands are suggested to be contaminators. Among the 92 ``contaminators'', 13 are observed with spectroscopy and confirmed as young stars, thus excluded as  the contaminators. Among the remaining 79 sources without spectroscopy, 66 ``contaminators''  show the SEDs characteristic shape of classI/Flat YSOs, and 13  are classified as Class\,II sources. Thus, statistically about 6\% (79/1273) of our YSO candidates could be extragalactic contaminators. This, however, is a conservative estimate: a few of these ``contaminators'' are located at the dense regions of the L1641 cloud, thus could be real YSOs. The foreground stars can also contaminate our YSO catalog. These stars show detectable X-ray emission, thus being wrongly classified as Class\,III sources. The typical X-ray luminosity of field main-sequence stars is several$\times10^{27}$\,\ergps\ for solar-type to M-type main sequence stars\citep{2004A&ARv..12...71G}. According to the 5-$\sigma$ upper limit of the 0.2--2\,keV X-ray luminosity in the XMM survey of L1641 estimated with FLIX, XMM observations can detect the field stars within the distance$\lesssim$90\,pc. Using the Besan\c{c}on\ model of stellar population synthesis of the Galaxy \citep{2003A&A...409..523R}, we obtain small numbers ($\sim$15) of stars within the  distance$\lesssim$90\,pc in the direction of L1641, suggesting the faction of contaminators from foreground stars in our YSO catalog is about 1\%.}

  {\lrev Recently \citet{2012A&A...547A..97A} proposed that there is a large cluster, NGC\,1980, in front of Orion~A cloud. They separated foreground stars and young stars in the Orion~A region  based on an assumption that young stars in Orion~A are reddened and foreground stars have negligible amount of extinction. They found that the foreground sources are mainly centered on NGC\,1980. The extent of their suggested foreground population reach northern part of L1641, and therefore could contaminate our YSO sample in L1641. The age of NGC 1980 cluster is estimated to be $\sim$4--5 Myr, which is much older than the median age ($\sim$1 Myr) of the stellar population in our L1641 sample \citep[][this work]{2009A&A...504..461F}.  The Kinematics of the stellar population in NGC\,1980 is similar to that of young stars and molecular gas in Orion~A, suggesting NGC\,1980 may be physically associated with Orion~A. It is possible that our sample could be contaminated with older population of NGC\,1980, which could result in older median age and lower disk fraction. We inspect the contamination of these ``old'' sources into our YSO census by comparing the ages of sources with $A_{\rm V}$$<$0.1 and those with $A_{\rm V}$$>$0.5. The median ages of the two samples are 1.6\,Myr and 1.3\,Myr, respectively. Thus we expect that the contamination in our YSO sample is not significant. }

\subsubsection{Infrared spectral slopes}
Figure~\ref{fig:alpha} shows the distributions of the dereddened spectral infrared slopes,  $\alpha$$_{2-8}$,  $\alpha$$_{3.6-8}$, $\alpha$$_{2-24}$, and $\alpha$$_{3.6-24}$, for the YSOs in L1641. {\newrev The disked and diskless objects appear in two distinct peaks in four panels in Fig.~\ref{fig:alpha}, although only a few diskless YSOs are detected at 24 microns.} As a comparison, in  Fig.~\ref{fig:alpha} we also show the peak positions of such distributions for YSOs in Taurus, which is at the similar age to L1641.  Both star-forming regions show similar peak positions for the disk population, suggesting that the disks are at a similar evolutionary stage.

Using the spectral slopes  $\alpha$$_{3.6-8}$ and $\alpha$$_{3.6-24}$, we classify our sources into Class\,III, Class\,II, Flat-spectrum, and Class\,I and list their  types in  \tab~\ref{tab:allYSO_L1641_infrared}, according to the criteria described in \sect\ref{Sec:ana_SED_classification}.  We have {\newnewrev 507}  Class\,III sources, {\newnewrev 533}  Class\,II sources, {\newnewrev 131}  Flat-spectrum sources, and {\newnewrev 143} Class\,I  sources. In Fig.~\ref{Fig:yso_dis}, we show their spatial distributions.  The ratios of Class\,I to Class\,II, and Flat-spectrum to Class\,II  are 0.27 and 0.25, respectively, comparable to those ($\sim$0.27 and $\sim$0.20, respectively) in the Spitzer c2d Legacy survey.

\begin{figure}[htbp]
\begin{center}
\includegraphics[width=0.9\columnwidth]{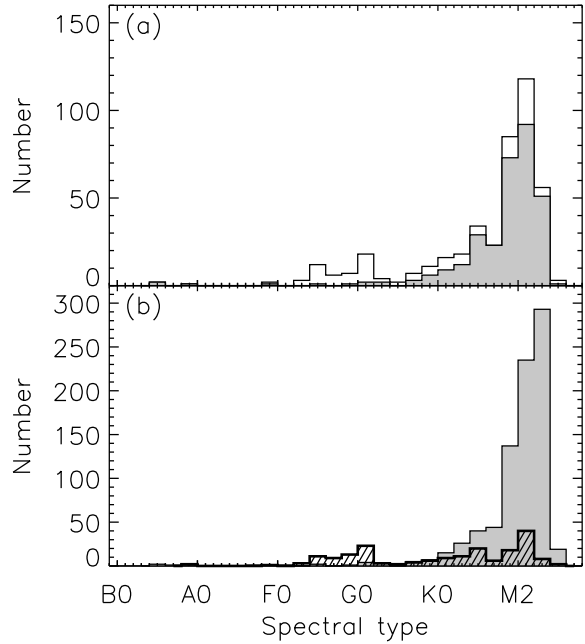}
\caption{\label{fig:spt_dis} (a) The spectral-type distribution of our sample observed with Hectospec in L1641. The open histograms show the distribution of all the stars with reliable spectral types. The filled histograms display the YSO distribution. (b) The grey-color filled histograms show the spectral-type distribution of {\newnewrev all known YSOs with spectral types in L1641}. The hatched histograms display the spectral-type distributions of field stars identified in both works.}
\end{center}
\end{figure}

\begin{figure}
\begin{center}
\includegraphics[width=\columnwidth]{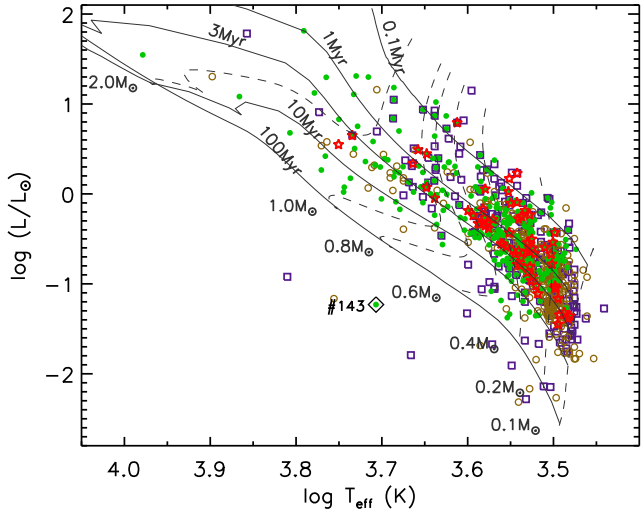}
\caption{The HR diagram for the YSOs in L1641. The filled circles are for the YSOs in our spectroscopic survey with Hectospec,  the open boxes for the YSOs in Paper\,I, {\newnewrev and the open circles for the YSOs in \citet{2012ApJ...752...59H}}. The {\lrev star symbols} represent the  TDs in L1641. The open diamond marks the new subluminous object (See \sek~\ref{sec:results:underlum}). The PMS evolutionary tracks are from \citet{2008ApJS..178...89D}.}\label{Fig:HRD}
\end{center}
\end{figure}

\subsection{Spectroscopic sample}
We have obtained optical spectra for a large sample of the YSOs in L1641 with Hectospec and Hectochelle. In this section, we will present the results from these observations.

\subsubsection{Spectral types}
We determined the spectral type of each star in our spectroscopic sample as described in \sect\ref{sec:analysis_spt}. We have $\sim$400 stars with reliable estimates of their spectral types. Among them, $\sim$300 sources are identified as young stars according to the criteria in \sect\ref{sec:analysis:YSO_selection_cirteria}, and others as field stars. The spectral types of the YSOs are listed in \tab~\ref{Tab:spt}. In \fig\ref{fig:spt_dis}(a) we show a histogram of spectral types for all stars, including field stars and YSOs, in our Hectospec survey. The distribution is bimodal, with peaks around spectral type G0 and mid-M. In this figure, we also show the spectral-type distribution of the identified YSOs. These YSOs are typically K or M type with few stars earlier than K0 type. We combined the YSOs and field stars with spectral types {\newnewrev in the field of L1641}, and  display their spectral-type distributions in \fig\ref{fig:spt_dis}(b). The field stars show a bimodal distribution with peaks around spectral type G0 and early~M, while the YSOs peak around early~M.

\subsubsection{Stellar masses and ages}
\begin{figure}
\begin{center}
\includegraphics[width=\columnwidth]{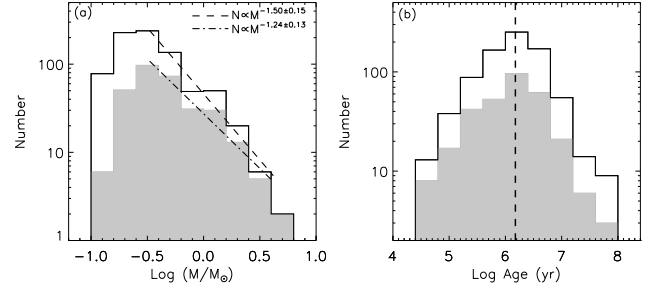}
\caption{(a): the mass distribution of the YSOs with spectral types (filled histograms) in this work, and all the YSOs (open histograms) in L1641 with spectral types. A linear regression fit of the mass spectra in the range of $-$0.5\,$\leq$\,log($M_{\star}$/M$_{\odot}$)\,$\leq$\,0.5 gives slopes of $-$1.24$\pm$0.13 and $-$1.50$\pm$0.15 for each sample, respectively. (b): The age distribution of the YSO sample in L1641. The filled and open histograms are for the same YSO samples shown in the left panel. The dashed line shows the median age for all the YSOs in L1641 with spectral types.}\label{Fig:mass_age}
\end{center}
\end{figure}

With the determined  effective temperatures and bolometric luminosities as described in \sect\ref{Sec:ana_stellar_properties}, we can place the young stars  in the HR diagram. In Fig.~\ref{Fig:HRD}, we show the HR~diagrams for the YSOs with spectral types. Most of our YSOs lie between the 0.1 and 3\,Myr isochrones. We use distinct symbols (open star symbols) for objects with SEDs typical for TDs (see also \sect\ref{sec:results:transition_disk}), as well as a new ``exotic'' object that is apparently under-luminous(see \sect\ref{sec:results:underlum}). In \fig\ref{Fig:mass_age}(a), we show the mass distribution (filled histograms) for the YSOs in this work, and the mass  distribution (open histograms) when adding the YSOs {\newnewrev with spectral types in the literature}. The median masses are $\sim$0.40~M$_{\odot}$  and 0.30\,M$_{\odot}$ for the two populations, respectively.  A linear regression fit to the mass functions within the mass bin $-$0.5\,$\leq$\,log($M_{\star}$/M$_{\odot}$)\,$\leq$\,0.5 yields slopes of $-$1.24$\pm$0.13 and $-$1.50$\pm$0.15 for the two populations, respectively.  In \fig\ref{Fig:mass_age}(b), we show the age distribution of the sample two populations as in \fig\ref{Fig:mass_age}(a). The median age for the entire sample is 1.5\,Myr.

\begin{figure}
\begin{center}
\includegraphics[width=\columnwidth]{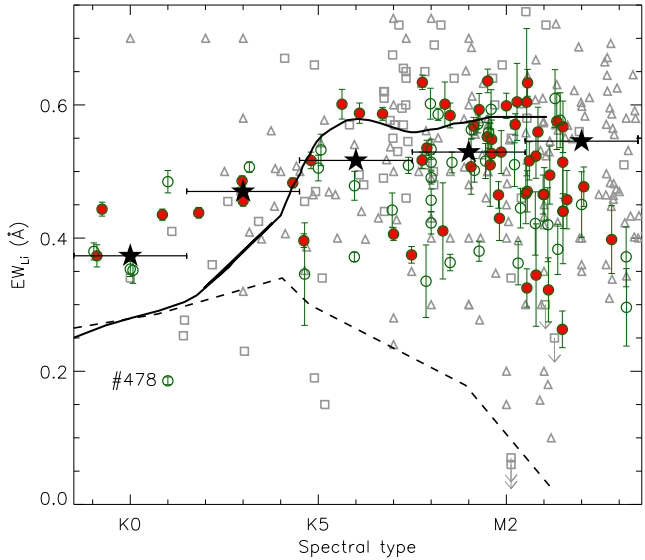}
\caption{ \LiI\ $EW$ versus spectral type. The open circles and the red-filled  circles are for the CTTSs and the WTTSs, respectively,  in this work. The uncertainties of the $EWs$ is shown with error bars. The open triangles show the YSOs in L1630N and L1641 (Paper\,I). The open boxes display the YSOs in Taurus \citep{2008A&A...487..965S}. The  {\lrev star symbols} show the median $EW$ for individual spectral-type bins in the three regions and the bar centered on each {\lrev star symbol} shows the bin size. The solid line shows the model locus for the YSOs with undepleted Li \citep{2002A&A...382..563B,2009MNRAS.400..317J}. The dashed line displays the boundary for the Tucanae-Horologium association with an age of 27$\pm$11\,Myr \citep{2008ApJ...689.1127M}.}\label{Fig:Li_spt}
\end{center}
\end{figure}

Lithium depletion trends in young PMS stars can be used to estimate the stellar ages  at sub-solar masses \citep{2008ApJ...689.1127M,2009MNRAS.400..317J}. In Fig.~\ref{Fig:Li_spt} we show the equivalent width ($EW$) of the \LiI\ absorption line as a function of spectral type for the YSOs in our sample. The  $EW$ of the \LiI\ absorption line for each YSO is derived from the Hectochelle spectra. Some YSOs have been observed at multiple epochs with Hectochelle and for these YSOs, we show their mean $EWs$ weighted by the uncertainty in each spectral epoch.

In Fig.~\ref{Fig:Li_spt}, the $EWs$ of CTTSs are generally smaller than those of WTTSs with similar spectral types due to the veiling effect in which continuum emission from accretion shocks on the surface of the star can reduce the measured $EW$ of CTTSs. In  Fig.~\ref{Fig:Li_spt} it can be also noted that the $EW$ of \LiI\ in WTTSs with similar spectral types shows a large scatter, which may be due to the age spread in these WTTSs. As a comparison, Figure~\ref{Fig:Li_spt} also shows the  $EWs$ of  \LiI\ for the YSOs in Taurus \citep{2008A&A...487..965S}, and in L1630N and L1641 presented in Paper\,I. The L1641 star-forming region is roughly the same age as Taurus and L1630N, thus it is expected to show a lithium depletion trend similar to those in Taurus and L1630N. Using the $EW$ of the \LiI\ absorption line for YSOs in the three regions, we estimate the median $EW$  in each spectral-type bin (filled star symbols in Fig.~\ref{Fig:Li_spt}). For the YSOs with spectral types later than K5, the median $EWs$ generally agree with the theoretical locus for  YSOs with  undepleted Lithium \citep{2002A&A...382..563B,2009MNRAS.400..317J}. The median $EW$  and theoretical locus deviate for the YSOs with spectral type earlier than K5. The median $EWs$ for the 1\,Myr old population are much larger than those in the Tucanae-Horologium stellar association, which has an age of 27$\pm$11\,Myr \citep{2008ApJ...689.1127M}.

\begin{figure}
\begin{center}
\includegraphics[width=\columnwidth]{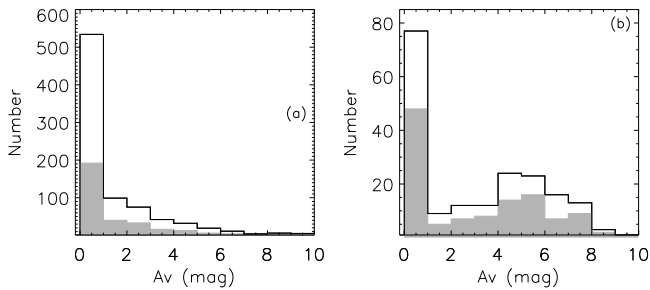}
\caption{(a): The visual extinction distribution of the YSOs (filled histograms) observed with Hectospec, and {all known YSOs with spectral types (open histograms)}. (b): The visual extinction distribution of the non-members in L1641 (filled histograms) identified in this work, and all the non-members  (open histograms)  in both this work and Paper\,I.}\label{Fig:AV_dis}
\end{center}
\end{figure}

\begin{figure*}
\centering
\includegraphics[width=2\columnwidth]{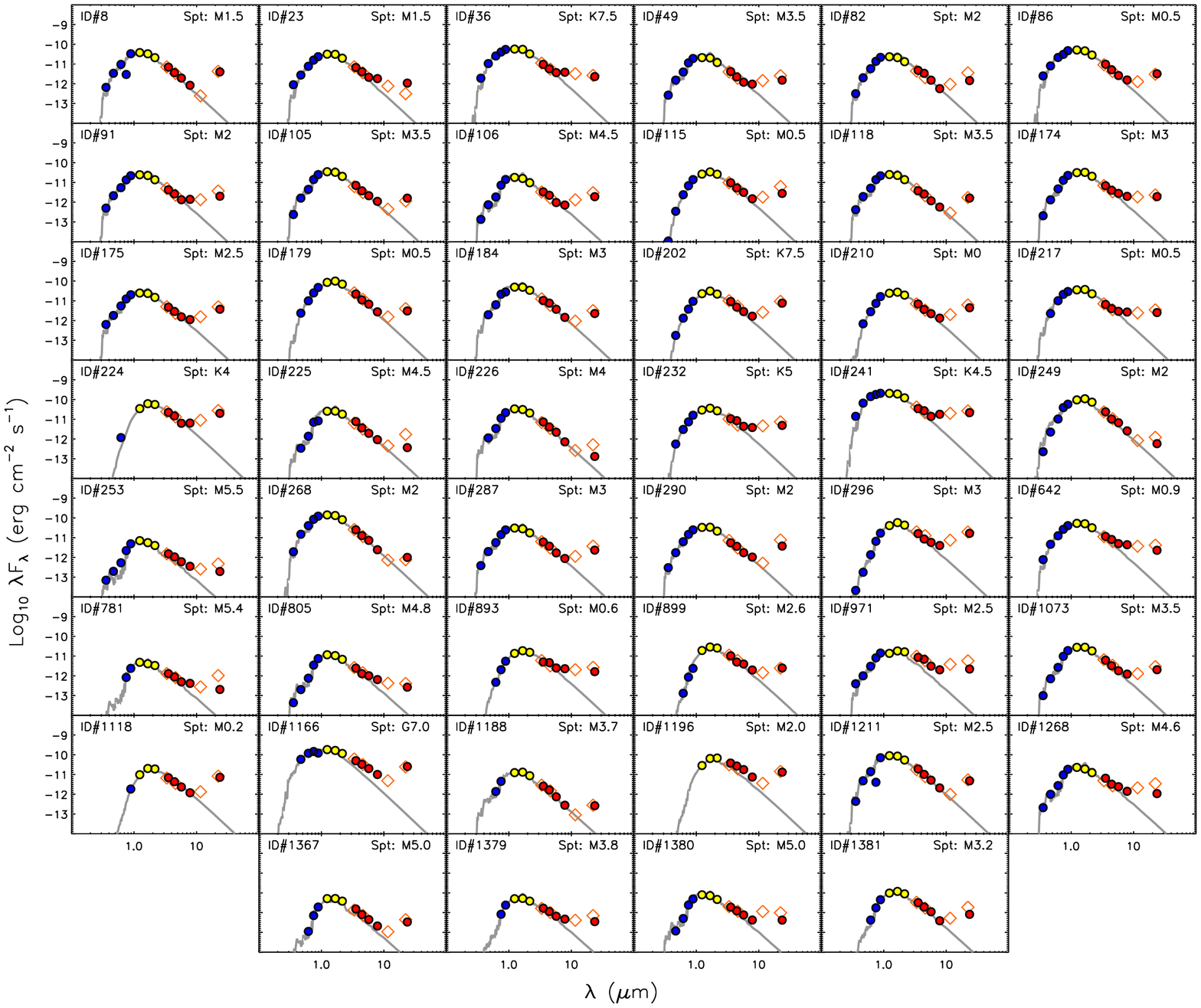}
\caption{\label{fig:TD_SED} The SEDs of newly confirmed YSOs with transition disks in L1641. The circles show the photometry from SDSS/LAICA (blue), 2MASS  (magenta), and Spitzer (red). The open diamonds are the photometry from WISE. The photospheric emission level is indicated with a grey curve in each panel. }
\end{figure*}

\subsubsection{Extinction}
We estimated the extinction for each individual object using the method outlined in \sect\ref{Sec:ana_stellar_properties}. {\newrev The typical uncertainty for the extinction values is less than 0.1  magnitude.} In \fig\ref{Fig:AV_dis}(a), we show the \AV\ distribution of the YSOs (filled histograms) from this works, as well as all YSOs (open histograms) with spectral types in L1641. Both populations have a similar \AV\ distribution  with a peak at very low extinction (\AV$<$1\,mag) and a gradual decrease of YSO numbers towards higher \AV. In \fig\ref{Fig:AV_dis}(b), we show the field stars (filled histograms) identified in this work, and all the field stars (open histograms) identified in both this work and Paper\,I. The two populations of field stars show a similar \AV\ distribution, but different from the \AV distribution of the YSO samples (see \fig\ref{Fig:AV_dis}(a)). The  \AV\ distribution of the field stars has a peak at \AV$\lesssim$1\,mag consistent with foreground and background objects along sight lines with low extinction, and  a relatively flat distribution between $\sim$1 and $\sim$8 magnitudes of \AV\ with a peak around \AV=4--6\,mag.

\begin{figure}
\centering
\includegraphics[width=\columnwidth]{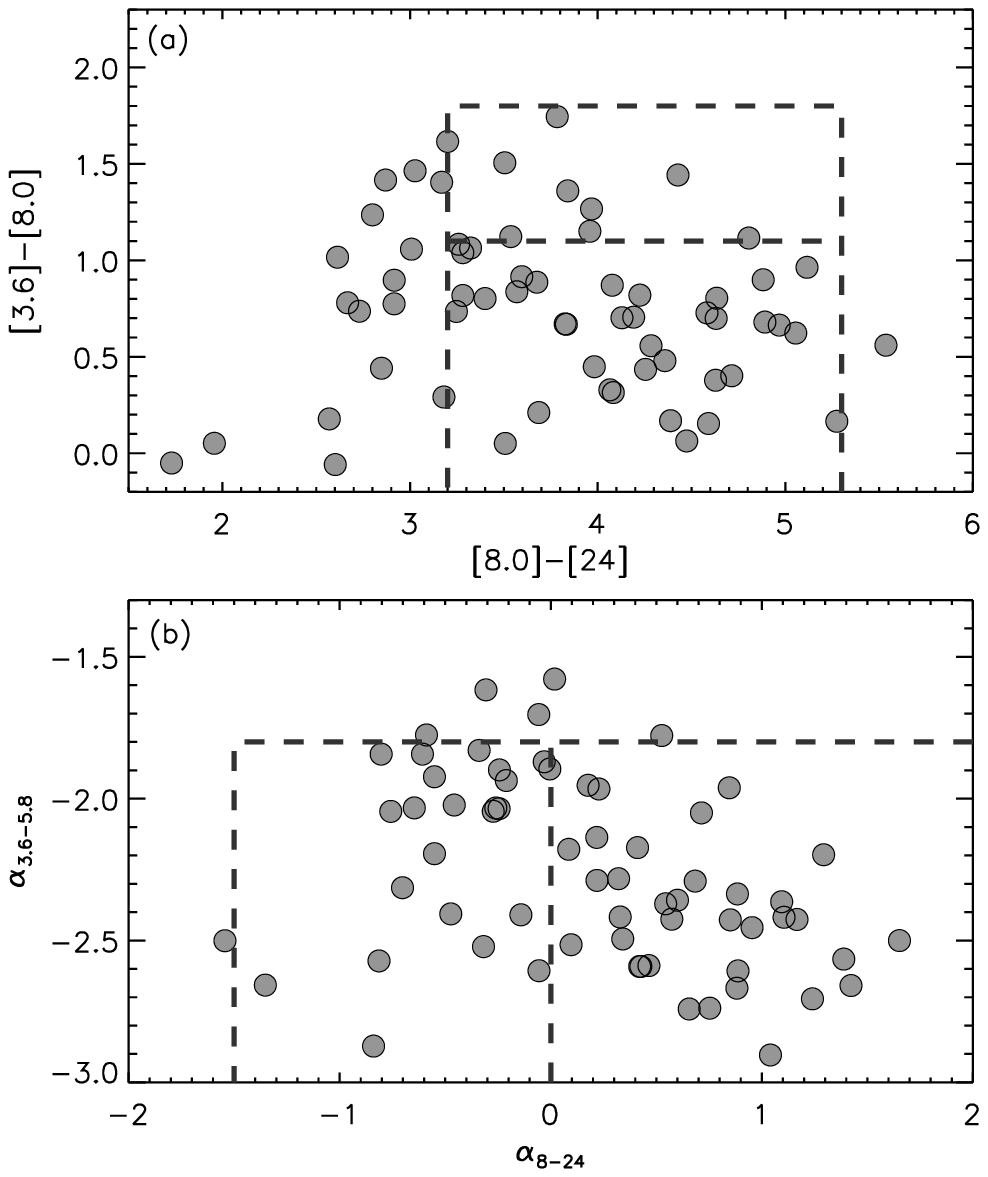}
\caption{\label{fig:TD_criteria} The TD selection criteria from (a)  \citet{2010ApJ...718.1200M}, and (b)  \citet{2010ApJ...708.1107M}. The filled circles are our confirmed TDs in L1641 from both Paper\,I and this work. In panel (a), classical TDs (with inner holes) lie within 0.0$<$[3.6]$-$[8.0]$<1.1$ and 3.2$<$[8.0]$-$[24]$<$5.3, and TDs with weak excess emission in Spitzer IRAC bands lie within 1.1$<$[3.6]$-$[8.0]$<$1.8 and 3.2$<$[8.0]$-$[24]$<$5.3. In panel (b), the region with $-$1.5$\le$$\alpha$$_{8-24}$$\le$0 and $\alpha$$_{3.6-5.8}$$\le$$-$1.8 are for weak excess TDs, and the region with $\alpha$$_{8-24}$$>$0 and $\alpha$$_{3.6-5.8}$$\le$$-$1.8 for classical TDs.}
\end{figure}

\subsection{Disks}
\subsubsection{Transition disks}\label{sec:results:transition_disk}

 {\newnewrev Our main TD selection criteria are based on colors as shown in Fig.~\ref{fig:allccmap}(b): [8.0]$-$[24]$\geq$2.5 and $K_{\rm s}$$-$[5.8]$\leq$0.56$+$([8.0]$-$[24])$\times$0.15. For the YSOs with spectral types, we use the spectral types of the central stars to constrain the photospheric emission. We compare their SEDs with their model photospheric emissions to look for the sources showing very weak or no infrared excess at near infrared wavelengths and shorter IRAC bands, but strong excess emission at mid-infrared and longer wavelengths.}

In Fig.~\ref{fig:TD_SED}, we show the SEDs of the {\newnewrev 46} newly confirmed TDs in this work. {\newnewrev Among them, 40 TDs are selected based on our TD selection criteria (see Fig.~\ref{fig:allccmap}(b)). The additional 6 TDs are selected by the comparison of their SEDs and photospheric emission. Among the 6 TDs, two(\#226 and 249) of them show nearly photospheric emission at wavelength out to $\sim$8\mum, and moderate excess at 24\mum\ ([8.0]$-$[24]$<$2). The other 4 TDs (\#971, 1196, 1268, and 1379), showing near-infrared excess emission significantly reduced from normal T~Tauri stars, are located close to the TD selection boundary.}

In the literature, there are several criteria to  select TDs, e.g. Paper\,I, \citet{2010ApJ...718.1200M}, \citet{2010ApJ...708.1107M}, etc. Our selection criteria, based on $K_{\rm s}$$-$[5.8] and [8.0]$-$[24] colors, is displayed in Fig.~\ref{fig:allccmap}(b). \citet{2010ApJ...718.1200M} provide different criteria based on [3.6]$-$[8.0] and [8.0]$-$[24] colors: (1) 0.0$<$[3.6]$-$[8.0]$<$1.1 and 3.2$<$[8.0]$-$[24]$<$5.3 for classical TDs (with inner holes); (2)  1.1$<$[3.6]$-$[8.0]$<$1.8 and 3.2$<$[8.0]$-$[24]$<$5.3 for TDs with weak excess emission in Spitzer IRAC bands. \citet{2010ApJ...708.1107M} employ two infrared spectral slopes to select TD candidates: (1) $-$1.5$\le$$\alpha$$_{8-24}$$\le$0 and $\alpha$$_{3.6-5.8}$$\le$$-$1.8 for weak excess TDs; (2) $\alpha$$_{8-24}$$>$0 and $\alpha$$_{3.6-5.8}$$\le$$-$1.8 for classical TDs (with inner holes). Our TD sample in this work is mainly selected using the criteria in Paper\,I. Including the TDs in  Paper\,I,  we have 65 TDs with spectral-type estimates in L1641. Among them, 64 TDs are detected in all four IRAC bands and at MIPS 24\mum. {\newrev We use these 64 TDs to compare our TD criteria with those from \citet{2010ApJ...718.1200M} and \citet{2010ApJ...708.1107M}.} In Fig.~\ref{fig:TD_criteria}, we show how these TDs fit to the selection  criteria from \citet{2010ApJ...718.1200M} and \citet{2010ApJ...708.1107M}. According to the criteria of \citet{2010ApJ...718.1200M}, 38 sources are classified into their Group\,(1), and 9 sources are classified into Group\,(2). Among the remaining 17 sources which are excluded as TDs, 16 objects do not show strong enough excess emission at 24\mum, and one object shows too strong excess emission at 24\mum\ (see Fig.~\ref{fig:TD_criteria}(a)). These rejected sources fall within the criteria of \citet{2010ApJ...708.1107M}.  According to their criteria, 23 are categorized into their Group\,(1), 35  into their Group\,(2), and 6 sources excluded as TDs but very close to the selection boundaries. {\newrev Thus, our selection criteria are generally consistent with those from \citet{2010ApJ...718.1200M} and \citet{2010ApJ...708.1107M}.}

\subsubsection{Globally depleted disks}\label{sec:results:Homologously_depleted_disk}

\begin{figure*}
\begin{center}
\includegraphics[width=2\columnwidth]{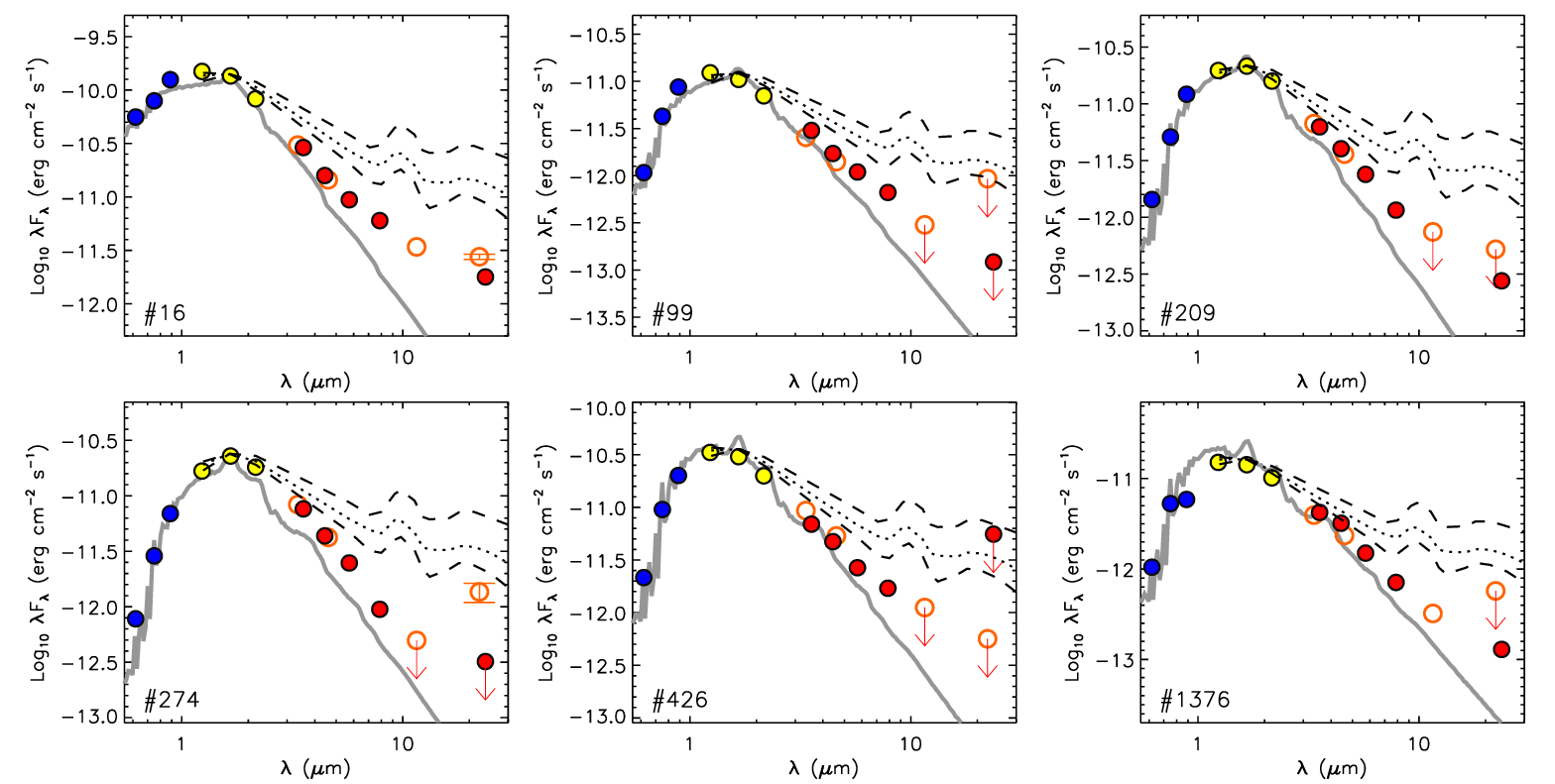}
\caption{The SEDs of globally depleted disk candidates in L1641. The filled circles show the photometry from 2MASS and Spitzer. The open circles are the photometry from WISE. The photospheric emission level is indicated with a grey curve in each panel. In each panel, the dotted line shows the median Taurus SED with upper and lower quartiles shown with dot-dashed lines \citep{2006ApJS..165..568F}.}\label{Fig:Homo_disk_SED}
\end{center}
\end{figure*}

In Fig.~\ref{Fig:Homo_disk_SED}, we show the SEDs of the globally depleted disk candidates in L1641, as well as the median SED of T~Tauri stars in Taurus \citep{2006ApJS..165..568F}. Compared with the median SED of Taurus, these YSOs in Fig.~\ref{Fig:Homo_disk_SED} show uniformly depleted SEDs with an infrared excess emission level that has been reduced by a similar factor across 3.6--24\mum. Such SED types can be produced when disks are deficient in small dust grains \citep{2009ApJ...698....1C,2011ApJ...742...39S}. For these disks, their effective disk height, i.e. the height above the disk midplane where the disk becomes optically thick to the stellar radiation, can be substantially reduced, causing a smaller fraction of the stellar energy to be absorbed and reprocessed by the disk, which produces an infrared excess emission of reduced magnitude.  Dust coagulation can lead to a deficiency of small dust grains, and the larger grains couple less efficiently with the gas, allowing them to settle towards the midplane.

\begin{figure}
\begin{center}
\includegraphics[width=1\columnwidth]{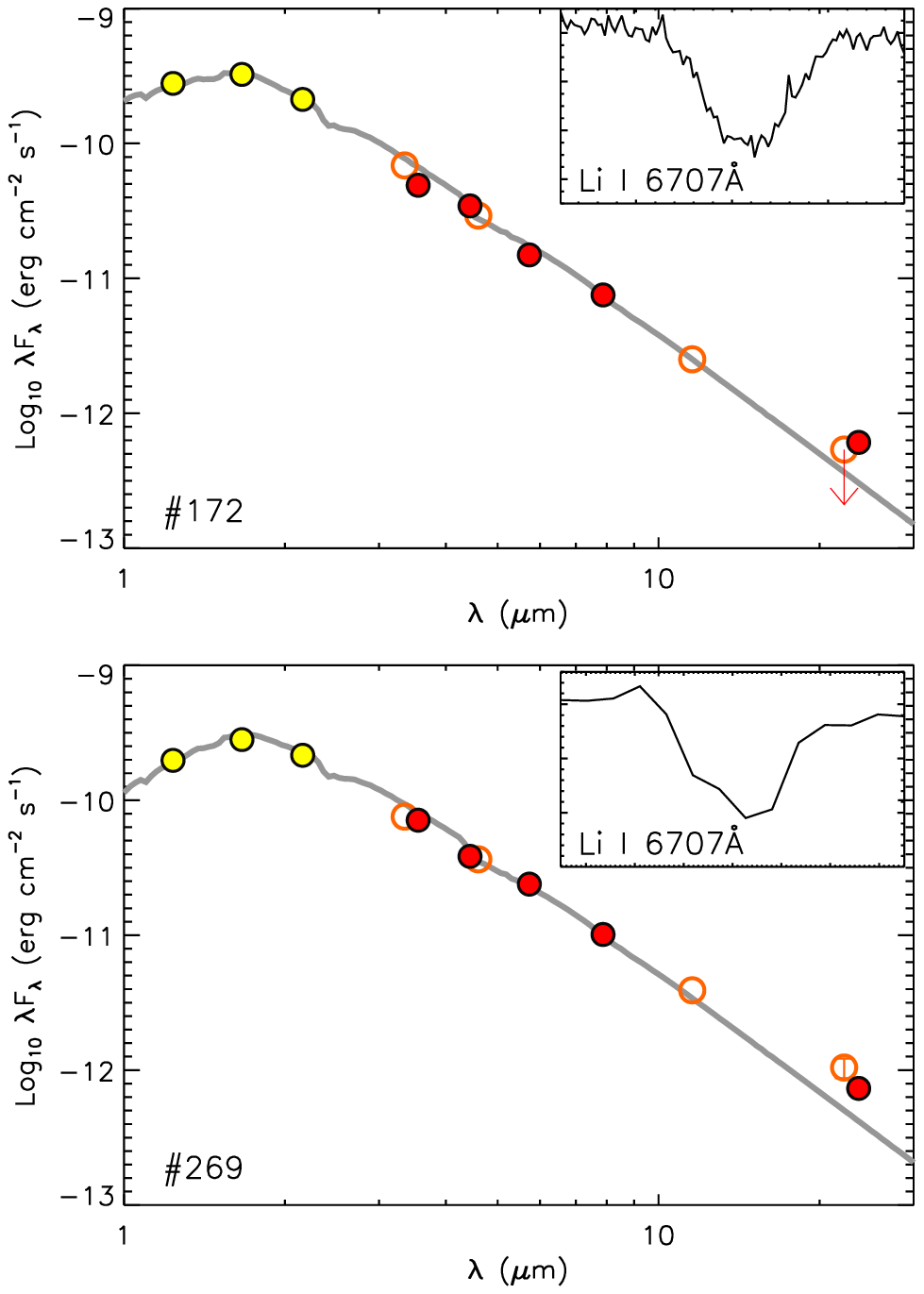}
\caption{The SEDs of debris disk candidates in L1641. The filled circles show the photometry from 2MASS and Spitzer. The open circles are the photometry from WISE. The photospheric emission level is indicated with a grey curve in each panel. The inset in each panel shows the \LiI\ absorption line. The  Li absorption line is from the Hectochelle spectra for the star \#172, and from the Hectospec spectra for the star \#269.}\label{Fig:SED_debris}
\end{center}
\end{figure}

\subsubsection{Young debris disks or ``evolved'' TDs?}\label{sec:results:debris_disk}
Figure~\ref{Fig:SED_debris} shows the SEDs of two objects \#172 and 269. These two sources show essentially photospheric emission levels at $\lambda$$\lesssim$10\mum, but display weak excesses at 24\mum. This SED shape is typical of ``debris disks'' \citep[see, e.g.][]{2009ApJS..181..197C}. Such disk system possess relatively small amounts of circumstellar material which are produced in the collisional grind-down of larger bodies (``planetesimals''). In Fig.~\ref{Fig:SED_debris}, both of the stars have the spectral type K6, and show X-ray emission (see \tab~\ref{tab:allYSO_L1641_optical}). Their ages, estimated from different sets of PMS evolutionary tracks, are all less than 1\,Myr (see \tab~\ref{Tab:mass_age}). The youth of the stars \#172 and \#269 can be further confirmed by our spectroscopic observation. The star \#172 has been observed with both Hectochelle and Hectospec and both spectra clearly show  \LiI\ absorption line. In its high-resolution Hectochelle spectrum, the $EW$ of the \LiI\ absorption line is 0.53\,\AA. The star \#269 has only been observed with Hectospec. In its Hectospec spectrum, the $EW$ of  the \LiI\ absorption line is $\sim$0.5\,\AA. The Li $EWs$ of both stars suggest they  are as young as other YSOs in L1641 (see Fig.~\ref{Fig:Li_spt}). Furthermore, the star \#172 has a  radial velocity $V_{\rm LSR}$=5$\pm$2\,km/s (relative to the local standard of rest) similar to the local molecular gas in L1641 \citep{1987ApJ...312L..45B}, suggesting the star \#172 is physically associated with the L1641 cloud.

In the case that the disks are optically thin through 24\mum\ we can estimate the timescale for small dust grains removed  by Poynting-Robertson drag using the method described in \citet{2009ApJ...698....1C}.  For 10\mum\ dust grains with a  volume density ($\rho_{\rm s}$) $\sim$1\,g\,cm$^{-3}$, and an absorption coefficient ($\langle Q_{\rm abs}\rangle$) $\sim$1, the Poynting-Robertson drag can remove these grains on timescales$\lesssim$6$\times$10$^{5}$\,yr. When the dust grains are porous, the drag time could be much shorter \citep{2009ApJ...698....1C}. For the two stars, the typical grain sizes, below which the radiation pressure can remove dust from the system, are estimated to be $\sim$3--8\mum\  using the Equation~(4) in \citet{2009ApJ...698....1C}  with $\rho_{\rm s}$$\sim$1\,g\,cm$^{-3}$, and $\langle$$Q_{\rm abs}$$\rangle$$\sim$1. However, whether the Poynting-Robertson drag can work on  dust grains also depends on the coupling of dust to the gas. If the gas density in circumstellar disks is low, the dust grains emitting the 24\mum\ excess emission can be quickly removed at the above dynamical timescale without the replenishment by the collision of larger planetesimals. Thus the disks around the two stars \#172 and 269 could be debris if they are at ages of several \,Myr. However, the ages of the stars \#172 and 269, estimated from different sets of PMS evolutionary tracks, are all younger than several 10$^{5}$\,yr. If their isochrone ages are reasonable, it is also possible that the two sources are relatively evolved TDs with much larger inner holes (several ten AU) in the disks than normal TDs. Such type of evolved disks could be still gas rich, and only produce weak excess emission at 24\mum.

\begin{figure}
\begin{center}
\includegraphics[width=1\columnwidth]{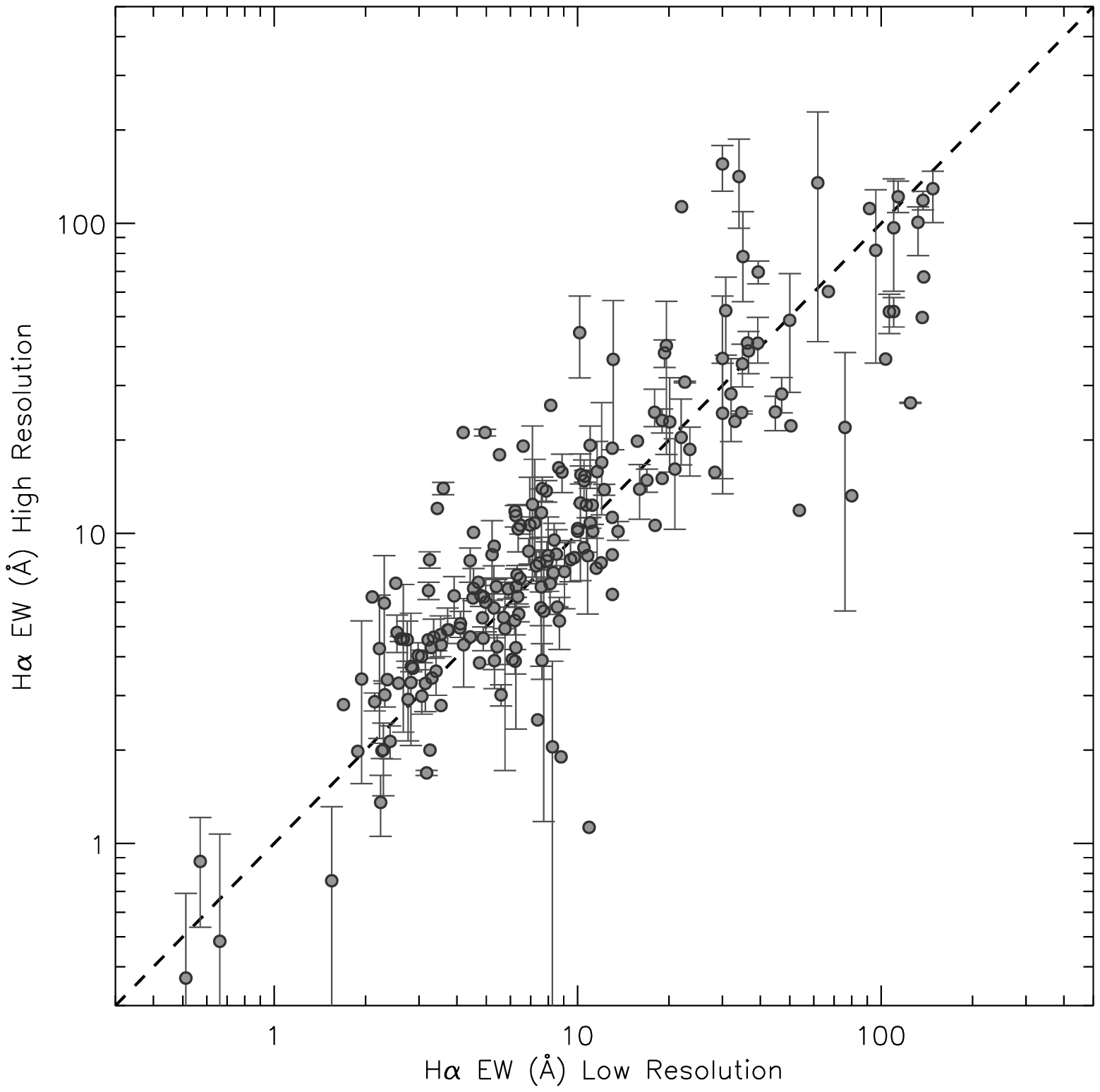}
\caption{Comparison of the H$\alpha$ $EW$ measured from medium-resolution (VIMOS and Hectospec, in Paper\,I and this work) and high-resolution (Hectochelle) spectra. Note that the high resolution and low resolution spectra are taken at different times. The dashed line marks where the two $EWs$ are equal. For the sources which have been observed at more than one epoch, their average $EW$ is used for the plot and the error bar represents their minimum and maximum $EWs$.}\label{Fig:com_EW}
\end{center}
\end{figure}

\begin{figure*}
\begin{center}
\includegraphics[width=2\columnwidth]{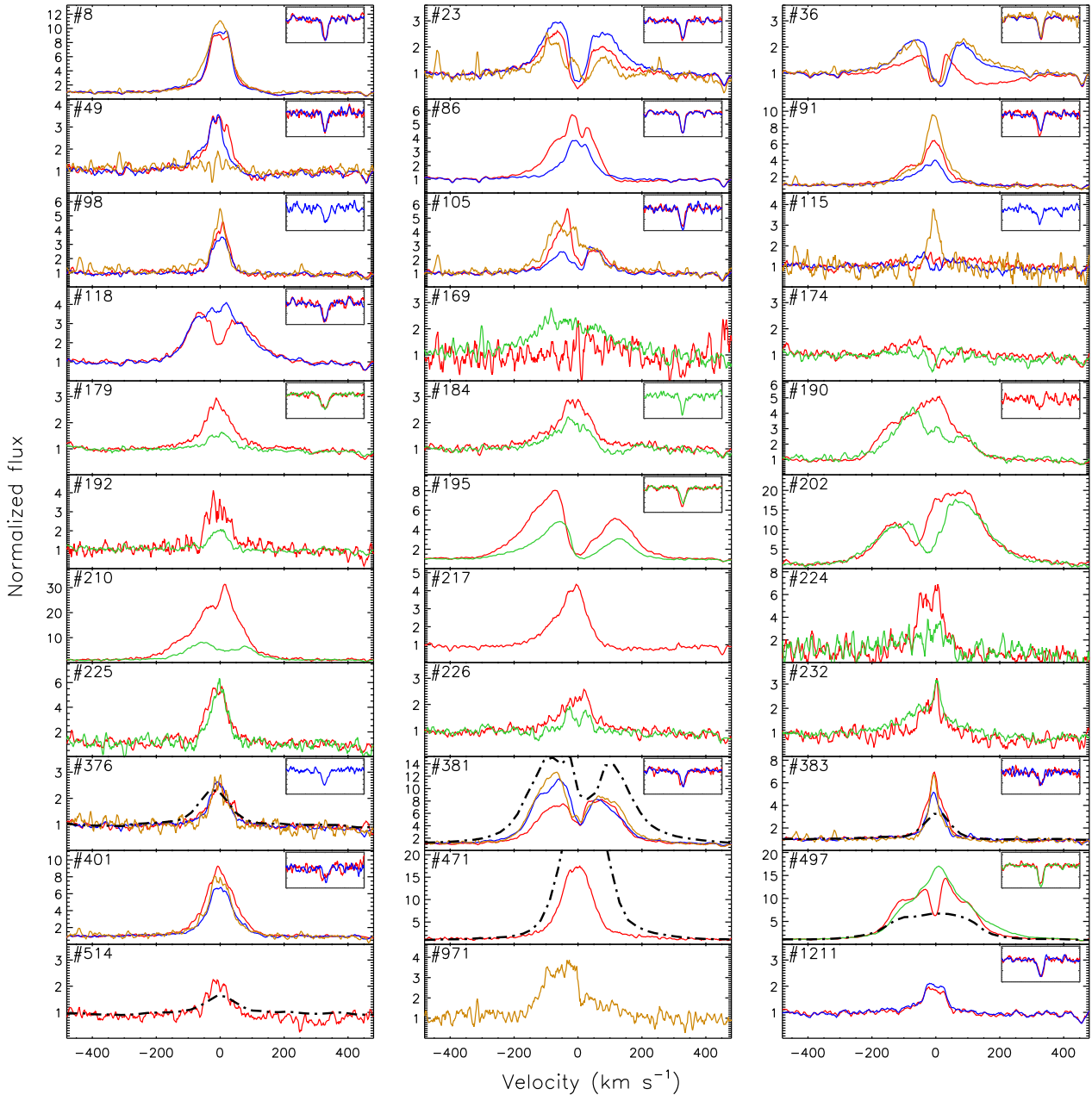}
\caption{The H$\alpha$ profiles at several epochs for TDs in L1641. The observation dates have been encoded as different colors: red for 2010 Feb 5, blue for 2010 Mar 3, yellow for 2010 Nov 29, and green for 2011 Oct 19. The dash-dotted lines display the H$\alpha$ profiles (R$\sim$2500) from Paper\,I, which was observed with  VIMOS. The insets show the \LiI\ absorption line  at several epochs. {\newrev In each panel, the spectra have been binned up by a factor of 5 for clarity.}}\label{Fig:Halpha_SED_TD}
\end{center}
\end{figure*}

\begin{figure*}
\begin{center}
\includegraphics[width=2\columnwidth]{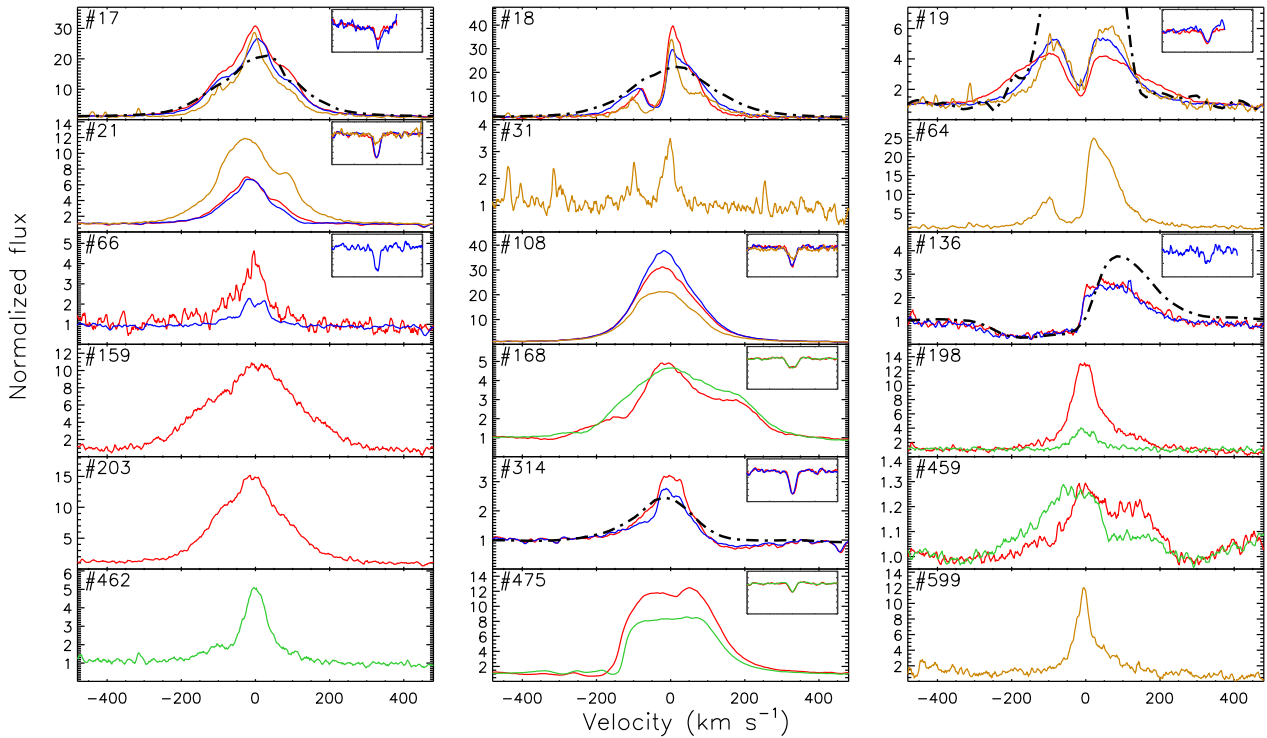}
\caption{The H$\alpha$ profiles for the disk population with $\alpha$$_{3.6-8}$$\ge$$-$1.0 in L1641. The H$\alpha$ profiles have been monitored at several epochs. The observation dates have been encoded as different colors: red for 2010 Feb 5, blue for 2010 Mar 3, yellow for 2010 Nov 29, and green for 2011 Oct 19. The dash-dotted lines display the H$\alpha$ profiles (R$\sim$2500) from Paper\,I, which is observed with  VIMOS. The insets show the \LiI\ absorption line at several epochs. In each panel, the spectra have been binned up by a factor of 5 for clarity.}\label{Fig:Halpha_SED_alpha1}
\end{center}
\end{figure*}

\begin{figure*}
\begin{center}
\includegraphics[width=2\columnwidth]{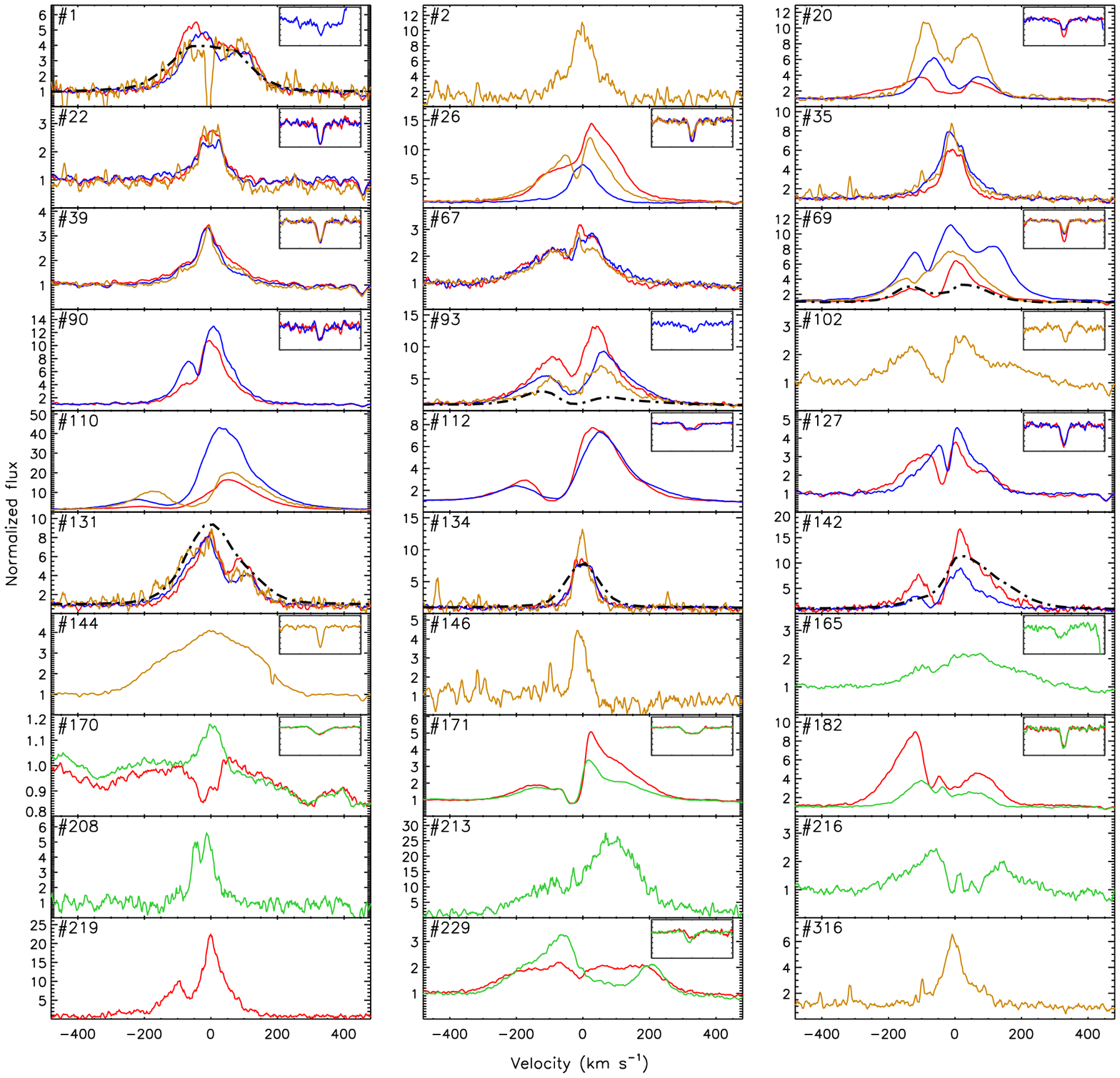}
\caption{Similar to Fig.~\ref{Fig:Halpha_SED_alpha1} but for the disk population with $-$1.0$>$$\alpha$$_{3.6-8}$$\ge$$-$1.8 in L1641.}\label{Fig:Halpha_SED_alpha2}
\end{center}
\end{figure*}

\begin{figure*}
\begin{center}
\includegraphics[width=2\columnwidth]{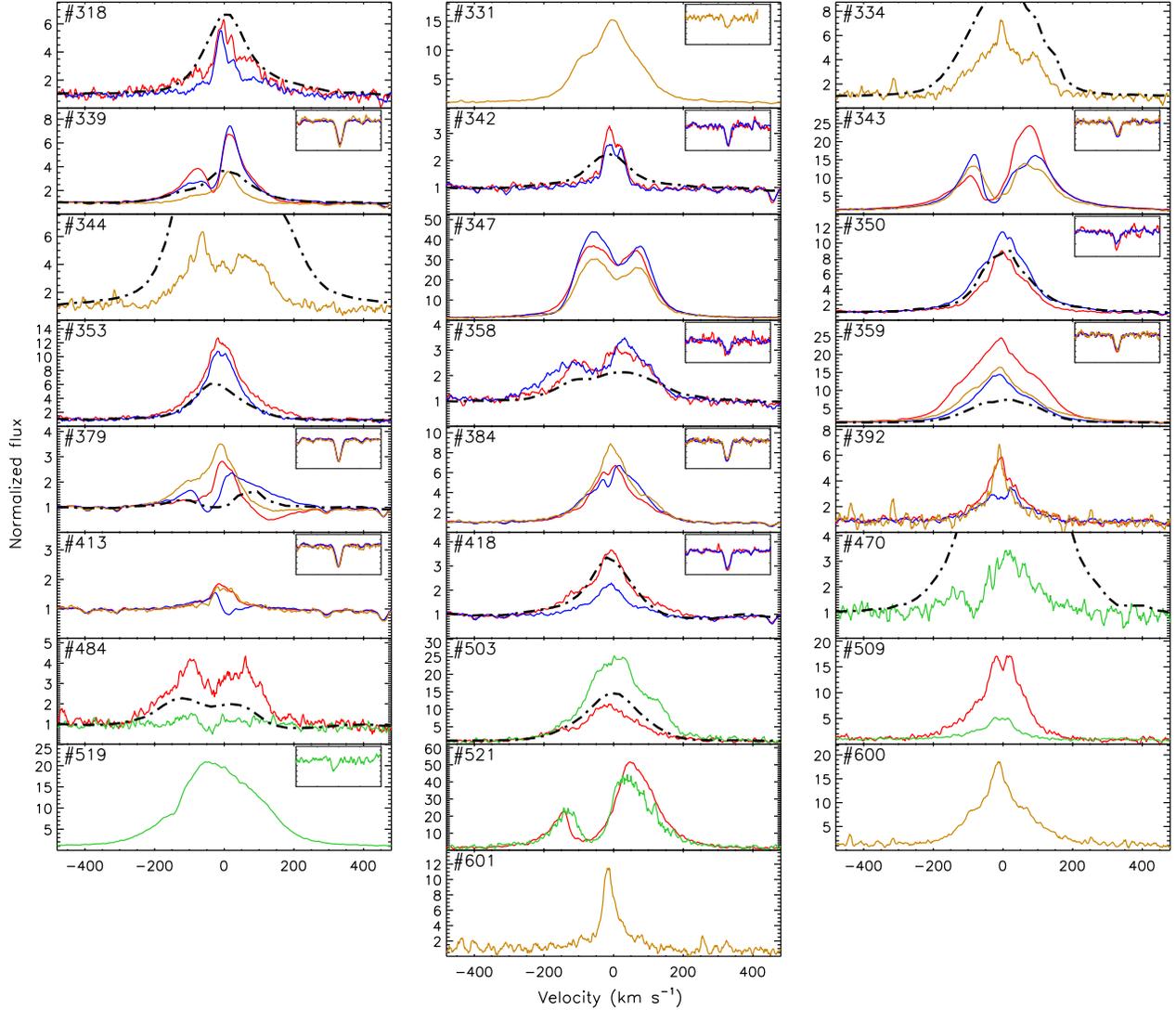}
\caption{Similar to Fig.~\ref{Fig:Halpha_SED_alpha2}, but for different targets.}\label{Fig:Halpha_SED_alpha3}
\end{center}
\end{figure*}

\begin{figure}
\begin{center}
\includegraphics[width=\columnwidth]{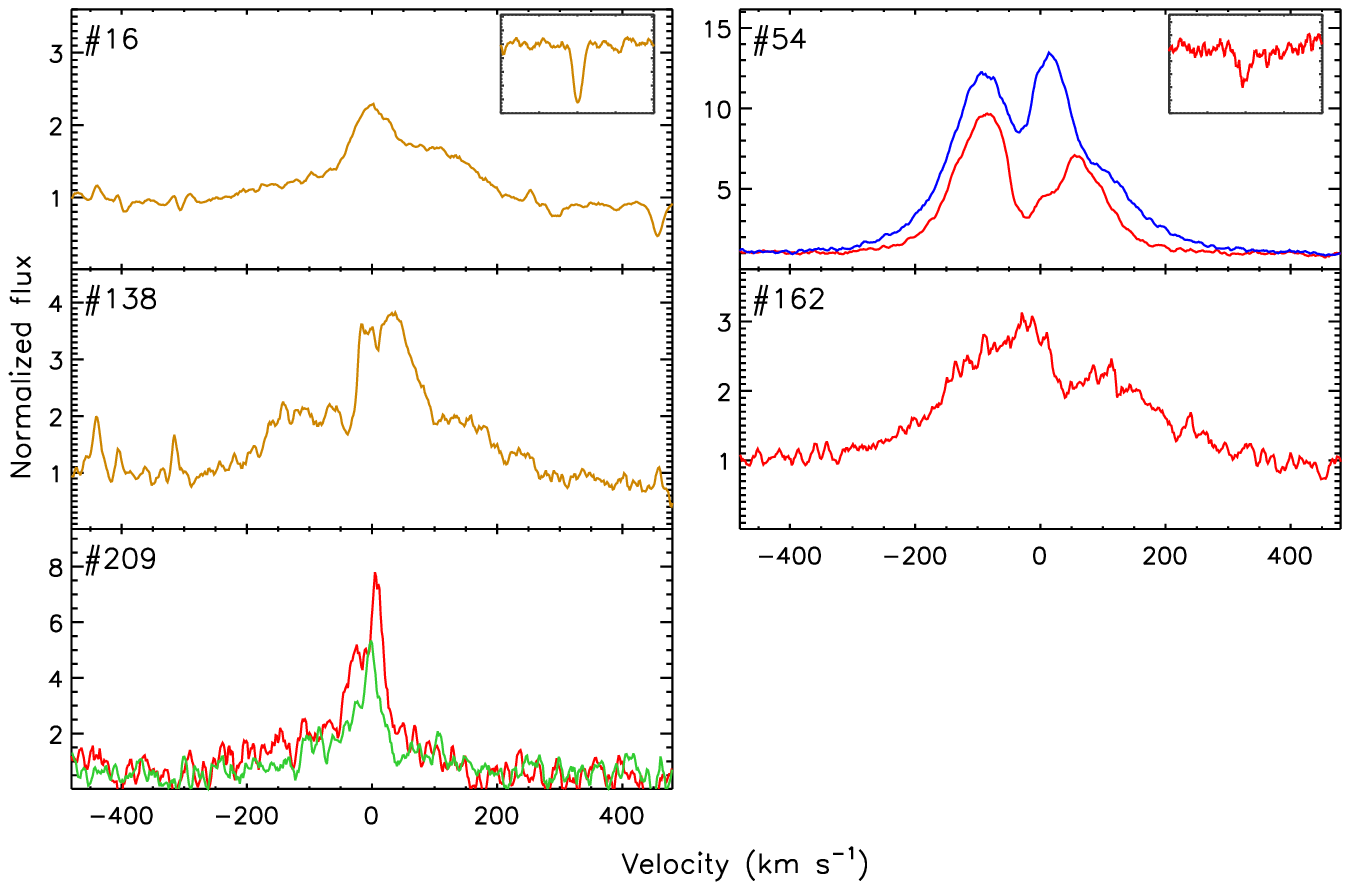}
\caption{Similar to Fig.~\ref{Fig:Halpha_SED_alpha2} but for the disk population with $\alpha$$_{3.6-8}$$<$$-$1.8 in L1641.}\label{Fig:Halpha_SED_alpha4}
\end{center}
\end{figure}

\subsection{Accretion}
The Balmer emission lines in the spectra of PMS stars can be  produced in two ways: chromospheric activity and  magnetospheric accretion. The former mechanism usually yields weak and narrow emission lines, while the magnetospheric accretion process can produce very broad and strong emission lines  \citep{1994ApJ...426..669H,2001ApJ...550..944M,2003ApJ...582.1109W}. Thus, the strength and width of the  Balmer emission lines, especially for the H$\alpha$ line, are used to  characterize the accretion activity of young stars. Our Hectospec spectra cover both the H$\alpha$ and H$\beta$ lines with a medium spectral resolution, and the Hectochelle spectra  cover the H$\alpha$ line with a high spectral resolution. We will use all these data to study the accretion properties of young stars.

\subsubsection{Characterizing accretion activity with medium-resolution spectroscopy}
We classified the YSOs into weak T~Tauri stars (WTTS) or CTTSs based on their H$\alpha$ $EW$ in the Hectospec spectra using the spectral type dependent criteria described in Paper\,I. A star is considered to be a CTTS if $EW$(H$\alpha$)$\geq3$\,\AA\ for K0--K3 stars, $EW$(H$\alpha$)$\geq$5\,\AA\ for K4 stars, $EW$(H$\alpha$)$\geq7$\,\AA\ for K5--K7 stars, $EW$(H$\alpha$)$\geq9$\,\AA\ for M0--M1 stars, $EW$(H$\alpha$)$\geq11$\,\AA\ for M2 stars, $EW$(H$\alpha$)$\geq15$\,\AA\ for M3--M4 stars, $EW$(H$\alpha$)$\geq18$\,\AA\ for M5--M6 stars, and $EW$(H$\alpha$)$\geq20$\,\AA\ for M7--M8 stars. These criteria are based on the  H$\alpha$ line $EWs$, which makes them useful for spectra with  low spectral resolution. The disadvantage of this definition is that it may misclassify some accretors as WTTSs. In our spectroscopic sample observed with Hectospec, a few of the YSOs (e.g. ID\#23 and 36) have been observed with Hectochelle, and show broad H$\alpha$ line profiles, indicating an ongoing accreting activities despite the small H$\alpha$ $EW$. The  H$\alpha$ line $EWs$ of these sources are below the thresholds to be classified as CTTSs because of the strong absorption around their H$\alpha$ line centers. In this case, we also use the velocity width at 10\%\ maximum height ($FW_{H\alpha, 10\%}$) of the Hectochelle spectra to distinguish the CTTSs and WTTSs. As discussed in the Appendix~\ref{Appen:Halpha}, a YSO can be classified as a CTTS if its $FW_{H\alpha, 10\%}$$>$250\kms.

In table~\ref{Tab:spt}, we list the $EWs$ of  H$\alpha$ and H$\beta$ from the Hectospec spectra for our YSO sample, and their accretion properties, i.e. CTTS or WTTS. For the CTTSs, we estimated their accretion rates from the H$\alpha$ and H$\beta$ line luminosity using the methods described in \sect\ref{sec:ana_acc}. The results are also listed in \tab~\ref{Tab:spt}. {\newrev The typical uncertainties are a factor of $\sim$5 for the accretion rates derived from H$\alpha$ line luminosity, and a factor of $\sim$3 for those from H$\beta$.} For the WTTSs with disks, only upper limits on the accretion rate, estimated with  H$\alpha$ and H$\beta$  line luminosity as done for CTTSs, are given since their  Balmer emission is mainly due to the chromospheric activities. In table~\ref{Tab:spt}, the disk property of each YSO is also listed. For the normal full disks, the disks categorized into  optically thick disks  if $\alpha$$_{3.6-8}\ge-$1.8, or optically thin disks when $\alpha$$_{3.6-8}<-$1.8.

\subsubsection{High-resolution H$\alpha$ spectroscopy}
We obtained the H$\alpha$ emission lines for {\newnewrev 235} young stars in L1641 with Hectochelle. {\newnewrev In this sample, 8 objects are Class\,I/Flat sources, 102 are  Class\,II sources, and 124 are Class\,III sources. One object (\#385) in the sample cannot be classified due to the lack of photometric data in infrared bands.} During 2010-2011, a major fraction of these YSOs have been observed at two or three epochs  In \tab~\ref{Tab:EW} we list the $EW$ and $FW_{H\alpha, 10\%}$ of the H$\alpha$ emission line, $EW$ of \LiI\ absorption line, types (CTTS or WTTS) of T~Tauri stars, and the accretor probabilities. The accretor probabilities are estimated from $FW$$_{H\alpha, 10\%}$ using the method described in Appendix~\ref{Appen:Halpha}. The accretion rates estimated from the $FW$$_{H\alpha, 10\%}$ are also listed in \tab~\ref{Tab:EW}.

Many of the YSOs in our Hectochelle sample have been observed in our medium-resolution spectroscopic survey with VIMOS and Hectospec (in Paper\,I and this work). In Fig.~\ref{Fig:com_EW}, we compare the  H$\alpha$ $EWs$ from Hectochelle and those from VIMOS or Hectospec. We find that there is a better agreement between these observations for H$\alpha$  $EW\lesssim$10\,\AA\ than for H$\alpha$  $EW\gtrsim$10\AA. This can be attributed to the accretion variability of accreting YSOs which usually show  H$\alpha$~$EW>$10\,\AA.

\vspace{0.3cm}
\begin{center}
\textit{4.4.2.1. H$\alpha$ emission profiles} \\
\end{center}
\vspace{-0.15cm}

In Fig.~\ref{Fig:Halpha_SED_TD}, \ref{Fig:Halpha_SED_alpha1}, \ref{Fig:Halpha_SED_alpha2},  \ref{Fig:Halpha_SED_alpha3}, \ref{Fig:Halpha_SED_alpha4}, we display the Hectochelle H$\alpha$ line profiles for the disk population in our spectroscopic sample.  For the sources which have been observed at multiple epochs, all the data are shown as a comparison. In general, most of these line profiles can be classified into the four groups according to the schemes in \citet{1996A&AS..120..229R}: (I) profiles with a single peak and no or very weak absorption features; (II) profiles with two relatively equal peaks; (III) profiles with one primary strong peaks and one weak secondary peak; and (IV) P\,Cygni or inverse  P\,Cygni profiles. In our YSO sample, the type\,I H$\alpha$ line profiles are either narrow or broad, while the type\,II and III profiles are typically broad. We only have one source (ID\#136, see the detail description of this sources in \sect\ref{sec:results:Fuori}) showing a P\,Cygni profile (Type IV). One source (ID\#69) shows a triple-peak profile on 2010-03-03 and cannot be classified into any of the four groups in \citet{1996A&AS..120..229R}.

In our spectroscopic sample observed with Hectochelle, there are 33 TDs. Their H$\alpha$ line profiles are shown in Fig.~\ref{Fig:Halpha_SED_TD}. Among the 33 TDs, $\sim${\newnewrev 42\%}  show broad H$\alpha$ line profiles ($FW$$_{H\alpha, 10\%}$$>$250\kms), characteristic of ongoing accretion.  Among the TDs showing broad H$\alpha$ line profiles, 5 sources  (ID\#23, 36, 195, 202, and 381) display H$\alpha$ line profiles with double peaks from the data observed at all epochs. As defined in this work, TDs show very weak or no excess emission in shorter IRAC bands. The absence of excess emission at such wavelengths can be explained if (1) TDs  have dissipated  most or all of the hot inner disks, or (2) in the inner regions of TDs there are opacity holes due to a lack of {\rev small dust grains}, while gas still exists in these regions. For the accreting TDs in our sample, the scenario (1) can be excluded.

In Fig.~\ref{Fig:Halpha_SED_alpha1}, we show the line profiles for the disk population with $\alpha$$_{3.6-8}$$\ge-$1.0. These sources have strong excess emission in the IRAC bands, thus are expected to show strong and broad  H$\alpha$ line profiles. In the figure, $\sim$83\% of the sources display the broad line profiles ($FW$$_{H\alpha, 10\%}$$>$250\kms) at one epoch. Five sources (ID\#31, 66, 314, 462, and 599) exhibit relatively narrow line profiles at  and these sources may have more quiescence accretion activities.

In Fig.~\ref{Fig:Halpha_SED_alpha2} and \ref{Fig:Halpha_SED_alpha3}, we show the line profiles for the sources with $-1.0$$>$$\alpha$$_{3.6-8}$$\ge-$1.8. These sources are typical Class\,II sources with optically-thick inner disks. In this sample, $\sim$80\% of the sources show broad  H$\alpha$ line profiles($FW$$_{H\alpha, 10\%}$$>$250\kms) at one epoch. In Fig.~\ref{Fig:Halpha_SED_alpha4}, we display the  H$\alpha$ line profiles of the YSOs with $-1.8>$$\alpha$$_{3.6-8}$$\ge$$-$2.5. The YSOs with such $\alpha$ values are considered as having optically thin inner disks. However, four sources (ID\#16, 54, 138, and 162) still show very broad  H$\alpha$ line profiles ($FW$$_{H\alpha, 10\%}$$>$250\kms), suggesting that they  are still accreting. 

\vspace{0.3cm}
\begin{center}
\textit{4.4.2.2. H$\alpha$ line variability} \\
\end{center}
\vspace{-0.15cm}

H$\alpha$ line profile fluctuations are very common in CTTSs, and can be due to   the variability of accretion flow or outflow\citep{1995ApJ...449..341J,1996A&A...314..835G,2001AJ....122.3335A,2002ApJ...571..378A}. In our sample, $\sim$90 diskless YSOs and  $\sim$80 disked YSOs have had their H$\alpha$ line profiles monitored at multiple epochs with Hectochelle. The H$\alpha$ emission-line variations  of WTTSs are mainly related to the chromospheric activity, and are {\rev less variable than the CTTS }.  In this work, we will only focus  on the CTTSs. In our sample, most of CTTSs show  clear variations in their H$\alpha$ line profiles. Among them, several sources, e.g sources ID\#20, 26, 69, etc, exhibit extreme line variations.  The H$\alpha$ line fluctuations (see Fig.~\ref{Fig:Halpha_SED_TD}, \ref{Fig:Halpha_SED_alpha1}, \ref{Fig:Halpha_SED_alpha2},  \ref{Fig:Halpha_SED_alpha3}, and \ref{Fig:Halpha_SED_alpha4}) can be mainly grouped into several types: (1) H$\alpha$ line profiles vary from single peak to double peak, or even to triple peaks, e.g. the sources ID\#26, 69, 339, etc., (2) the peak positions vary, e.g. the sources ID\#93, 110, 195, etc., (3) the ratio between two peaks of H$\alpha$ lines varies, e.g. the sources ID\#343 and 381, (4) H$\alpha$ line profiles broaden or narrow, e.g. the sources ID\#19, 21, 86, etc.  The H$\alpha$ line variation of one YSO can usually be classified into more than one group. For the YSOs which can be categorized into the group (4), their H$\alpha$ line variations are mainly due to their accretion variations. {\newrev Especially,  two sources in our sample (ID\#26 and \#86) show large variability from CTTS  ($FW$$_{H\alpha, 10\%}$$>$250\kms) to WTTS ($FW$$_{H\alpha, 10\%}$$\le$250\kms).} 

A few of the sources in our Hectochelle sample have been observed with VIMOS  in January or March, 2008. These data have been presented in Paper\,I. As a comparison, we show these observations in Fig.~\ref{Fig:Halpha_SED_TD}, \ref{Fig:Halpha_SED_alpha1}, \ref{Fig:Halpha_SED_alpha2}, \ref{Fig:Halpha_SED_alpha3}, and  \ref{Fig:Halpha_SED_alpha4}. The spectral resolution of the VIMOS spectra  is $\sim$2500 and the data can only resolve relatively broad H$\alpha$ line profiles. The comparison shows that the sources  ID\#69, 344, and 470, exhibit large variations in H$\alpha$ emission line.

\subsection{Exotic objects}\label{sec:results:exotic_objects}

\subsubsection{Extremely embedded sources}\label{sec:results:blue_source}

\begin{figure*}
\begin{center}
\includegraphics[width=1.8\columnwidth]{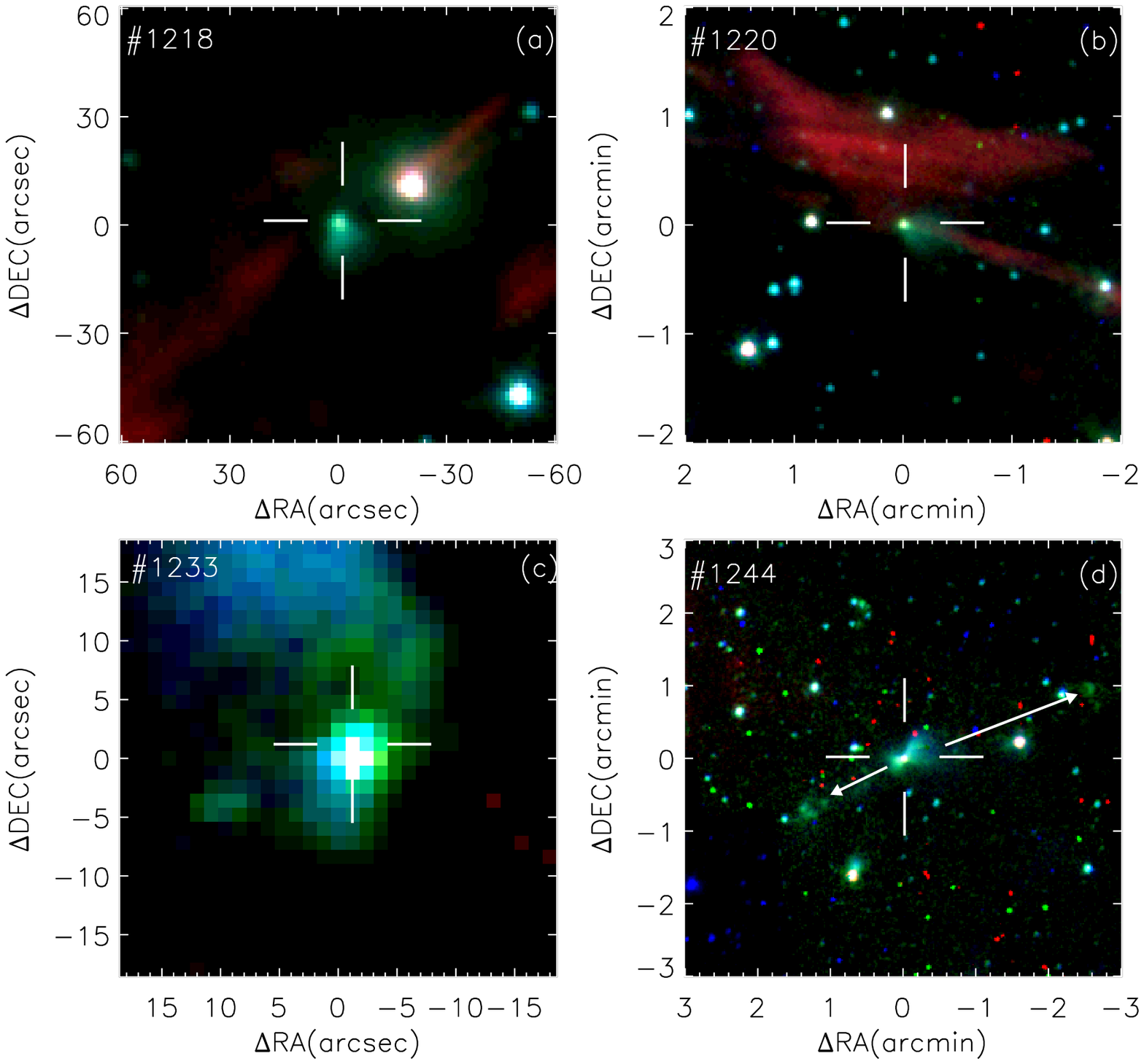}
\caption{\label{fig:weird} The Spitzer color images (red: 8\mum, green: 4.5\mum, and blue: 3.6\mum) centered on the four sources ID\#1218 (a), 1220 (b), 1233 (c), and 1244 (d) with [3.6]$-$[4.5]$\sim$0 and [5.8]$-$[8.0]$>$1. The arrows in panel(d) show the directions of a bipolar jet.}\end{center}
\end{figure*}

In  the [3.6]$-$[4.5] vs. [5.8]$-$[8.0] color-color diagram (see Fig.~\ref{fig:allccmap}(a)), five objects (ID\#1074, 1218, 1220, 1233, and 1244) have colors distinct from the other sources. Their colors ([5.8]$-$[8.0]$\sim$0 and [3.6]$-$[4.5]$>$1.0) make them appear as diskless stars with extremely high extinction ($A_{\rm K}\gtrsim 10$). Among the five sources, four (ID\#1074, 1220, 1233, and 1244) are in the FOVs of the XMM surveys, and only ID\#1074 is detected. In the Spitzer IRAC images, four sources (ID\#1218, 1220, 1233, and 1244) are surrounded by extended emission with fan-shaped structures (see Fig.~\ref{fig:weird}), and only the source ID\#1074  appears as a point source without any nebulosity. The source ID\#1074 is not detected in the MIPS 24\mum\ images. Thus we consider it as a  highly reddened diskless stars. The four sources associated with fan-shaped nebulae are all detected at  24\mum, and show strong excess emission in this band with [8.0]$-$[24]$\gtrsim$7, indicating they could be at a very early stage in the star-formation process. The fan-shaped nebulae around these sources are most easily explained by the scattering of photons from the central star out of cone-shaped cavities which are carved out of the dense envelope by energetic outflows. Furthermore, near ID\#1244 we find bipolar jets traced by the Spitzer 4.5\mum\ emission \footnote{The Spitzer 4.5\mum\ band  cover several H$_{\rm 2}$ lines ($\mu$=0--0,S(9, 10, 11)), which could be excited by the interaction of the outflows with the ambient medium. Thus the Spitzer 4.5\mum\ image are used to  search for the jets from young stars \citep[see e.g.][]{2009AJ....138.1830Z}.}. The morphology of the jets suggests ID\#1244 could be its driving source.

\begin{figure}
\begin{center}
\includegraphics[width=1.\columnwidth]{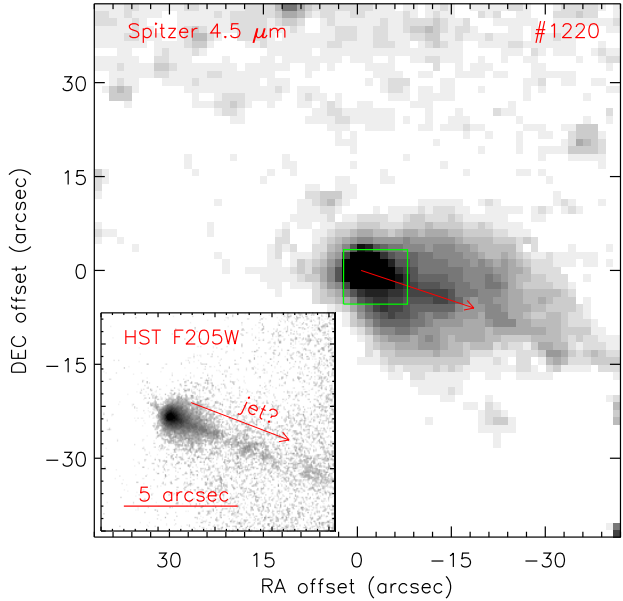}
\caption{\label{fig:HST} The Spitzer 4.5\mum\ image centered on the source ID\#1220. The inset shows the HST $F205W$ image of ID\#1220 with the sample FOV as that shown with the box on the 4.5\mum\ image. The arrows show the direction of a probable jet from the source \#1220 detected in the HST image.}\end{center}
\end{figure}

We have searched for the five sources in the Hubble space telescope (HST) data archive.  Two objects (\#1233 and 1244) have been observed with  HST/WFC3 at $F160W$ band, and are not detected in these images. The source \#1220 has been observed with HST/NIC2 at $F160W$ and $F205W$ bands.  In addition to continuum emission, the broad bands $F160W$ and $F205W$ principally transmit the [\FeII]\,1.644\mum\ and H$_{2}$\,2.122\mum\ lines, respectively both of which can originate from  shocked gas. In the HST images at both $F160W$ and $F205W$ bands, the object   \#1220 shows a cometary shape with a bright head and a long tail. Fig.~\ref{fig:HST} shows the  $F205W$ image of \#1220. Two scenarios can explain the shape of \#1220 in the HST images: (1) a young stellar system photoevaporated by nearby massive stars, e.g. like proplyds \citep{1993ApJ...410..696O,2000AJ....119..292B,2012A&A...539A.119F}, or (2) a YSO producing a jet which appears as a long tail. For the object \#1220, the former scenario can be excluded since there are no massive stars near  \#1220. The latter scenario could be a promising explanation for the shape of \#1220 in the HST images.

\subsubsection{A new subluminous object}\label{sec:results:underlum}
\begin{figure*}
\begin{center}
\includegraphics[width=1.8\columnwidth]{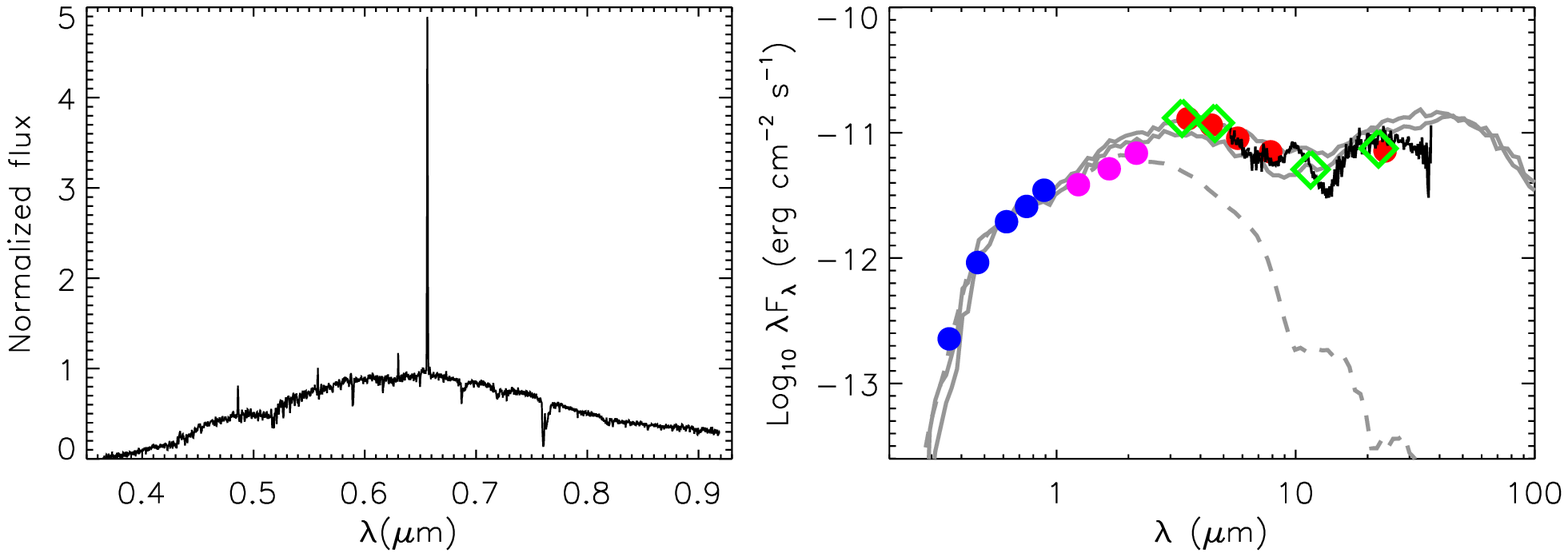}
\caption{\label{fig:exotic}Left panel: The optical spectra of the source  ID\#143. Right panel: the SED of the source ID\#143. The filled circles show the photometry from SDSS, 2MASS, and Spitzer. The open diamonds present the photometry from WISE. The dark line show the IRS spectra from Spitzer. The gray solid lines are the model SEDs of \citet{2007ApJS..169..328R} which best fit to the observed SED of ID\#143. The gray dashed line shows the photospheric level.}
\end{center}
\end{figure*}

 In the HR diagram (see Fig.~\ref{Fig:HRD}), one source (ID\#143) appears to be subluminous by a factor of $\sim$110 with respect to a 1\,Myr PMS star of a similar spectral type. Its  optical spectrum and SED is shown in Fig.~\ref{fig:exotic}. In its  optical spectrum, the H$\alpha$ emission line seems to be quite strong. Objects with similar properties have been found in many regions e.g. L1630N, Lupus 3 dark cloud, and Taurus \citep[Paper\,I,][]{2003A&A...406.1001C,2004ApJ...616..998W}. Typically these exotic objects show abnormally large Balmer line $EWs$, and normal $EWs$ of \HeI\,5876~\AA, and the \CaII\, near-infrared triplet (8498, 8542, and 8662~\AA) \citep[see Paper\,I;][]{2003A&A...406.1001C}. One likely scenario for these  exotic objects is that they are harboring flared disks with high inclinations. In this case, the stellar photospheric light is largely absorbed by the material in the cold, flared outer disk. The light that we can see mainly comes from the photon scattering off the disk surface,  therefore much reduced. The optical emission lines with large $EWs$, e.g. Balmer lines and  forbidden oxygen emission lines, may arise in an outflow or disk wind whose scale is much larger than the central star allowing at least part of the line flux to reach us relatively unscreened. Emission lines like  \HeI\,5876~\AA, and the \CaII\, near-infrared triplet (8498, 8542, and 8662~\AA) that mainly form in the magnetospheric infall flows, close to the central stars, should be as obscured as the photospheric continuum.

We employ the SED fitting tool of \citet{2007ApJS..169..328R} to fit the SED of ID\#143. We excluded models with stellar spectral types  more than two subclass different from the observation, and defined the good-fit models as the models with $\chi^{2}$$-$$\chi^{2}$$_{\rm best}$$<$$2n$$_{\rm data}$, where $\chi$$^{2}$$_{\rm best}$ is the $\chi^{2}$ of the best-fitting model, and $n$$_{\rm data}$ is the number of data points for the fits.  We obtained two good-fit models (Model IDs: 3002401 and 3002775) which are shown in Fig.~\ref{fig:exotic}(b). The disk inclination angles of both models are $\sim$87$^{\circ}$ which supports the hypothesis that the subluminous objects are harboring disk systems with high inclinations.  The SED fits also give a better constraint on  the stellar  luminosity of ID\#143. From the stellar  luminosity, in combination with the stellar effective temperature from the optical spectrum, its mass and age is estimated to be 2.0--2.2\,\Msun\ and 0.4--0.6\,Myr from D08 tracks, 2.6--2.7\,\Msun\ and 1.3--1.8\,Myr from S00, 1.8--2.0\,\Msun\ and 0.4--0.6\,Myr from DM97, and 1.9\,\Msun\ and 1.7--2.4\,Myr from Pisa11. These age estimates are comparable to the ages of other YSOs in L1641 from each PMS evolutionary model.

Though the two good-fit models generally follow the shape of the observed SED of ID\#143, they fail to fit the 10\mum\ silicate feature (see Fig.~\ref{fig:exotic}(b)). The observed 10\mum\ silicate feature of ID\#143 is seen in emission, suggesting that its disk inclination should be substantially different from 90 degrees (edge-on) since the edge-on disk system usually show the 10\mum\ silicate feature in absorption \citep{2005ApJ...622..463P}. More detailed models as well as observations are required to better constrain the nature of the object ID\#143.

\subsubsection{The FU~Ori type object candidate}\label{sec:results:Fuori}

The source ID\#136 is one of the most interesting YSOs in our sample. It shows a P\,Cygni profile in the H$\alpha$ line in all of our spectra observed with VIMOS, Hectochelle, and Hectospec (see  Fig.~\ref{Fig:Halpha_SED_alpha1}).  In Paper\,I, we  proposed this object as an FU~ori candidate. This object has been further investigated in \citet{2012A&A...538A..64C}. It shows a 10\mum\ silicate feature in absorption \citep{2012A&A...538A..64C}, suggesting it is probably a Class\,I protostar. This source also shows a blue-shifted He\,I line in absorption \citep{2012A&A...538A..64C}, a characteristic feature of FU~ori objects \citep{2010AJ....140.1214C}. However, the non-detection of absorption features of the CO band-head lines longward of 2.29 \mum\ make this source less likely to be of the FU~Ori type \citep{2012A&A...538A..64C}. We have collected the near-infrared photometric data of ID\#136 in the literature  \citep{1989ApJS...71..183S,1994ApJS...90..149C,2006AJ....131.1163S}. The data span epochs from 1985/1986 to 2009 during which, the  $K$ magnitudes of the source range from 8.14 to 8.39. Thus it seems that  ID\#136 did not show big change on its brightness, which furthermore argue against it as a FU~Ori type object.

\section{Discussion}

\subsection{Disk Evolution}\label{sec:discussion:disk_frequency}

\subsubsection{Disk frequency}

 In the YSO catalog of L1641, {\newnewrev 1314} sources can be grouped into one of the four classes, i.e. Class\,I, Flat-spectrum, Class\,II, and Class\,III. {\lrev The TDs in our sample are considered as Class\,II sources.} We calculate the  disk fraction as $N({\rm II})$/$N({\rm II+III})$, where $N({\rm II})$ and $N({\rm III})$ are numbers of Class\,II and Class\,III sources, respectively, as done for Taurus by \citet{2010ApJS..186..111L}. The disk fraction of  our YSO sample is estimated to be {\newnewrev $\sim$51$\pm$2\% (533/1040). If we only count the YSOs with X-ray emission, the disk fraction is $\sim$36$\pm$3\% (166/466). The lower disk fraction among the X-ray emitting YSOs, compared with the whole YSO sample, may be due to that the XMM observations detect more Class\,III sources than Class\,II sources since X-ray observation is more sensitive to the former  than the latter \citep{1999ARA&A..37..363F}. As shown in Fig.~\ref{Fig:yso_dis}, there is a slight difference on the FOVs of the XMM survey and Spitzer survey. If we only count the YSOs within the common observed regions of both surveys, the disk fraction is $\sim$51$\pm$2\% (448/884) for all the catalogued YSOs, and disk fraction is $\sim$49$\pm$3\% (278/567) for ``bright'' YSOs ($K_{\rm s}$$\le$12.5\,mag). As discussed in  \sect\ref{Sec:completeness}, the census of diskless population with XMM data is relatively complete for Class\,III sources with $K_{\rm s}$$\le$12.5\,mag, and become very incomplete for fainter sources. Thus, we may expect a much higher disk fraction for all the catalogued YSOs than the one for the  brighter YSOs ($K_{\rm s}$$>$12.5\,mag). However, we do not see a significant difference on the disk fractions for the two populations. In this work, $\sim$460 Class\,III sources are identified from several spectroscopic surveys  \citep[][Paper\,I and this work]{1995PhDT..........A,2012ApJ...752...59H}. These surveys have selected  YSO candidates for observations partially from the optical color-magnitude diagrams. In these surveys, more than 200 Class\,III sources without X-ray emissions have been identified, which alleviate the incompleteness of Class\,III sources in our YSO census suffered from the XMM survey. In general, we conclude that disk fraction in L1641 is $\sim$50\%.}   

{\newnewrev The disk fraction in L1641 is slightly lower than that in Taurus (59$\pm$4\%) which is at a similar age to L1641 \citep{2010ApJS..186..111L}.  In Taurus, most of YSOs have been observed with mid-IR spectroscopy, which can be used to distinguish the YSOs with envelopes from those without envelopes \citep{2006ApJS..165..568F}. Using these data, 59\% of the ``Flat-spectrum'' YSOs are classified into  Class\,II \citep{2010ApJS..186..111L}. Assuming that the same fraction of the ``Flat-spectrum'' YSOs in L1641 is actually Class\,II sources, the disk fraction in L1641 would increase to 54$\pm$2\%, which is generally consistent with that in Taurus (59$\pm$4\%).} 

Furthermore, in L641  we have found {\newnewrev 73} TDs. Among them, {\newnewrev 65} have been confirmed and 8 are  TD candidates. The TDs count for {\newnewrev 14\%} of the disk population (Class II) in our YSO sample. Under the assumption that  Class\,II lifetime is 2\,Myr, the lifetime of TDs is estimated to be  $\sim$0.3\,Myr.

\subsubsection{Disk frequency as a function of stellar masses}

\begin{figure}
\centering
\includegraphics[width=\columnwidth]{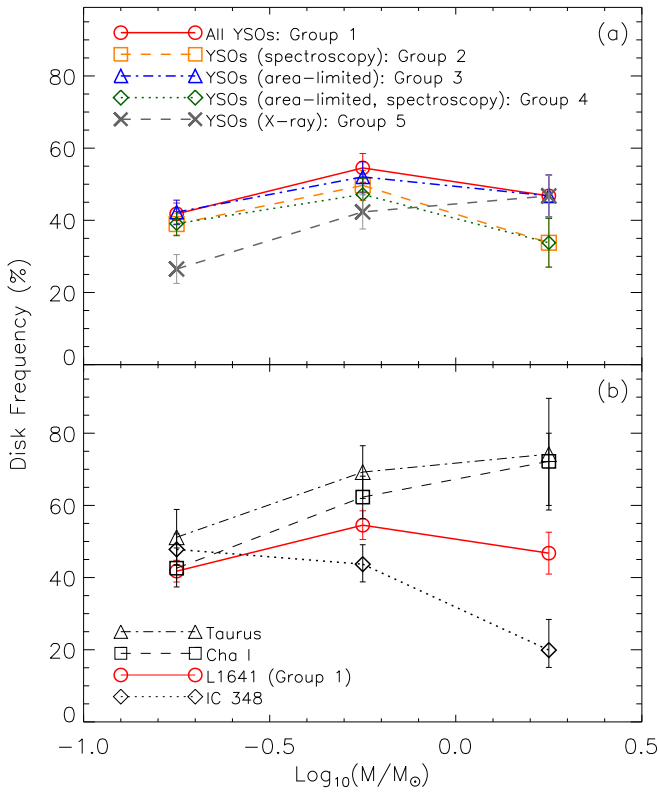}
\caption{\label{fig:df_mass} (a) The disk frequency  as a function of stellar mass  for all YSOs with masses$\ge$0.1\,\Msun\ in L1641 (open circles), for all YSOs with spectroscopy (open squares), YSOs within the common regions covered by the XMM and Spitzer surveys (area-limited, open triangles), area-limited YSOs with spectroscopy (open diamonds), and YSOs with X-ray emission (crosses). (b) The disk frequency  as a function of stellar mass  for all YSOs with masses$\ge0.1$\,\Msun\ in L1641 (open circles), Taurus \citep[open triangles, ][]{2010ApJS..186..111L},  Chamaeleon~I \citep[open squares, ][]{2008ApJ...675.1375L}, IC~348 \citep[open diamonds, ][]{2006AJ....131.1574L}.}
\end{figure}

{\newnewrev In our sample young stars, more than 800 sources have been observed with spectroscopy, thus their mass estimates are relatively reliable. For the other $\sim$200 YSOs without spectral types, we derived masses from their dereddened $J$-band photometry via the theoretical relation between stellar mass and the $J$-band photometry for PMS stars at an age of 1.5\,Myr (the median age of YSOs in L1641) from \citet{2008ApJS..178...89D}. In \fig\ref{fig:df_mass} we show the disk frequency as a function of stellar mass for five groups of YSOs\footnote{The mass bins used are, in log($M_*$/\Msun), $-$1 to $-$0.5, $-$0.5 to 0, and $>$0.}: (1) all YSOs with estimates of stellar masses from either spectroscopy or $J$-band photometry, (2) all YSOs observed with spectroscopy, (3) all YSOs within the common regions covered by the XMM and Spitzer surveys (See Fig.~\ref{Fig:yso_dis}), (4) the YSOs among Group 3 with spectroscopy, and (5) the YSOs with X-ray emission. As shown in  \fig\ref{fig:df_mass}, the trends of disk frequency with stellar mass for the YSOs of groups 1-4 are similar, and  their disk frequencies are almost constant as a function of stellar masses with a slight peak at log($M_*$/\Msun)$\approx-$0.25. The disk frequencies of the YSOs with X-ray emission are lower within the sub-solar mass range than those of Group 1--4, and generally increases with the increasing stellar masses, which may be because that the X-ray observations are very inefficient in detecting the very low-mass disked YSOs.}

The trend of disk frequency with stellar mass in this work is different from our finding in Paper\,I, in which the disk frequency increases with increasing stellar mass. In Paper\,I, we used a limited sample ($\sim$260) of YSOs, only including the spectroscopic YSOs, to construct the disk frequency vs. stellar mass relation. In this work, the combination of the XMM and Spitzer imaging surveys make our study less biased. As a comparison, in \fig\ref{fig:df_mass} we also show relations between the disk frequency and stellar mass for YSOs in IC~348, Chamaeleon~I, and Taurus \citep{2006AJ....131.1574L,2008ApJ...675.1375L,2010ApJS..186..111L}. The trend in L1641 that we identify is quite different from that seen in Taurus\citep{2010ApJS..186..111L}, or in  2\,Myr old  Chamaeleon~I \citep{2008ApJ...675.1375L}. In these regions, the disk frequency increases with the stellar mass.  In the somewhat older IC~348 (see \fig\ref{fig:df_mass}), the disk frequency shows a decreasing trend with stellar mass. {\lrev A similar trend has also been found in the Cep\,OB3b cluster \citep[3--5\,Myr][]{2012ApJ...750..125A}.} The trend of disk frequency with stellar mass in L1641 is different from any of the above regions. To interpret the different trends of disk frequency with stellar mass in these regions, one may need to know the disk properties around young stars with different masses, the star formation history, and the local environment in each region.

\subsubsection{Disks around very low mass stellar/substellar objects}

\begin{figure}
\begin{center}
\includegraphics[width=1.\columnwidth]{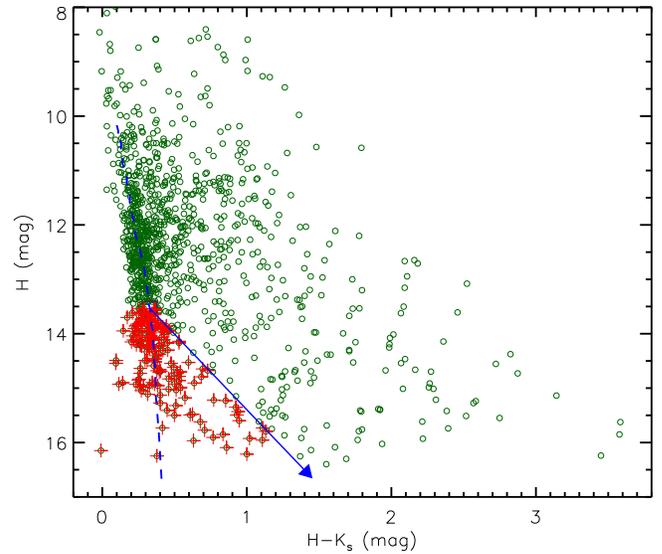}
\caption{\label{fig:BD} $H$ vs. $H$$-$$K_{\rm s}$ color-magnitude diagram for the YSOs (open circles) in L1641.  The dashed line shows the 2\,Myr PMS isochrone from \cite{1998A&A...337..403B}. The open box show the location of a  2\,Myr old {\rev low-mass PMS star with a mass of 0.1\,\Msun}. The pluses mark the {\newnewrev very low-mass stellar/substellar candidates} (M$_*\le$0.1\,\Msun). The arrow  shows the reddening vector with a length of A$_{K}$=2 \citep{1985ApJ...288..618R}.}
\end{center}
\end{figure}

In our spectroscopic sample, there are 3 young brown dwarfs (\#340, 396, and 402)  according to the evolutionary tracks from  \citet{1998A&A...337..403B}. Among them, the source \#404 harbors a circumstellar disk. For the sources without spectral types, we use the  $H$ vs. $H$$-$$K_{\rm s}$ color-magnitude diagram (Fig.~\ref{fig:BD}) to select very low mass stellar/substellar objects  by comparing with the theoretical PMS isochrone.  The ages of spectroscopic YSOs in L1641, estimated with the evolutionary tracks from \citet{1998A&A...337..403B}, show a broad distribution with a median age $\sim$2\,Myr.  In the figure, we also display  the 2\,Myr PMS isochrone \citep{1998A&A...337..403B}. We select very low mass  object candidates which are below the reddening vector of a 2\,Myr old,  0.1\,\Msun\ PMS star in the $H$ vs. $H$$-$$K_{\rm s}$ color-magnitude diagram, and find {\newnewrev 161} sources. In this sample {\newnewrev 115} sources have spectral-type estimates in the literature, and {\newnewrev 50\% of them have spectral types later than M5, and 83\% of then are later than M4}, suggesting a major fraction of our selected candidates are really very low mass objects. Among the {\newnewrev 161} very low mass stellar/substellar object candidates, the disk fraction is {\newnewrev 47\% (76/161)}, which is consistent with that we found for all the YSOs in L1641. 

\begin{figure}
\begin{center}
\includegraphics[width=1.\columnwidth]{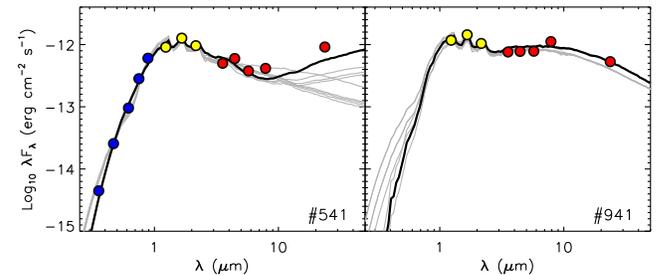}
\caption{\label{fig:BD_SED} The SEDs of two very low-mass brown dwarf candidates with disks. The solid lines in each panel shows the top 10 best-fitting model SEDs using the online fitting tool from \citet{2012MNRAS.423.1775M}, and the dark solid line is the best-fit model.}
\end{center}
\end{figure}

 Figure~\ref{fig:BD_SED} shows the SEDs of two  brown dwarf candidates (\#541 and 941) with disks. The masses of the two sources are estimated to be  $\sim$0.04\,\Msun using the 2\,Myr PMS isochrone from \cite{1998A&A...337..403B} in Fig.~\ref{fig:BD}. The SEDs of the two sources are fitted using the online fitting tool provided by \citet{2012MNRAS.423.1775M}. The fitting tool employs a set of parameters including stellar mass, accretion rate, disk mass,  etc. \citep[see detailed description in ][]{2012MNRAS.423.1775M}. When fitting the SEDs, we fix the stellar masses to be 0.04\,\Msun, and free other parameters. In  Fig.~\ref{fig:BD_SED} we show the 10 best fitting model SEDs for each source. Among the top 10 best-fitting models, the fitted ages of the two sources are both 1\,Myr, and the disk radii are  around 100-300\,AU. From the fitting, the disk around the source \#541 may have a higher inclination angle  (40$^{\circ}$-60$^{\circ}$) and lower mass (4$\times$10$^{-4}$--4$\times$10$^{-5}$\,\Msun) than that of  the source \#944 in which the disk is almost face-on, and the disk mass is $\sim$4$\times$10$^{-4}$\,\Msun.

\subsubsection{Ages of the different populations}
According to traditional low-mass star formation theory, CTTSs evolve into WTTSs as circumstellar disks dissipate. Thus, one would expect that WTTSs are older than CTTSs.  However, the observations of many star forming regions do not support this theory \citep{1998ApJ...497..736H,2001AJ....121.1030H,2002AJ....123..304H,2005AJ....129..829D}, and there are only a few cases in which the WTTSs are found to be older than CTTSs \citep{1995ApJ...452..736H,2007A&A...473L..21B}. In Paper\,I, we showed that the WTTSs without disks are statistically older than CTTSs in L1641, though both show age distributions with large spreads. In this work, instead of classifying the YSOs into CTTSs or WTTSs when comparing their age distributions, we divide the YSOs into three groups: YSOs with optically-thick disks, TD objects, and diskless YSOs. In this analysis we include both YSOs from this work and from Paper\,I.

\begin{figure}
\begin{center}
\includegraphics[width=\columnwidth]{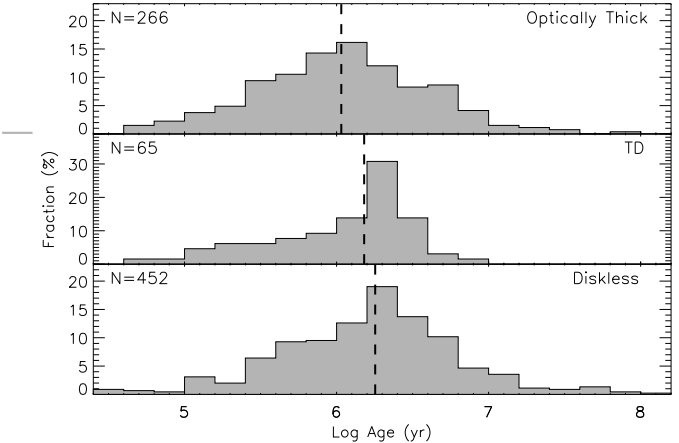}
\caption{Histograms showing the age distribution for  YSOs with  optically thick disks, transition disks, and without disks, respectively.}\label{fig:age_dis}
\end{center}
\end{figure}

The age distributions of the three populations are shown in \fig\ref{fig:age_dis}.  The median ages of the YSOs with optically-thick disks, transition disk objects, and diskless YSOs are $\sim$1.1\,Myr, $\sim${\newnewrev 1.5}\,Myr, and  $\sim${\newnewrev 1.8}\,Myr, respectively. Though the median ages of different populations  increase  with their evolutionary stages, each of them shows a very broad distribution. A Kolmogorov-Smirnov (KS) test reveals a  relatively low  probability {\newnewrev ($P$\,$\sim$3$\times$10$^{-6}$)} for the optically-thick disks  and diskless YSOs to be drawn randomly from the same age distribution. The probability that the  TDs  and  diskless YSOs are drawn randomly from the same population is also low ($P$\,$\sim$0.02), whereas the age distributions of  the  optically-thick disks  and TDs are less indistinguishable {\newnewrev ($P$\,$\sim$0.2)}.

\fig\ref{fig:age_dis} shows that 34\% of the diskless YSOs have ages less than 1\,Myr. Here, the presence of a disk around a YSO is determined by the excess emission at infrared wavelengths $\lesssim$24\mum, which is only sensitive to the disk  at several AU to tens of AU, depending on the effective temperature of central stars. A few of these ``diskless'' YSOs may still have disks but with large inner holes. If these YSOs are really diskless or harbor disks with big inner holes, then there must be an efficient mechanism to dissipate the inner disks. In Fig.~\ref{fig:age_dis}, the age distribution of the TDs hint that there may be bimodality with one relatively flat distribution at age $\lesssim$1\,Myr, and one peak at $\sim$2\,Myr. The two populations of TDs are more clearly distinguished in the HR diagram (see Fig.~\ref{Fig:HRD}). The ``young'' TDs  ($<$1\,Myr)  require a fast  mechanism to dissipate the inner regions of their disks. One  promising  mechanism could be the interaction between the disk and a close binary \citep{1993prpl.conf..749L}. The tidal interaction between the disk and a close companion can quickly clear away the material from the binary orbit, form a gap in the disks, and finally terminate the accretion from the outer disk to the inner disk \citep{1993prpl.conf..749L}, thus accelerating the disk destruction process. Surveys toward nearby field G and M type dwarfs have found about 50\% of them have companions, with separation peaking at  tens of AU \citep{1992ApJ...396..178F}. Furthermore, surveys towards star forming regions have revealed the binary fraction to be even higher than in the field \citep{1993AJ....106.2005G,1993A&A...278..129L,1997ApJ...481..378G,2008ApJ...683..844L}. In general, such a high fraction of  close binaries can explain these young diskless YSOs or TDs. However, the recent near-infrared interferometric observations of five TDs in Taurus have excluded the possibility of having a companion with a flux ratio $\ge$0.05, and a separation ranging from 0.35--4\,AU \citep{2010ApJ...710..265P}. {\newrev \citet{2009ApJ...704..989A} have proposed that young TD formation can be due to planet formation, while older ones could be formed due to photoevaporation, which works in disks with substantial evolution. \citet{2011ApJ...742...39S} show that many TDs in the 4 Myr old cluster, Tr 37, can be due to grain evolution. Thus, a bimodal behavior in the age distribution of TDs is expected.}



\subsubsection{The evolution of IRAC spectral slope}
\begin{figure}
\begin{center}
\includegraphics[width=\columnwidth]{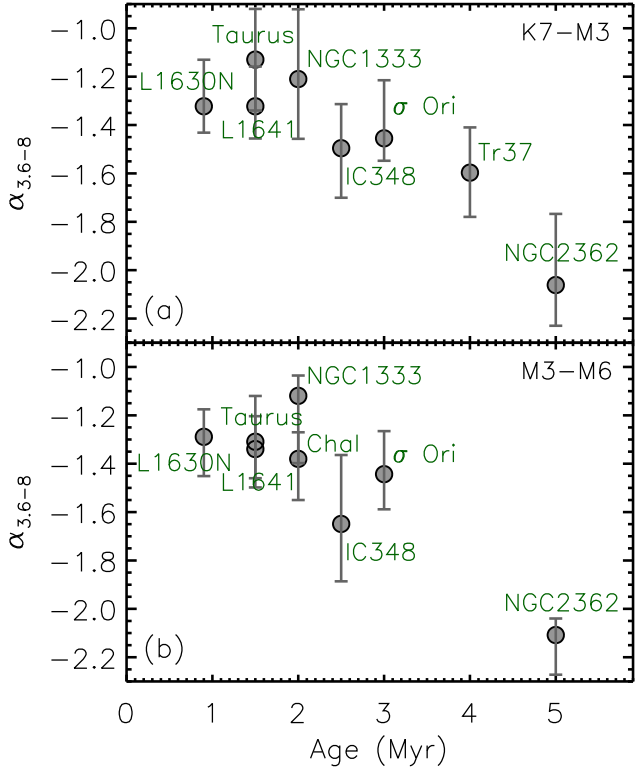}
\caption{The mean spectral slope $\alpha_{3.6-8}$ of the disk population in  different star-forming regions vs. their ages.  Panel (a) is for young stars with spectral type of K7-M3, and Panel (b) is for young stars with spectral type of M3-M6. Cha\,I and Tr\,37 are missing from panel (a)  and (b), respectively, due to the lack of members within those spectral type ranges.}\label{Fig:alpha_evolve}
\end{center}
\end{figure}

In Fig.~\ref{Fig:alpha_evolve}, we show the average  $\alpha_{3.6-8}$ of the disk population in several star-forming regions vs. their ages. The regions include  NGC\,2068/2071 ($\sim$1\,Myr, Paper\,I), L1641 ($\sim$1.5\,Myr, Paper\,I and this paper), Taurus \citep[$\sim$1.5\,Myr,][]{2010ApJS..186..111L}, {\lrev NGC\,1333 \citep[$\sim$2\,Myr,][]{2009AJ....137.4777W, 2010AJ....140..266W}}, Cha\,I \citep[$\sim$2\,Myr,][]{2008ApJ...675.1375L}, IC\,348 \citep[$\sim$2.5\,Myr,][]{2006AJ....131.1574L,2007AJ....134..411M}, $\sigma$\,Ori \citep[$\sim$3\,Myr,][]{2007ApJ...662.1067H}, Tr\,37 \citep[$\sim$4\,Myr,][]{2006ApJ...638..897S}, and NGC\,2362 \citep[$\sim$5\,Myr,][]{2009ApJ...698....1C}. We separate the disk population into two groups according to the spectral types of their central stars. For $\sigma$\,Ori, NGC\,2362, and NGC\,1333, there are no available spectral types in the literature for most of their members and we use their $J$-band magnitude to estimate their spectral types using the isochrone at their corresponding ages. In Fig.~\ref{Fig:alpha_evolve} 
a clear trend of $\alpha_{3.6-8}$ decreasing with age can be noted for each spectral type range, as noticed in \citet{2006ApJ...638..897S}.  The average $\alpha_{3.6-8}$ decreases slowly over the first 3--4\,Myr with most of disk population harboring optically-thick inner disks ($\alpha_{3.6-8}$$>-$1.8). A fast decrease in the average  $\alpha_{3.6-8}$ can be noted for the cluster NGC\,2362  at an age of $\sim$5\,Myr.

\subsubsection{The two paths of protoplanetary disk evolution}

Spitzer surveys of young clusters, such as IC\,348, NGC\,2362, $\eta$\,Cha,  the Coronet cluster, and Tr\,37  suggest that there are qualitatively two evolutionary paths for disks progressing from a primordial disk to a debris disk: (1) radially depleted disks, and (2) globally depleted disks\citep{2006AJ....131.1574L,2008ApJ...687.1145S,2009ApJ...701.1188S,2011ApJ...742...39S,2009AJ....138..703C,2009ApJ...698....1C}.  Each path can be distinguished based on their SEDs. The radially depleted disks show little or no excess emission in the shorter IRAC bands, but strong excess emission at 24\mum, suggesting that they are  dissipating in an inside-out fashion.  The globally depleted disks show more or less uniformly reduced infrared excess emission compared to primordial disks over all wavelengths out to 24\mum. This indicates that there is a reduction in their effective disk height {\lrev for lack of small dust grains in disks \citep{2011ApJ...732...24C,2013A&A...551A..34S}}. The reduced disk height can decrease the fraction of the stellar energy that is absorbed and reprocessed by the disk, which results in an SED with reduced infrared excess. 

\begin{figure}
\begin{center}
\includegraphics[width=\columnwidth]{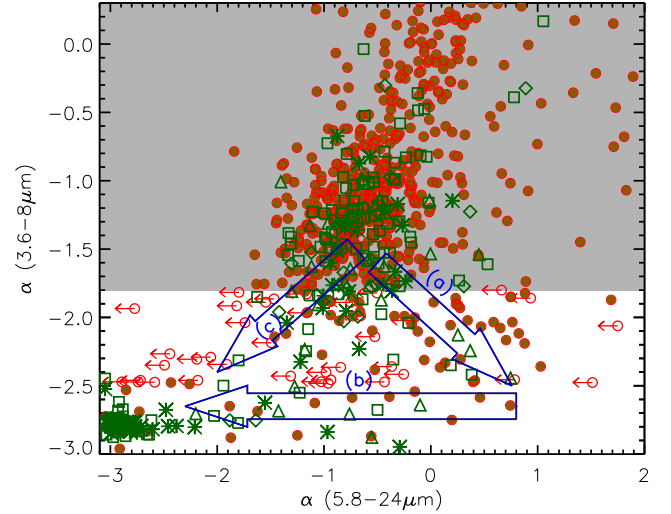}
\caption{Spitzer infrared slopes of YSOs.  The filled circles are the YSOs in L1641 with 24\mum\ detections. The open circles with arrows are YSOs in L1641 without 24\mum\ detections.  The upper limits at 24\mum\ were used to calculate  $\alpha_{5.8-24}$. The YSOs in the relatively older clusters $\sigma$\,Ori (open boxes), NGC\,2362 (open triangles), Ori~OB1 Association (open diamonds), and $\lambda$\,Ori (crosses) are also shown. The big arrows mark two possible paths for disk evolution. The arrows (a) and (b) show the path for radially  depleted evolution, and the arrow (c) display the path for globally depleted evolution. The disks that appear in the grey-shaded area harbor optically thick disks($\alpha_{3.6-8}$$>-$1.8).}\label{Fig:disk_evolve}
\end{center}
\end{figure}

In Fig.~\ref{Fig:disk_evolve},  we show the dereddened spectral slopes $\alpha_{3.6-8}$ and  $\alpha_{5.8-24}$ for the YSOs in L1641. For the sources which are undetected at 24\mum, the 24\mum\ upper limits are used to calculate $\alpha_{5.8-24}$. As a comparison, we also show the YSOs in several older clusters,  $\sigma$\,Ori \citep[$\sim$3\,Myr,][]{2007ApJ...662.1067H} ,NGC\,2362 \citep[$\sim$5\,Myr][]{2007AJ....133.2072D},  $\lambda$\,Ori \citep[$\sim$5\,Myr,][]{2010ApJ...722.1226H}, and Ori~OB1 Association \citep[$\sim$5-10\,Myr][]{2007ApJ...671.1784H}.   As discussed above, the two disk evolution processes can be distinguished by their SEDs, which can be quantified by their  spectral slopes. For the radially depleted disks, one would expect $\alpha_{3.6-8}$ decreases while  $\alpha_{5.8-24}$ increases due to the reduced fluxes at 5.8 and 8.0\mum. With the dissipation of the inner disk regions, $\alpha_{3.6-8}$ will be photospheric, and then as the outer disk is removed $\alpha_{5.8-24}$ starts to decrease. In globally depleted disks $\alpha_{3.6-8}$ and $\alpha_{5.8-24}$ decrease simultaneously. In  Fig.~\ref{Fig:disk_evolve}, the evolved disks ($\alpha_{3.6-8}$$<-$1.8) are distributed around the triangle (marked by three arrows), which can clearly explained by the two paths of disk evolution as described above.


\subsection{Accretion}

\subsubsection{Why are accretion rates  related to stellar masses?}

\begin{figure}
\begin{center}
\includegraphics[width=\columnwidth]{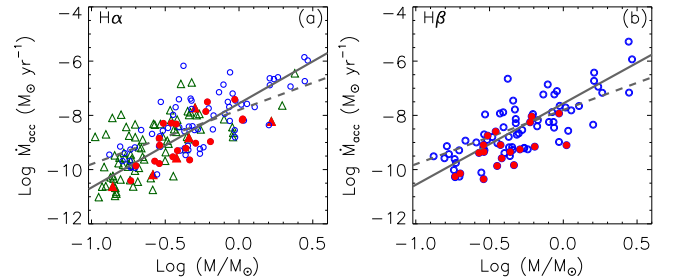}
\caption{The relation between accretion rates deduced from the H$\alpha$(a) and H$\beta$(b) line luminosity and stellar mass for the CTTSs in this paper. In panel (a), the (open and filled) circles are for the YSOs in our Hectospec survey, and the (open and filled) triangles are for the YSOs using the data from \citet{2012ApJ...752...59H}. The filled circles and triangles are for TDs. In Panel (b), the symbols are the same as in Panel (a). The slopes for the $\dot{M}_{\rm acc}\propto M_*^2$  relation (the dash line), and for  $\dot{M}_{\rm acc}\propto M_*^3$ relation (the solid line) are presented. Note that these lines are merely indicative and not fits to the data in each panel. }\label{Fig:acc_mass}
\end{center}
\end{figure}

Figure~\ref{Fig:acc_mass} shows the relation between the mass accretion rate, derived from the H$\alpha$ and H$\beta$ line luminosity, and stellar mass for the YSOs in L1641, which is consistent with the  $\dot{M}_{\rm acc}\propto M_*^{\alpha}$ relation with $\alpha\sim$2$-$3 as suggested in the literature \citep{2004AJ....128.1294C,2005ApJ...625..906M,2005ApJ...626..498M,2006A&A...452..245N,2006A&A...459..837G,2008ApJ...681..594H,2008A&A...481..423G}.  A large scatter in this relation is also  noted. Figure~\ref{Fig:acc_mass} also shows the  $\dot{M}_{\rm acc}$  vs. $M_*$ relation for  $\sim$20 TDs. These objects generally follow the same trend as the normal CTTSs. {\lrev Furthermore, TDs in Fig~\ref{Fig:acc_mass}(b) seems to show slightly lower accretion rates than the normal disk populations at the same stellar masses, which contradicts those in Fig~\ref{Fig:acc_mass}(a). To understand it, we need to distinguish the better tracer for accretion from H$\alpha$ and H$\beta$, which is beyond the scope of this paper.}

\begin{figure}
\begin{center}
\includegraphics[width=\columnwidth]{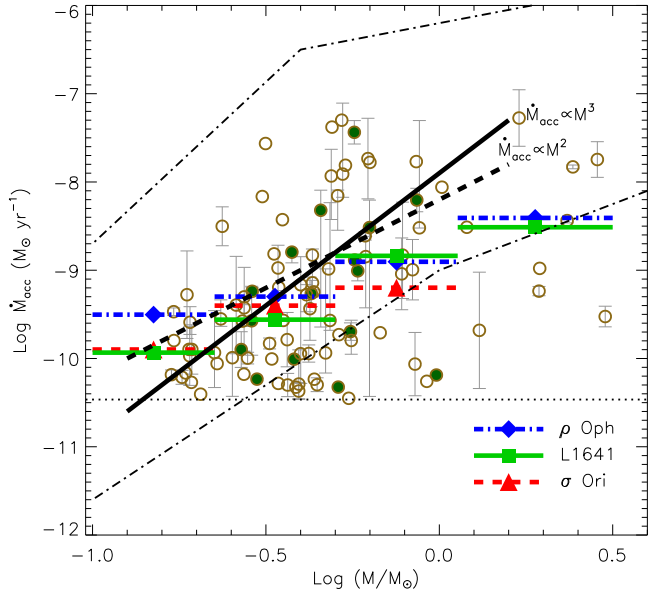}
\caption{The relation between accretion rates deduced from the $FW_{H\alpha, 10\%}$ of the H$\alpha$ line profiles and stellar mass. The filled circles are for the TDs.  The open circles represent the other disk population in L1641. The sources are considered accretors when $FW_{H\alpha, 10\%}$$>$250\kms, which  corresponds to an accretion rate of 3.4$\times$10$^{-11}$\,M\accunit\ marked with the dotted line in the figure. For the sources with multiple observations, the median accretion rates are used for the plot with error bars showing their minimum and maximum accretion rates. The YSOs are divided into four mass bins. The thick solid lines centered on filled boxes show the median accretion rate within these mass bins. As a comparison, the median accretion rates for YSOs in $\rho$\-Oph and $\sigma$\,Ori are shown as the thick dashed lines centered on filled diamonds, and dash-dotted lines centered on filled triangles, respectively \citep[]{2011A&A...525A..47R}. The slopes for the  $\dot{M}_{\rm acc}\propto M_*^2$  relation (the dark dash line), and for  $\dot{M}_{\rm acc}\propto M_*^3$   relation (the dark solid line) are presented. The dash-dotted lines show the accretion rate detection limits for the T~Tauri stars in the literature (see Paper\,I).}\label{Fig:FW_acc_mass}
\end{center}
\end{figure}

In Fig.~\ref{Fig:FW_acc_mass}, we present another $\dot{M}_{\rm acc}$  vs. $M_*$  plot for the YSOs in L1641. The accretion rates here are  estimated from the $FW_{H\alpha, 10\%}$ of the high-resolution H$\alpha$ line profiles in the Hectochelle spectra. Similar to the result in Fig.~\ref{Fig:acc_mass}  and in the literature, the accretion rates generally increase with stellar masses with a large scatter in $\dot{M}_{\rm acc}$ at similar masses.  We divided the YSOs into four mass bins, and estimated the median accretion rates in each mass bin as did in \cite{2011A&A...525A..47R}. For comparison we show the median accretion rates within the same mass bins  for YSOs in $\rho$~Oph and $\sigma$~Ori \citep{2011A&A...525A..47R}. The L1641 region has an age between the $\rho$~Oph ($<$1\,Myr) and $\sigma$~Ori($\sim$3\,Myr) regions.  Within each mass bin, the median accretion rate in $\rho$~Oph is higher than that in  $\sigma$~Ori, indicating a trend of decreasing accretion rate with age. The YSOs in L1641 show median accretion rates similar to those in  $\rho$~Oph within the bins for masses\,$>$0.5\,\Msun, and similar to those in $\sigma$~Ori within the bins with masses\,$<$0.5\,\Msun, which may indicate that accretion rate decreases faster with age in the low-mass range than in the high-mass range.

\begin{figure}
\begin{center}
\includegraphics[width=\columnwidth]{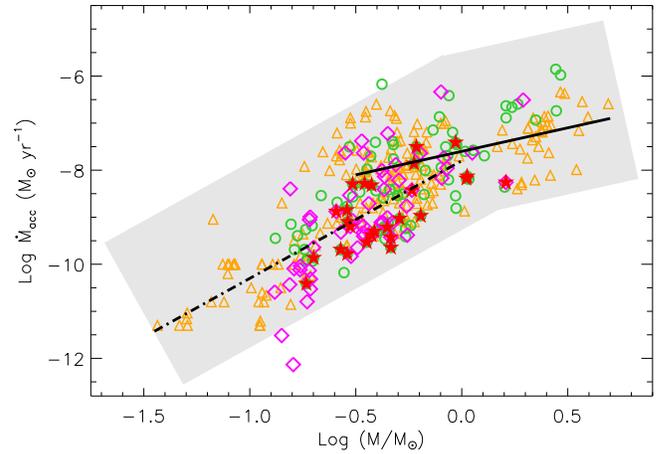}
\caption{The relation between accretion rate from H$\beta$  emission line luminosity and stellar mass for the CTTSs in this work (open  circles), Paper\,I (open diamonds), and other literature (open triangles). The filled {\lrev star symbols} are the TDs in L1641 and L1630N. We also include the  $\dot{M}_{\rm acc}\propto M_*$  relation (black solid line), and the    $\dot{M}_{\rm acc}\propto M_*^{2.5}$  relation (black dash-dotted line). Note that these lines are merely indicative and not fits to the data in each panel.}\label{Fig:acc_mass_all}
\end{center}
\end{figure}

\citet{2006ApJ...648..484H} discussed the angular momentum transport in disks. The  magneto-rotational instability \citep[MRI,][]{1998RvMP...70....1B} is the key mechanism for the  angular momentum transport. For the MRI to take effect, ionization in the disks is required. There are different sources that can ionize the gas in disks, including cosmic rays or stellar UV and X-rays photons impinging on the disk surface \citep{1996ApJ...457..355G,1997ApJ...480..344G}, and  thermal ionization driven by either viscous dissipation within the disk or irradiation of the disk surface by the central star \citep{2006ApJ...648..484H}. \citet{2006ApJ...648..484H} considered  the two limiting cases: (1) a lower-mass disk, in which the entire column is ionized for an active magneto-rotational instability (MRI), which is subject to full viscous evolution, resulting in $\alpha\approx$2.5; and (2) a higher-mass disk, in which thermal ionization by irradiation from the central star is dominant, causing an MRI-active surface layer resulting in $\alpha\approx$1. To test this scenario, we have collected a large set of YSOs with accretion rate estimates from the literature \citep[][Paper\,I]{1998ApJ...492..323G,2003ApJ...582.1109W,2003ApJ...592..266M,2003ApJ...583..334H,2004AJ....128.1294C,2005ApJ...625..906M,2006A&A...452..245N,2005ApJ...626..498M,2006A&A...459..837G,2008ApJ...681..594H,2008AJ....136..521D}, within the star forming regions Taurus-Aurigae, IC\,348, $\rho$\,Oph, $\lambda$\,Ori, Orion OB,  upper Scorpius, and L1641, and L1630N. All these data are shown in Fig.~\ref{Fig:acc_mass_all}. The dependency of  $\dot{M}_{\rm acc}$ on $M_{\star}$ is steeper for the low-mass disks around low-mass stars and brown dwarfs than for higher mass disks around solar to intermediate mass stars, which may support the scenario  proposed by \citet{2006ApJ...648..484H}. {\lrev However, Given the large scatter in Fig.~\ref{Fig:acc_mass_all}, the difference on dependency of  $\dot{M}_{\rm acc}$ on $M_{\star}$ in different mass ranges is only marginally significant. Furthermore, it is possible that the ages of intermediate-mass stars are older than the sub-solar mass stars in Fig.~\ref{Fig:acc_mass_all}. However, There are a large uncertainty in estimating ages of intermediate-mass stars from HR diagram \citep{1999NewAR..43....1H}, which complicates the  comparison.}

\subsubsection{Can accretion variability  explain the scattering in $\dot{M}_{\rm acc}$ vs. $M_*$?}


\begin{figure}
\begin{center}
\includegraphics[width=1\columnwidth]{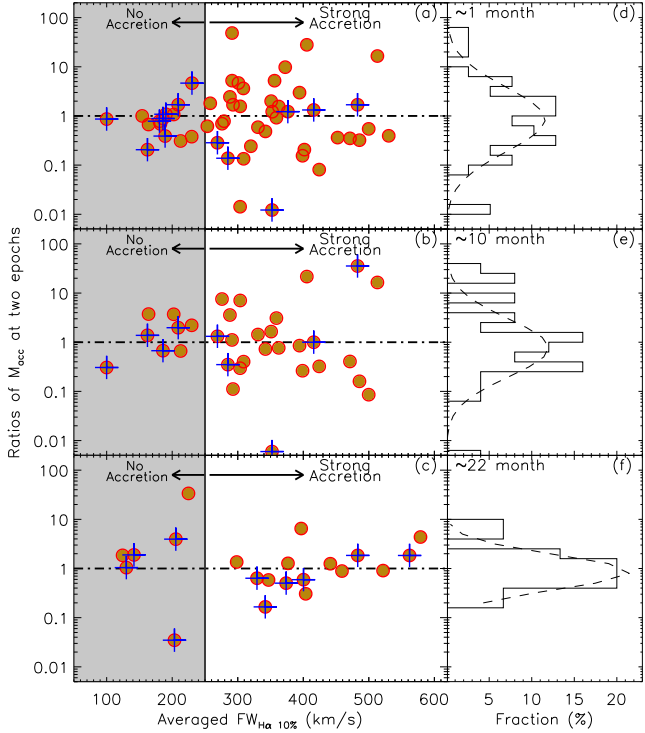}
\caption{Left Panels: The ratio of accretion rates ($\dot{M}_{\rm acc}$) observed at two epochs separated by $\sim$1, 10 , and 22 months vs. averaged $FW_{H\alpha, 10\%}$ for the disk population (filled circles) in L1641.  The pluses mark the TDs. In the grey filled region, the YSOs have $FW_{H\alpha, 10\%}$$<$250\kms, thus considered as not accreting. Right Panels: The distribution of ratios of $\dot{M}_{\rm acc}$ for the accretors ($FW_{H\alpha, 10\%}$$>$250\kms) shown in the left panels. The dashed line shows the fit to the distribution using the Gaussian function with a standard deviation $\sim$0.6 for (d) and (e), and  $\sim$0.3 for (f).}\label{Fig:Acc_var}
\end{center}
\end{figure}

 Though the correlation between accretion rate and stellar mass is obvious for a large sample of young stars, individual objects with similar ages and masses commonly scatter around the average relation by up to 2 orders of magnitude. Observations show that the amplitude of accretion variation for CTTSs is large\citep{1995ApJ...449..341J,1996A&A...314..835G,2001AJ....122.3335A,2002ApJ...571..378A,2010ApJ...710..597S}. Thus, the temporal accretion variation is proposed as one potential explanation for the large scattering in the $\dot{M}_{\rm acc}$ vs. $M_*$ plot. To test this, we have performed a multi-epoch high-resolution spectroscopic survey of YSOs with Hectochelle covering baselines of  $\sim$1, 10, and $\sim$22 months. A total of 52 disked YSOs were observed at two epochs separated by an interval of  $\sim$1\,month, 33 YSOs were observed at two epochs separated by  $\sim$10\,months, and 21 YSOs were observed at two epochs  separated by $\sim$22\,months. Figure~\ref{Fig:Acc_var}(a,b,c) shows the  the ratio of accretion rates ($\dot{M}_{\rm acc}$) estimated from $FW_{H\alpha, 10\%}$ at epochs separated by $\sim$1, 10, and 22 months, respectively,  against their averaged $FW_{H\alpha, 10\%}$. The accretion rates are estimated from $FW_{H\alpha, 10\%}$  via the relation between the accretion rate and $FW_{H\alpha, 10\%}$ from \citet{2004A&A...424..603N}. Figure~\ref{Fig:Acc_var}(d,e,f) shows the distribution of accretion rate ratios for the  accreting YSOs ($FW_{H\alpha, 10\%}>$250\,km/s )  which are shown in left panels of Fig.~\ref{Fig:Acc_var}, respectively.  The Gaussian fits to these distributions have standard deviations in the logarithm of accretion rate ratio of $\sim$0.66$\pm$0.09, 0.59$\pm$0.12, and 0.35$\pm$0.06, respectively. The small standard deviation for the 22 month interval could be due to the low number statistics.  Thus, our data suggest that the accretion variability cannot explain the two orders of magnitude scatter in accretion rates at similar masses.  Our result confirms the finding in \citet{2009ApJ...694L.153N} and \citet{2010ApJ...710..597S}. \citet{2009ApJ...694L.153N} have investigated the accretion variability towards a sample of  40 young stars in Taurus and Cha\,I based on the Ca\,II fluxes, and concluded that  the intrinsic source variability cannot be the dominating factor in the large scatter of the $\dot{M}_{\rm acc}$ versus $M_*$ relationship. 

{\lrev The uncertainties in calculating $\dot{M}_{\rm acc}$ can  induce a scatter of the $\dot{M}_{\rm acc}$ versus $M_*$ relationship. \citet{2008ApJ...681..594H}  discuss the possible uncertainty contributors, including distance, stellar parameters (extinction, effective temperature, mass, and radius), assumption of accretion geometry, conversion from UV excess continuum emission to accretion luminosity. They concluded these uncertainties can lead to a $\sim$0.6\,dex uncertainty in $\dot{M}_{\rm acc}$ for calculating $\dot{M}_{\rm acc}$ of young stars in Taurus from UV excess continuum emission. When $\dot{M}_{\rm acc}$ is calculated from emission lines, the uncertainties ($\sim$0.8--1.0\,dex) in $\dot{M}_{\rm acc}$ could be bigger than those from  UV excess emission due to large uncertainties in the conversions of line quantities (line luminosities or $FW_{H\alpha, 10\%}$) to accretion quantities ($L_{{\rm acc}}$ or $\dot{M}_{\rm acc}$), which are constructed mostly based on non-simultaneous data \citep[Paper\,I;][]{1998AJ....116.2965M,2004A&A...424..603N,2008ApJ...681..594H,2012A&A...548A..56R}. The combination of uncertainty (0.6--1.0\,dex) in measurement and accretion variability (0.5\,dex) may account for the  large scatter of the $\dot{M}_{\rm acc}$ versus $M_*$ relationship. 
}

\begin{figure}
\begin{center}
\includegraphics[width=1\columnwidth]{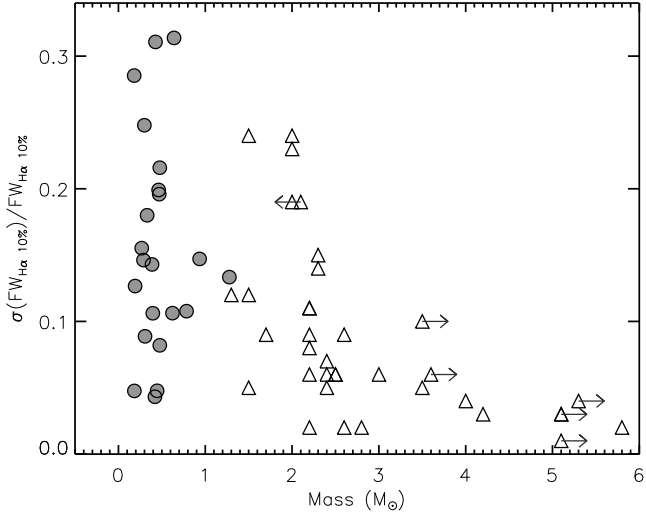}
\caption{ Relative variability of $FW_{H\alpha, 10\%}$ vs. stellar mass. The filled circles show the disk population with three-epoch Hectochelle data. The triangles present the sources from \citet{2011A&A...529A..34M}. The left pointing and right pointing arrows mark the lower and upper limits for the stellar masses.}\label{Fig:var_FW_mass}
\end{center}
\end{figure}

 We estimate the relative variability of $FW_{H\alpha, 10\%}$ ($\sigma/<FW_{H\alpha, 10\%}>$)  for the disk population with three-epoch Hectochelle data, and plot this against stellar masses in Fig.~\ref{Fig:var_FW_mass}. As a comparison,  we also show a sample of intermediate-mass stars collected from \citet{2011A&A...529A..34M}. In the figure,  the relative variability of $FW_{H\alpha, 10\%}$ show a large scatter for stars less than 3\,\Msun, and the scatter increases with decreasing mass. The values of  $\sigma/<FW_{H\alpha, 10\%}>$ are much smaller and show less scatter for the Herbig\,Be stars ($M_*$$>$3\,\Msun). \citet{2011A&A...529A..34M}  suggest that the distinction between  $\sigma/<FW_{H\alpha, 10\%}>$ for  Herbig\,Ae stars and  Herbig\,Be stars is due to the different formation mechanisms of the H$\alpha$ emission line. For Herbig\,Ae stars, H$\alpha$ line fluxes may be originate partially  from the magnetospheric accretion flow, and thus are sensitive to accretion variations.  In Herbig\,Be stars H$\alpha$ emission may come from the disk surface \citep{2002MNRAS.337..356V,2007MNRAS.377.1363M} and thus be less sensitive to accretion variations. In this work, we extend the study of \citet{2011A&A...529A..34M} to the sub-solar mass regime, and find that the mean $\sigma/<FW_{H\alpha, 10\%}>$ in the low-mass range is larger than that of Herbig\,Ae stars. This may be due to two factors: (1) the accretion of sub-solar mass stars is more variable than Herbig\,Ae stars or (2) accretion rate variability is independent of mass, but the H$\alpha$ emission line in Herbig\,Ae stars includes a contribution from a disk wind, which is less susceptible to accretion rate fluctuations, diluting the fluctuations in the H$\alpha$ line.


\subsubsection{How does Accretion work in disks at different evolutionary stages?}
\begin{figure}
\begin{center}
\includegraphics[width=\columnwidth]{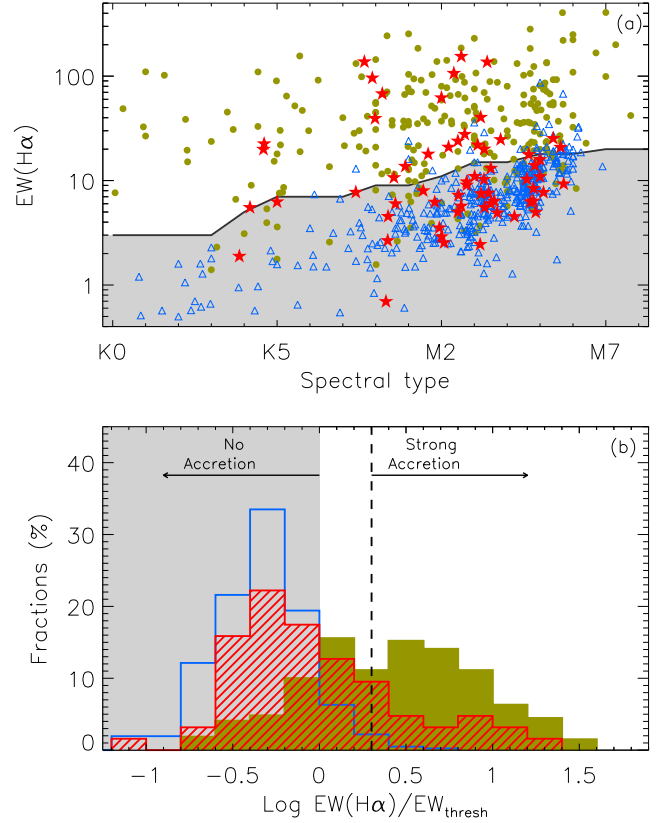}
\caption{(a): H$\alpha$ $EW$s vs. spectral type for known YSOs in L1641. The filled circles are for YSOs with optical thick disks, and triangles for YSOs without disks (in this work and Paper\,I). The {\lrev star symbols} show the TDs in L1641. (b): the distribution of the ratio between the observed H$\alpha$ $EW$ and the $EW$ threshold, separating CTTSs or WTTSs, for the three populations in the left panel: YSOs with optically thick disks (filled histogram), YSOs without disks (open histogram), and TDs (line-filled histogram). The dash line marks  $EW$(H$\alpha$)/$EW$$_{thresh}$=2. }\label{fig:EW_acc}
\end{center}
\end{figure}

In Fig.~\ref{fig:EW_acc}(a), we compare the H$\alpha$ $EW$s of our TDs with those of the YSOs with optically thick disks($\alpha_{3.6-8}$$\ge-$1.8) and diskless PMS stars. The YSOs with optically thick disks usually show strong  H$\alpha$ emission, while the PMS stars without disks present weak H$\alpha$ emission. The TDs straddle this boundary with a major fraction of TDs showing no active accretion, and a fraction of TDs showing strong accretion activity. In  Fig.~\ref{fig:EW_acc}(b), we show the distribution of logarithmic ratio between the observed H$\alpha$ $EW$ and the $EW$ threshold for the three populations in  Fig.~\ref{fig:EW_acc}(a). Here,  $EW$ threshold is the spectral type dependent threshold used to classify the YSOs into CTTSs or WTTSs (see Paper\,I). Using these $EW$ thresholds, {\newnewrev  79$\pm$5\%} of YSOs with optically thick disks are classified as CTTSs while only {\newnewrev 40$\pm$8\%} of TDs are CTTSs. And  {\newnewrev  21$\pm$6\%} of TDs show ``strong'' accretion activity, defined as having H$\alpha$ $EW$ of more than 2 times the $EW$ thresholds. For the YSOs with optically thick disks, this fraction is  {\newnewrev  60$\pm$5\%}. Therefore, there are significantly fewer strong accretors among TDs compared with optically thick disks. For the 6 globally depleted disks shown in Fig.~\ref{Fig:Homo_disk_SED}, only the source \#16 shows an H$\alpha$ $EW$ which is marginally above the threshold for classifying it as a CTTS. The active accretion in source \#16 is confirmed by the H$\alpha$ emission line profile ($FW_{H\alpha, 10\%}$$\sim$387\kms) observed with Hectochelle (see Fig.~\ref{Fig:Halpha_SED_alpha4}). Thus, statistically only 1/6 of YSOs with globally depleted disks are CTTSs, which is much lower than the YSOs with optically thick disks or TDs.

\begin{figure}
\begin{center}
\includegraphics[width=\columnwidth]{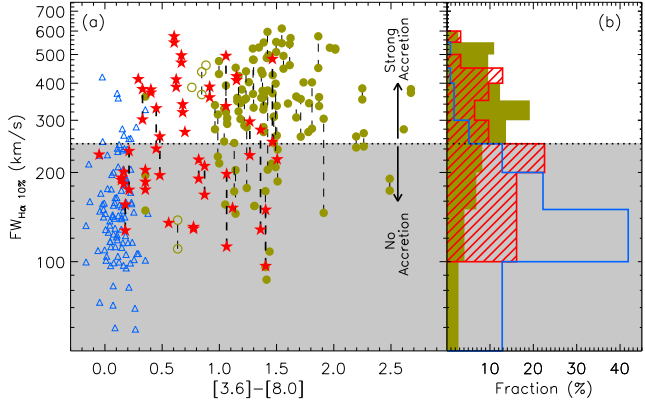}
\caption{(a) The  $FW_{H\alpha, 10\%}$ vs. [3.6]$-$[8.0] colors for YSOs in L1641. The open triangles show the sources without disks, the open circles display the YSOs with optically thin disks($\alpha_{3.6-8}$$<-$1.8), and the filled circles are for YSOs with optically thick disks($\alpha_{3.6-8}$$\ge-$1.8).  The {\lrev star symbols} show the TDs in L1641. A set of two symbols connected by a dashed line are the minimum and maximum $FW_{H\alpha, 10\%}$ when there are multi-epoch observations. (b) The distribution of $FW_{H\alpha, 10\%}$ for the diskless population (open histograms), the optically-thick disk population (filled histograms), and the TDs (line-filled histograms) in L1641. For the YSOs with two or three observations, the averaged $FW_{H\alpha, 10\%}$ is used to derive these  distributions. In the grey filled region, the YSOs have $FW_{H\alpha, 10\%}$$<$250\kms, thus considered as not accreting.}\label{Fig:FW_color}
\end{center}
\end{figure}

In Fig.~\ref{Fig:FW_color}(a) we show the $FW_{H\alpha, 10\%}$ vs. [3.6]$-$[8.0] colors for the YSOs in L1641. The diskless stars typically show small $FW_{H\alpha, 10\%}$ with 90$\pm$9\% of them showing $FW_{H\alpha, 10\%}$$<$250\kms. The diskless stars in L1641 show a  $FW_{H\alpha, 10\%}$ distribution, peaked at $\sim$125\kms, which is similar to that of diskless stars  collected in the literature (see Appendix~\ref{Appen:Halpha}). The optically-thick disk population have typically large $FW_{H\alpha, 10\%}$ with a smaller number of sources showing   $FW_{H\alpha, 10\%}$$<$250\kms. The TDs show  a broader and flatter distribution of  $FW_{H\alpha, 10\%}$ compared with the other two populations. If using $FW_{H\alpha, 10\%}$$>$250\kms\ as the criterion for active accretion, 45$\pm$12\% of TDs are accretors. The fraction of accretors increases to  77$\pm$10\% among the optically-thick disk population. The fractions of accretors for the two populations identified with $FW_{H\alpha, 10\%}$ are consistent with those ({\newnewrev 40$\pm$8\% vs. 79$\pm$5\%}) based on the H$\alpha$ $EWs$. Though the fraction of accretors among TDs is significantly lower than that in optically-thick disk population, the accretors in both populations show similar  median values of  $FW_{H\alpha, 10\%}$, which are  387, and 375\kms, respectively, thus corresponding to similar accretion rates \citep{2004A&A...424..603N}. This finding confirms our previous result in Paper\,I. \citet{2010ApJ...710..597S} have estimated the accretion rates of a sample of YSOs in the Tr\,37 cluster using $U$-band excess method, and also find that the median accretion rates of the accreting TDs and normal CTTSs are similar. This result is  inconsistent with what \citet{2007MNRAS.378..369N} find in Taurus, namely that the accretion rates of TDs are $\sim$10 times lower than those of normal accreting TTS at the same disk mass. Recently \citet{2012ApJ...747..103E} find the median accretion rate of full disks is $\sim$4 times higher than that of accreting TDs in NGC\,2068 and IC\,348. Thus, the relation between accretion processes in young stars and the age or evolutionary stage of their disks is still controversial, and may vary in different star-forming regions.

{\lrev In this work, we use the H$\alpha$ line as a tracer of accretion. The criteria that we use to classify CTTS and WTTS may affect our results. In Fig.~\ref{Fig:FW_color}, we consider the YSOs with $FW$$_{H\alpha, 10\%}$$<$250\kms\ as WTTSs. However, some YSOs with tiny accretion rates or with infall perpendicular to the line of sight may show $FW$$_{H\alpha, 10\%}$$<$250\kms. We could exclude more these YSOs among the TDs  than those among the optically-thick disks since more than half of TDs show $FW$$_{H\alpha, 10\%}$$<$250\kms, but only 20\% of the optically-thick disk population have $FW$$_{H\alpha, 10\%}$$<$250\kms\ (see the above discussion). If we change the criterion of $FW$$_{H\alpha, 10\%}$ to 200\kms\ for classifying CTTS and CTTS, the median values of  $FW$$_{H\alpha, 10\%}$ for accreting TDs and optically-thick disk population are 329, and 356\kms, respectively. And the corresponding accretion rate for TDs is $\sim$2 times lower than that of the optically-thick disks, which can slightly reconcile the contradiction between our results and the others in the literature.}

\begin{figure}
\begin{center}
\includegraphics[width=\columnwidth]{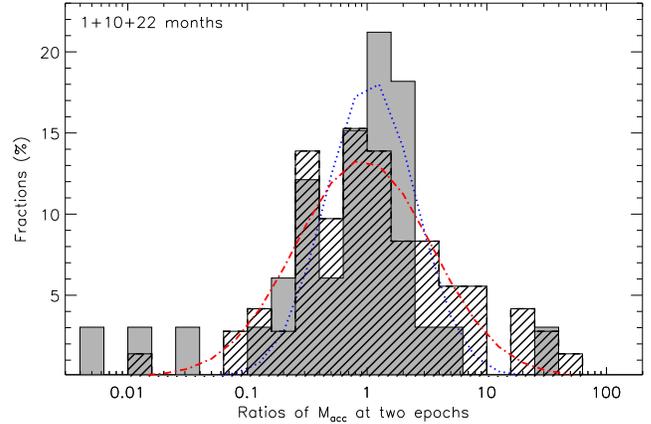}
\caption{The distribution of ratios of $\dot{M}_{\rm acc}$ for TDs (filled histograms) and normal disk population (line-filled histograms) with observations at two epochs separated by $\sim$1, 10, or 22 months. The dash-dotted  lines shows the Gaussian fit to the logarithm of the ratio for the normal disk population with  $\sigma\sim$0.56. The dotted  lines shows the gaussian fit to the distribution of  TDs  with $\sigma\sim$0.37. }\label{Fig:var_dis}
\end{center}
\end{figure}

In Fig.~\ref{Fig:var_dis}, we combine the three distributions (the  distribution of ratios of  $\dot{M}_{\rm acc}$ ) shown in Fig.~\ref{Fig:Acc_var}(d,e,f)  together, and separate it into two distributions, one of which is for TDs and the other for {\newnewrev optically-thick} disk population ($\alpha_{3.6-8}$$>-$1.8). We use the Gaussian function  to fit the  distribution in the logarithm of the ratio, and derive standard deviations of $\sim$0.37$\pm$0.05 and 0.56$\pm$0.07, respectively, suggesting that there is no significant difference between the accretion variability of TDs and normal disks. We have already found that the accreting TDs and normal disk population show similar median accretion rates. Here, we find their accretion variability is also similar. Thus, we conclude that the accretion activity of accreting TDs is similar to the normal disk population.

\subsection{Star formation modes in L1641}

\begin{figure}
\begin{center}
\includegraphics[width=\columnwidth]{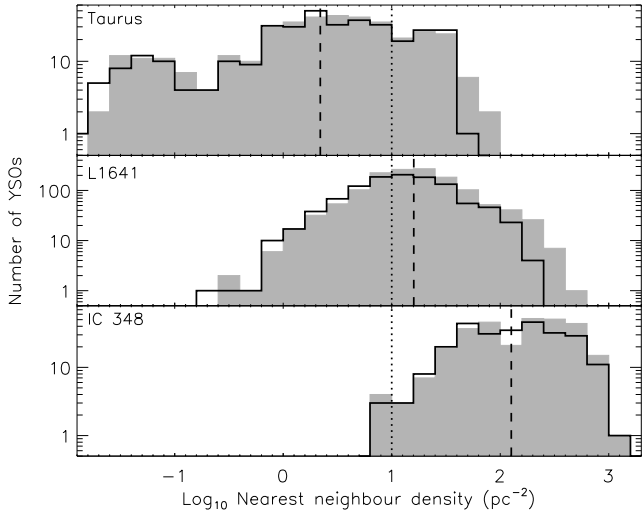}
\caption{The distribution of YSO nearest neighbor densities for Taurus, L1641, and IC\,348. The densities are calculated for each YSO using the equation density = $n$/($\pi r_{\rm n}^{2}$) where $r_{\rm n}$ is the distance to the $n$th nearest YSO and $n$ is set to 10. In each panel, the grey-filled histograms include  all known YSOs in each region, while the open histograms only contain the YSOs with masses larger than $\sim$0.05\,\Msun. In each panel, the dotted line marks the density of 10\,pc$^{-2}$, and the dashed line shows the median values for the open histograms in each region.}\label{fig:SDD}
\end{center}
\end{figure}

With the data from Spitzer, XMM, and spectroscopic survey, we have performed a census of YSOs in L1641 (see \tab~\ref{tab:allYSO_L1641_optical}, and ~\ref{tab:allYSO_L1641_infrared}). Based on this YSO catalog, we reexamine the issue of the star formation mode in L1641 (ie. clustered vs. isolated). We estimated the surface density (SD) using the nearest neighbor method, which is given with the equation density = n/($\pi$r$_{\rm n}^{2}$) where r$_{\rm n}$ is the distance to the $n$th nearest YSO and $n$ is set to 10. We calculate the surface density for all cataloged YSOs ($\sim$1390 sources)  in  \tab~\ref{tab:allYSO_L1641_optical}, and ~\ref{tab:allYSO_L1641_infrared}. We also calculate the surface density for known YSOs in Taurus \citep{2010ApJS..186..111L}, and IC\,348 \citep{2007AJ....134..411M}. Their SD distributions are shown in Fig.~\ref{fig:SDD}. The median SD is 2.8, {\newnewrev 15.7}, and 177.5~pc$^{-2}$ in Taurus, L1641, and IC\,348, respectively. To allow a comparative study of the SD distributions in these regions, we need to calculate the SD for  a selected sample of YSOs in each region down to a similar completeness limit. In L1641, we only select the YSOs with $K_{\rm s}<$14.3 (10$\sigma$ limit in the 2MASS survey), corresponding to a 0.05\,\Msun\ PMS star at the median age of L1641 \citep{1998A&A...337..403B}. We use this sample mass limit to select the YSOs in Taurus and IC\,348. We calculate the SDs for the selected YSOs in each region, and show their distributions in  Fig.~\ref{fig:SDD}. The median SDs are 2.6, 12.5, 141.4~pc$^{-2}$ in Taurus, L1641, and IC\,348, respectively. The result is consistent with the  common sense that  the star formation mode in Taurus is in isolation, and in IC\,348 is in cluster. Among the selected YSO sample in each region, 79\% of the YSOs in Taurus are located in regions with a SD less than 10~pc$^{-2}$, while in IC\,348 the fraction is only 1\%. In L1641, {\newnewrev 41\%} of YSOs live in regions with SD $<$10~pc$^{-2}$. In addition, in L1641 only {\newnewrev 6\%} of YSOs are in areas with SD less than  2.6~pc$^{-2}$ (the median SD in Taurus), and only 5\textperthousand\ of YSOs are in areas with SD larger than  141.4~pc$^{-2}$ (the median SD in IC\,348). The distribution of SDs in the L1641 cloud is between that of Taurus and IC\,348,  confirming that both of the two star formation modes (isolated and clustered) are at work in L1641 as suggested in the literature \citep[][Paper\,I]{1993ApJ...412..233S,1995PhDT..........A}.

We estimate the SDs for Class\,III, Class\,II, and Flat-spectrum+Class\,I sources in L1641. The median SDs are {\newnewrev 14.7}~pc$^{-2}$,  {\newnewrev 17.2}~pc$^{-2}$, and {\newnewrev 17.7}~pc$^{-2}$, for  Class\,III, Class\,II, and Flat-spectrum/Class\,I sources, respectively. In addition, 32\% of Class\,III sources, 28\% of Class\,II sources, and  32\% of Flat-spectrum/Class\,I sources are located in regions with SDs less than 10~pc$^{-2}$. Thus, our result suggests there is no significant difference in the SDs between the ``old'' Class\,III sources, and ``young'' Class\,II, or  Flat-spectrum/Class\,I sources.

 {\newnewrev We compare the ages of ``isolated'' YSOs (SD$\le$10\,pc$^{-2}$) with those of ``clustered'' YSOs (SD$>$10\,pc$^{-2}$) in L1641 using the sample young stars with spectral types. Both populations show very broad age distributions with  median ages of 1.6 and 1.5\,Myr, respectively. A KS test suggest that age distribution of both populations having been drawn from a same distribution has probability of $\sim$0.3.}

\begin{figure}
\begin{center}
\includegraphics[width=\columnwidth]{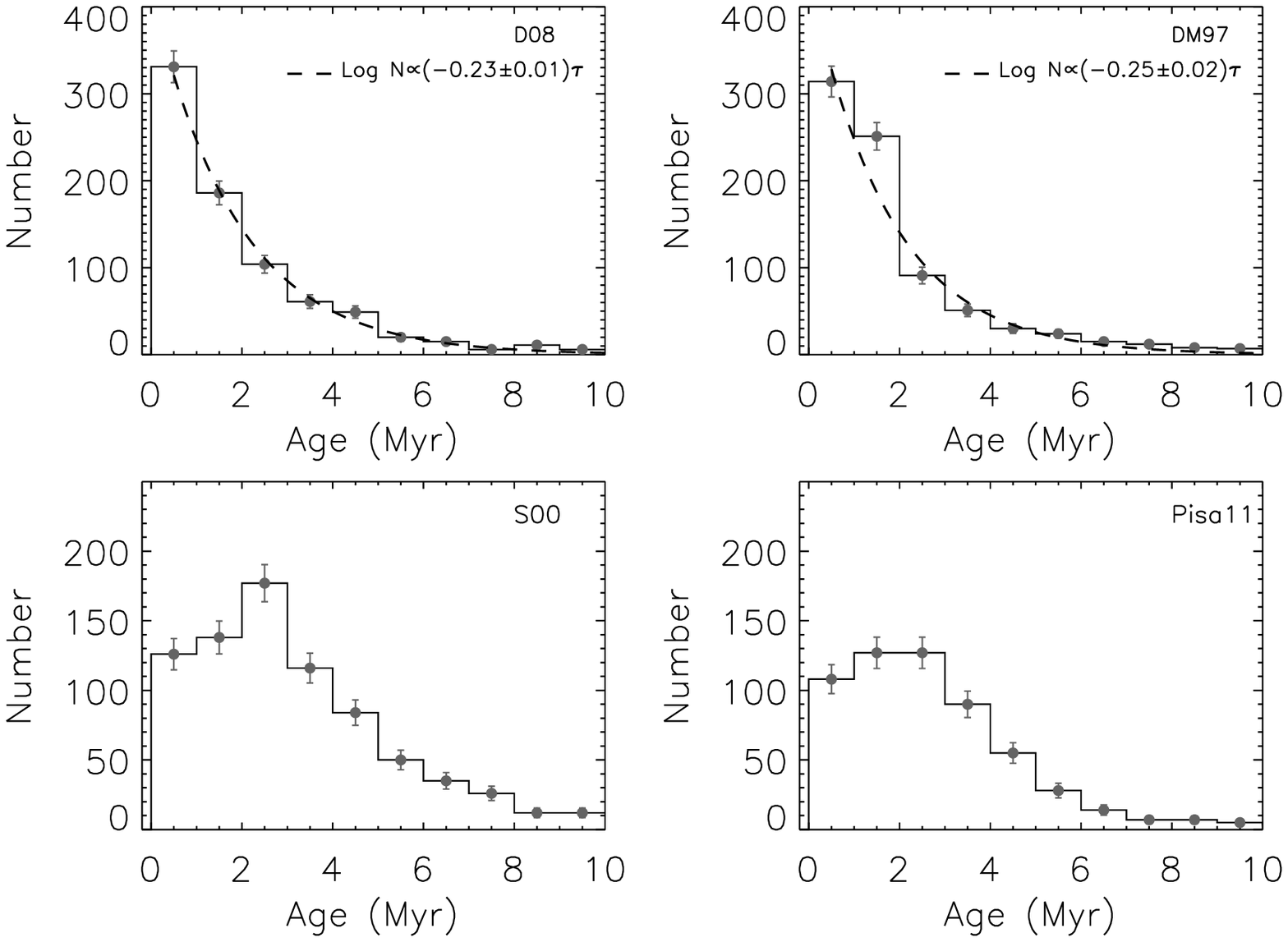}
\caption{The distribution of ages for the YSOs in our sample derived using different evolutionary tracks: i.e. \cite{2008ApJS..178...89D} (D08), \cite{1997MmSAI..68..807D} (DM97), \cite{2000A&A...358..593S} (S00), and \cite{2011A&A...533A.109T} (Pisa11).}\label{fig:all_age}
\end{center}
\end{figure}

\subsection{Star formation history in L1641, and the lifetime of Flat-spectrum and Class\,I sources}\label{sec:life_SFH}
In this work and Paper\,I, we have more than 800 young stars with spectral-type estimates. For these YSOs, we can derive their ages from the PMS evolutionary tracks,  and statistically investigate the star formation history in L1641. In Fig.~\ref{fig:all_age}, we show the distribution of ages for all these YSOs estimated from four sets of  PMS evolutionary tracks. The age distributions from D08 and DM97 suggest formation rate of stars is accelerating in L1641 starting  $\sim$10\,Myr ago.  The age distributions estimated from  D08 and DM97 can be fit using the following formulae:
\begin{equation}\label{Equ:SFH}
 Log\,N\propto\kappa\times \tau
\end{equation}
where N is the numbers of YSOs and  $\tau$ is the age in units of Myr. Fitting to the YSO ages derived from the D08 and DM97 evolutionary tracks results in similar $\kappa$ values $\sim-${\newnewrev 0.24}, an indicator of the accelerating star formation in L1641. The age distribution derived using the S00 tracks indicates that the star formation started $\sim$10\,Myr ago, and increased with time. The star formation rate peaked 2--3\,Myr ago. The age distribution derived using the Pisa11 tracks display an accelerating star formation rate starting $\sim$10\,Myr ago, similar to those from other PMS evolutionary tracks. However, the star formation rate has been constant from 3\,Myr ago until now. It is still being debated whether the stellar ages from PMS isochrone are accurate enough to constrain the star formation history in star-forming regions \citep{2000ApJ...540..255P,2001AJ....121.1030H,2008ASPC..384..200H}. Generally, though the age distribution varied when derived with different evolutionary tracks, but they show some common characteristic: in L1641 star formation was very inactive until 5\,Myr ago, and has been active for the past $\sim$2--3\,Myr.

\begin{figure}
\begin{center}
\includegraphics[width=\columnwidth]{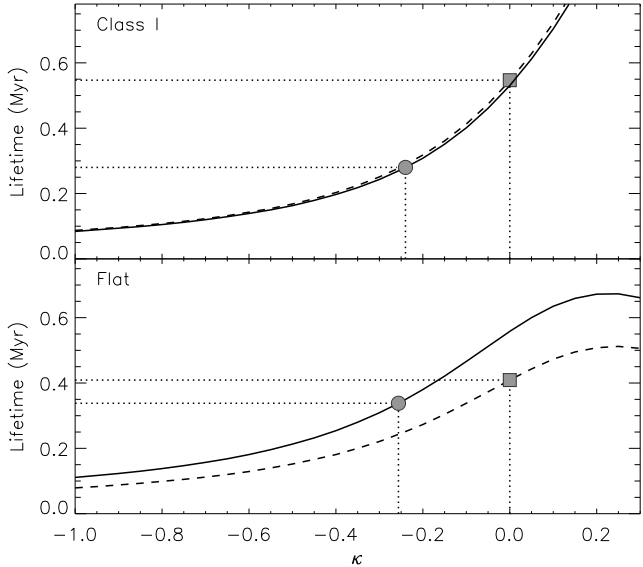}
\caption{The estimated lifetime of Class\,I (the upper panel) and Flat-Spectrum (the lower panel) sources as a function of $\kappa$ values. The solid lines are for YSOs in L1641, and the dashed line for YSOs from Spitzer c2d survey including  Cha\,II, Lupus, Serpens, Persus, and Ophiuchus star-forming regions. The parameter $\kappa$ is used to characterize the SFH (see \sek~\ref{sec:life_SFH}). The squares and circles mark the location on these curves that correspond to a constant star-formation history ($\kappa$=0) or an accelerating star-formation history that matches our observations ($\kappa$=$-$0.24).}\label{fig:life_SFH}
\end{center}
\end{figure}

 Observationally, the lifetimes of the  Class\,I or Flat-spectrum stages are statistically estimated using the ratio between the number of Class\,I and Flat-spectrum sources and Class\,II sources, assuming an age, e.g. 2$\pm$1\,Myr, for Class\,II sources, and a star formation history (SFH)  \citep{2009ApJS..181..321E}. Thus, the knowledge of the SFH is important in estimating the lifetimes of these sources. In the literature, the SFH of the studied region is usually assumed to be constant  \citep{2009ApJS..181..321E}. However, if the true star formation rate of the studied region is accelerating, the derived lifetimes of  the  Class\,I or Flat-spectrum sources  would be overestimated. To demonstrate this effect, we employ a simple model using  the YSOs from L1641 and from Spitzer c2d as examples. We use the function described in Equation~\ref{Equ:SFH} to characterize the SFH. Models with $\kappa>0$ correspond to a decelerating star formation rate, whereas $\kappa<0$ corresponds to an accelerating star formation rate and  $\kappa=0$ represents a constant star formation rate. In the model, we assume that Class\,II sources would evolve into Class\,III at an age of 3\,Myr. In L1641, $N({\rm I})/N({\rm II})$ and $N({\rm Flat})/N({\rm II})$ are 0.27 and 0.25, respectively, where $N({\rm I})$, $N({\rm II})$, and $N({\rm Flat})$ are numbers of Class\,I, Class\,II and Flat-spectrum sources, respectively. In the c2d survey, both fractions are $\sim$0.27 and $\sim$0.20, respectively, for five star-forming regions, i.e.  Cha\,II, Lupus, Serpens, Persus, and Ophiuchus.  In Fig.~\ref{fig:life_SFH}, we show how the estimated lifetimes of Class\,I and Flat-spectrum sources change with the different $\kappa$ values. If the star formation rate is assumed to be constant ($\kappa$=0), the lifetimes of Class\,I and Flat-spectrum sources  are  $\sim$0.54 and 0.50\,Myr, respectively, in L1641, and $\sim$0.55 and 0.41\,Myr, respectively, for the c2d survey \citep{2009ApJS..181..321E}. However, if $\kappa\sim-$0.24, an accelerating star formation rate, the lifetimes of both classes would decrease by a factor of two.

\section{Summary}\label{sec:summary}

We have performed a census of YSOs in  L1641 using the data from the Spitzer imaging surveys, the XMM surveys, and our spectroscopic surveys. We have spectroscopically surveyed a large sub-sample of YSOs with Hectospec and Hectochelle. We combined the Hectospec data with optical and infrared imaging data to estimate the stellar effective temperatures, luminosities and extinction values of individual YSOs. The masses and ages of individual YSOs are estimated by their placement in the HR diagram. Accretion activity is  characterized using the H$\alpha$ and H$\beta$ lines. The main results in this work are summarized as follows.

\begin{itemize}
 
\item[1.] We presented a catalog of $\sim$1390 YSOs in L1641, and constructed the SED from 0.4 to 24\mum\ for each YSO. Based on the infrared SEDs, we classified the YSOs into different classes. We have 143  Class\,I sources, 131 Flat-spectrum sources, 533 Class\,II sources, and 507  Class\,III sources in our YSO catalog. 

\item[2.] We have used the medium-resolution spectroscopic data from Hectospec in conjunction with the imaging data to derive spectral types,  line of sight extinction, estimates of the stellar mass/age, equivalent widths of H$\alpha$ and H$\beta$ lines, and accretion rates from the H$\alpha$ and/or H$\beta$ line luminosity for $\sim$300 YSOs with spectra from our observations, {\newnewrev as well as $\sim$300 YSOs with spectra types from \citet{2012ApJ...752...59H}}.

\item[3.] We have analyzed the high-resolution spectroscopic data observed with Hectochelle. For 235 YSOs we are able to examine the H$\alpha$ emission line profile, as well as look for \LiI\ absorption. In our sample, 65 YSOs have been observed at one epoch, 97 YSOs at two epochs, and 73 at three epochs, allowing us to trace accretion variability.

\item[4.] We find one new subluminous object in our spectroscopic sample. It is most likely a star with a nearly edge-on disk in which the cold outer parts of the disk can effectively screen the  central star but the 10\mum\ radiation of the warm inner disk can still reach us.

\item[5.] Based on our classifications, the disk fraction in L1641, defined as $N({\rm II})/N({\rm II+III})$, is {\newnewrev $\sim$50\%. The disk fraction is almost constant across the stellar mass spectrum with a slight peak at log($M_*$/\Msun)$\approx-$0.25.} 

\item[6.] We find that the ages of young stars with optically-thick disks, transition disks and without disks in L1641 all show very broad distributions but the median age of diskless YSOs is larger than that  of stars with optically-thick disks. The median age of stars with transition disks is intermediate between these two populations. 

\item[7.] Based on the infrared spectral slopes, we find that two types of disk dissipation processes, radially depleted and  globally depleted evolution, are at work in L1641.

\item[8.] The accretion rates of young stars in L1641 are consistent with the  $\dot{M}_{\rm acc}\propto M_*^{\alpha}$  relation with $\alpha\sim$2$-$3 with a large scatter in $\dot{M}_{\rm acc}$ for a given stellar mass. We compiled a set of mass and accretion rates from in this work and in the literature and find that this data support a scenario proposed by \cite{2006ApJ...648..484H}, in which the different disk properties  around the low-mass and brown dwarf stars, and intermediate-mass stars are responsible for the change of $\alpha$ between intermediate-mass stars and very low-mass stars.

\item[9.] Based on the multi-epoch data from Hectochelle, we investigated the accretion variability of young stars in L1641 and find typical accretion rate changes of $\sim$0.6\,dex. We confirm that intrinsic accretion variations cannot explain the two orders of magnitudes scatter in accretion rates for YSOs with similar masses. We find that the relative variability of $FW_{H\alpha, 10\%}$  for  YSOs with disks in the sub-solar mass range is larger than those in the intermediate-mass range. 

\item[10.] We confirmed  46 new transition disk objects in L1641, 6 globally depleted disks,  and provided 2 young debris disk candidates. We investigated the accretion properties of YSOs with  optically-thick disks and transition disks based on the H$\alpha$ $EWs$ and $FW_{H\alpha, 10\%}$. We find that the fraction of accretors among transition disks (40--45\%) is significantly lower than among young stars with  optically-thick disks (77--79\%). Among 6 globally depleted disks, only one source is accreting. We confirm our previous result that median accretion rate of accreting transition disks is similar to that of accreting stars with  optically-thick disks. We find that the two types of disks show similar fluctuations in their accretion rate.

\item[11.] We estimated the surface densities of YSOs in L1641 using the nearest neighbor method. We find that there is no significant difference on the  surface density distributions between Class\,III sources, Class\,II source, and Flat-spectrum/Class\,I sources. We compared the  surface density distribution of YSOs in L1641 with Taurus and IC\,348, and confirmed that both clustered and isolated star formation is ongoing in L1641.

\item[12.] We investigated the star formation history in L1641 based on $\sim$800 YSOs with spectral-type estimates. We found that the star formation is most active during the past 2--3\,Myr. We also show that the knowledge of star formation history is important for constraining the lifetime of Class\,I and Flat-spectrum sources.

\end{itemize}

\acknowledgements
We would like to thank Perry Berlind, Mike Calkins, and Nelson Caldwell at SAO for assisting hectospec and hectochelle observations, and the anonymous referee for comments that help to improve this paper. MF acknowledges the support by NSFC through grants 11203081 and 11173060. ASA acknowledges support by the Spanish "Ramon y Cajal" program. This research has made use of the SIMBAD database, operated at CDS, Strasbourg, France. This publication makes use of data products from the Two Micron All Sky Survey, which is a joint project of the University of Massachusetts and the Infrared Processing and Analysis Center/California Institute of Technology, funded by the National Aeronautics and Space Administration and the National Science Foundation. This work is in part  based  on observations made with the Spitzer Space Telescope, which is operated by the Jet Propulsion Laboratory, California Institute of Technology under a contract with NASA. This publication makes use of data products from the Wide-field Infrared Survey Explorer, which is a joint project of the University of California, Los Angeles, and the Jet Propulsion Laboratory/California Institute of Technology, funded by the National Aeronautics and Space Administration. This work is based in part on data obtained as part of the UKIRT Infrared Deep Sky Survey. This publication is in part based on data from the Sloan Digital Sky Survey project, whose many contributors can be found at the following web-page: http://www.sdss.org/collaboration/credits.html. This research is partially based on observations made with the NASA/ESA Hubble Space Telescope, and obtained from the Hubble Legacy Archive, which is a collaboration between the Space Telescope Science Institute (STScI/NASA), the Space Telescope European Coordinating Facility (ST-ECF/ESA) and the Canadian Astronomy Data Centre (CADC/NRC/CSA).

\appendix

\section{Accretion characterized by $FW_{H\alpha, 10\%}$}\label{Appen:Halpha}
\cite{2003ApJ...582.1109W} proposed to use the $FW_{H\alpha, 10\%}$ as an alternative to distinguish accretors from non-accretors, instead of the $EW$ of the H$\alpha$ line. They found that young stars with $FW_{H\alpha, 10\%}$ greater than 270 km\,s$^{-1}$ are accretors. Since a large sample of young stars with or without disks that have been observed with high-resolution H$\alpha$ spectroscopy exists in the literature, we can make statistically study how to use  $FW_{H\alpha, 10\%}$ to classify CTTSs and WTTSs. 

Our basic method is described as follows. We select from the literature young stars that show no evidence for  a significant infrared excess at $\lambda<$8\mum, i.e.  their (inner) disks have already dissipated. The vast majority of these stars will not be actively accreting, though the sample can contain a few TDs, some of which may still accrete. We then plot the frequency distribution of $FW_{H\alpha, 10\%}$ for all of these objects, and use a function to fit the distribution. We normalize the fitted function, and  use it as the probability function of  $FW_{H\alpha, 10\%}$ for non-accreting young stars.

\begin{figure}[h]
\begin{center}
\includegraphics[width=\columnwidth]{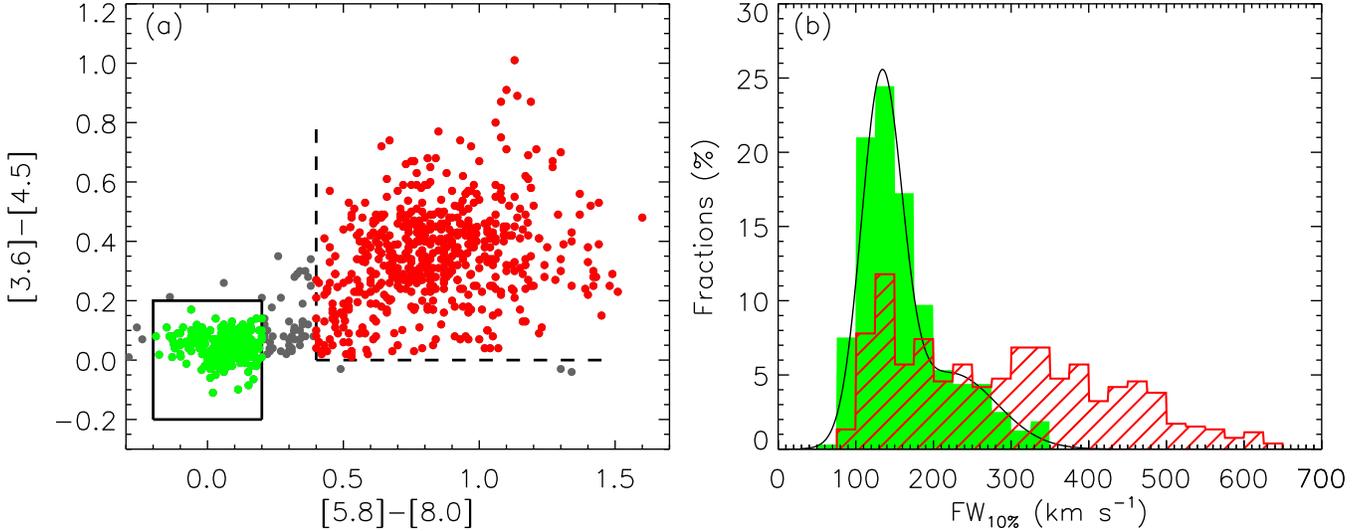}
\caption{(a): The [3.6]$-$[4.5] vs. [5.8]$-$[8.0] color-color diagram for young stars with high-resolution H$\alpha$ spectroscopy. The solid lines enclose the region containing diskless stars. We use the $FW_{H\alpha, 10\%}$ of these diskless stars to derive the accretion probability as a function of $FW_{H\alpha, 10\%}$. The dashed lines mark the region containing Class II YSOs. (b): The distribution of $FW_{H\alpha, 10\%}$ for diskless stars (filled histogram), and Class\,II YSOs (line-filled histogram).}\label{Fig:Accretion_prob}
\end{center}
\end{figure}

We collected young stars from the literature covering many different star-formation regions, i.e. Orion nebular cluster  \citep{2008ApJ...676.1109F,2009ApJ...697.1103T},  NGC\,2068/2071 \citep{2008AJ....135..966F}, Tr37 \citep{2006AJ....132.2135S,2006ApJ...638..897S}, $\eta$\, Cha, TW Hydrae, $\beta$ Pictoris, and Tucanae-Horologium groups \citep{2009ApJ...701.1188S,2006ApJ...648.1206J}, Cha\,I, and Taurus-Auriga star-forming regions \citep{2009ApJ...695.1648N}. In total, 334 stars satisfy our criteria for having no or weak infrared excess ($|$[3.6]$-$[4.5]$|$$\leq$0.2, and $|$[5.8]$-$[8.0]$|$$\leq$0.2). The inner disks around these stars have been mostly cleared away and the majority of these targets should no longer be actively accreting. In \fig\ref{Fig:Accretion_prob}, we show the [5.8]$-$[8.0] vs. [3.6]$-$[4.5] color-color diagram for these targets and the fractional distribution of $FW_{H\alpha, 10\%}$. As shown in Fig~\ref{Fig:Accretion_prob}(b), the distribution of $FW_{H\alpha, 10\%}$ for ``diskless'' YSOs has a sharp peak with a long tail towards large $FW_{H\alpha, 10\%}$. The long tail could be due to accreting TDs since we only use Spitzer IRAC bands to select the ``diskless'' YSOs and disks with large inner holes cannot be distinguished from diskless YSOs without data at longer wavelengths. We fit the distribution with two Gaussian functions and normalize the fitted function to a total area of 1. The normalized function is:

\begin{eqnarray*}
   \Phi(FW_{H\alpha, 10\%})=1.04\times10^{-2}exp[-(\frac{FW_{H\alpha, 10\%}-133.23}{36.96})^{2}]+2.20\times10^{-3}exp[-(\frac{FW_{H\alpha, 10\%}-223.04}{82.35})^{2}]
\end{eqnarray*}
\normalsize

We use the above function as the probability function of  $FW_{H\alpha, 10\%}$ for non-accretors. Thus, For a YSO with an estimate of $FW_{H\alpha, 10\%}$, the probability that it is a non-accretor is estimated to be $\int^{\infty}_{FW_{H\alpha, 10\%}}\Phi(x)dx$. And the probability that it is an accretor is  $prob(FW_{H\alpha, 10\%})=1-\int^{\infty}_{FW_{H\alpha, 10\%}}\Phi(x)dx$. A source with  $FW_{H\alpha, 10\%}$$>$250\kms, has a probability of 90\% of being an accretor. Thus we use $FW_{H\alpha, 10\%}$$>$250\kms\ as our criterion for distinguishing accretors from non-accretors. According to this criterion, 56\% of the Class\,II sources shown in Fig.~\ref{Fig:Accretion_prob} are accretors.

\begin{figure}[h]
\begin{center}
\includegraphics[width=0.5\columnwidth]{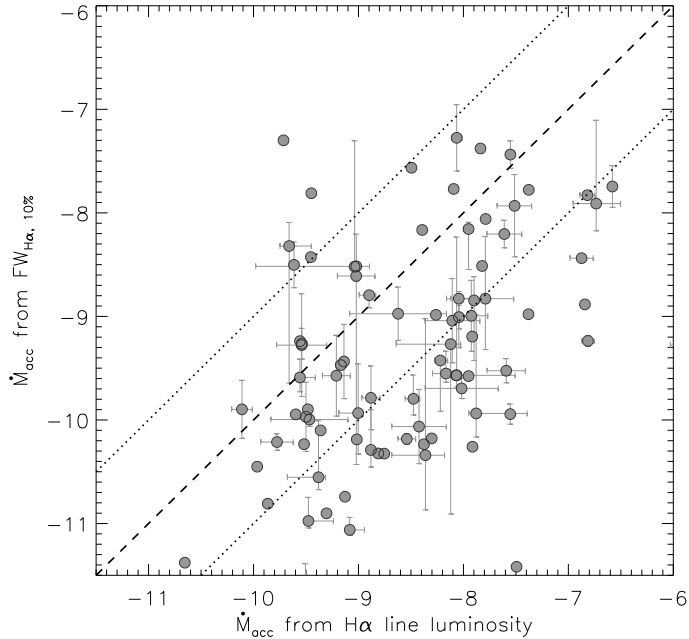}
\caption{Comparison of accretion rates of YSOs in L1641 derived from $FW_{H\alpha, 10\%}$ and H$\alpha$ line luminosity. For the sources with multiple observations, the median accretion rates are used for the plot with error bars showing their minimum and maximum accretion rates. The dashed line shows a 1:1 correlation, and dotted lines show a difference of 1 dex.}\label{Fig:com_acc}
\end{center}
\end{figure}

{\newnewrev
\section{Comparison of accretion rates from H$\alpha$ line luminosity and $FW_{H\alpha, 10\%}$}\label{Appen:Halpha_lum_FW}

In our sample young stars, there are $\sim$90 accreting sources which have been observed with Hectochelle. Using these data, we estimate the accretion rates from both $FW_{H\alpha, 10\%}$ and H$\alpha$ line luminosity as described in \sect\ref{sec:ana_acc}. We compare the accretion rates derived using the two methods in Fig.~\ref{Fig:com_acc}. In order to see whether there is any correlation between both types of accretion rates, we apply a Kendall $\tau$ test. If two datasets are fully correlated, this test returns a value of $\tau$=1. If they are anti-correlated, we get $\tau$=$-$1, and if they are independent, we obtain $\tau$\,$=$\,0. The Kendall $\tau$ test also returns a probability $p$, which is smaller when the correlation is more significant. We use the Kendall $\tau$ test to evaluate the possible correlation between the accretion rates from the two methods, which yields $\tau$=0.3 and $p$=1.5$\times$10$^{-5}$. Thus, though the scatter is large  in Fig.~\ref{Fig:com_acc}, the correlation between two kinds of accretion rates is existent. We also note that the accretion rates from $FW_{H\alpha, 10\%}$  are systematically lower than those from H$\alpha$ line luminosity. It is beyond the scope of the current work to distinguish the better tracer for accretion and calibrate the accretion tracer to the accretion rates. A serial of works have been done or are on-going with the data from VLT/X-shooter \citep[see e.g.][]{2011A&A...526L...6R,2012A&A...548A..56R}.
}

\section{Evaluating completeness of YSO census using color-magnitude diagram}\label{Appen:CMD}
\begin{figure}
\begin{center}
\includegraphics[width=0.8\columnwidth]{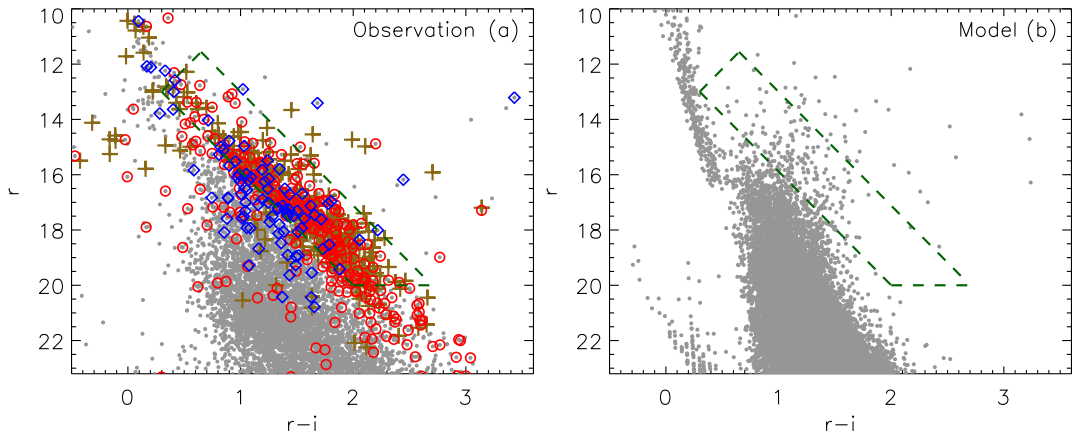}%
\caption{\label{fig:CMD_referee} (a) $r$ vs. $r$$-$$i$ color-magnitude diagram for observed sources in the field of L1641. The grey filled circles show all the sources found in the L1641 field (Paper\,I), The pluses mark the YSOs without disks, and the open circles are the YSOs with disks. The open diamonds are the stars identified as field stars in Paper I and in this work. Most of our YSOs are located in a narrow band which is enclosed with dashed lines. (b)  $r$ vs. $r$$-$$i$ color-magnitude diagram for model stars in the field of L1641, which are created with the Besan\c{c}on\ model of stellar population synthesis of the Galaxy\citep{2003A&A...409..523R}.}
\end{center}
\end{figure}

{\newnewrev In Fig.~\ref{fig:CMD_referee}(a), we show $r$ vs. $r$$-$$i$ color-magnitude diagram for $\sim$660 YSOs in L1641 with photometric data in both r and i bands. Among these sources, more than 70\% of them are located within a narrow band, in $r$ vs. $r$$-$$i$ color-magnitude diagram. We can use the locations of the known YSOs to select the potential YSO candidates. However the YSO candidates selected in this way could be contaminated by the reddened field stars. The fractions of contaminators can be estimated if we know how the field stars are distributed in $r$ vs. $r$$-$$i$ color-magnitude diagram. We create ten sets of model stars  in the direction of L1641 over the same area as we studied, using the Besan\c{c}on\ model of stellar population synthesis of the Galaxy\citep{2003A&A...409..523R}. We derive the extinction map of the L1641 cloud from the $^{13}$CO map (see Fig.~\ref{Fig:yso_dis}) using the relation between the $^{13}$CO intensity and visual extinction from Paper\,I. An extinction probability function is built from the extinction map of the L1641 cloud. For each model star with the distance$>$500\,pc, an extinction value, randomly sampled from the extinction probability function, is assigned to it to simulate the reddening effect on the background stars from the L1641 cloud. For each set of model stars from  Besan\c{c}on\ model, we sample 100 groups of extinction values from the extinction probability function. Totally, we obtain 1000 model  $r$ vs. $r$$-$$i$ color-magnitude diagrams for ten sets of model stars. In Fig.~\ref{fig:CMD_referee}(b), we show $r$ vs. $r$$-$$i$ color-magnitude diagram for one set of model stars.  The Besan\c{c}on\ model predicts that the field stars contaminate our YSO sample mostly within $1.0\lesssim r$$-$$i \lesssim1.5$ and $14.0\lesssim r\lesssim17.5$  in the $r$ vs. $r$$-$$i$ color-magnitude diagram, which is generally consistent with our observations (see Fig.~\ref{fig:CMD_referee}(a, b)).

We visually define a  region in  $r$ vs. $r$$-$$i$ color-magnitude diagram (enclosed by dashed lines in Fig.~\ref{fig:CMD_referee}) where most of these YSOs are located. We only study the sources with $r<$20\,mag, considering that the completeness of the imaging survey in $r$ band is around 20\,mag. Within our selected region, 190 sources are not included in our YSO sample. Among them, 39 sources have been identified as field stars in Paper I and this work. With the synthetic models, 194$\pm$12 field stars are predicted to be within our defined region in $r$ vs. $r$$-$$i$ color-magnitude diagram, suggesting that our YSO census is complete till $r\sim$20\,mag.}

\bibliographystyle{apj}
\bibliography{references}


\renewcommand{\tabcolsep}{0.05cm}

\input{tab2.tex}
\input{tab3.tex}
\input{tab4.tex}

\clearpage
\renewcommand{\tabcolsep}{0.05cm}

\input{tab5.tex}
\input{tab6.tex}


\end{document}

%% file: tab1.tex
\begin{table}[h]
\caption{Observation logs.}\label{Tab:obs_log}
\centering
\begin{tabular}{llccccccccccccccccccccccccccccccccc}
\hline\hline
        &          & RA      &DEC     & \multicolumn{2}{c}{Exposure}   &   \\
\cline{5-6}
        &          & (J2000) &(J2000) &Obj&sky &Offset\tablenotemark{a} \\
Config. & Obs-date & (h:m:s) &(d:m:s) &(min)&(min) &(arcsec) \\
\hline 
 \multicolumn{6}{c}{Hectochelle}\\
 \hline 
  1  & 2010 Feb 5 & 05:36:01 & $-$06:32:24 &3$\times$30  &1$\times$30&5\\
     & 2010 Mar 3 & 05:36:01 & $-$06:32:24 &3$\times$30  &1$\times$30&5\\
     & 2010 Nov 29 & 05:36:01 & $-$06:34:59 &4$\times$30  &1$\times$30&7\\
  2  & 2010 Feb 5 & 05:41:17 & $-$08:07:33 &3$\times$30  &1$\times$30&5\\
     & 2011 Oct 19 & 05:41:17 & $-$08:07:33 &8$\times$20  &1$\times$20&5\\
 \hline 
 \multicolumn{6}{c}{Hectospec}\\
 \hline 
  3 & 2011 Jan 23 & 05:36:01 & $-$06 32 33   &6$\times$20  &2$\times$20&10\\
  4 & 2011 Jan 24 & 05:41:21 & $-$08 06 10   &4$\times$10  &1$\times$10&5\\
  5 & 2011 Jan 25 & 05:39:12 & $-$07 24 46   &5$\times$10  &1$\times$10&6\\
\hline
\end{tabular}
\tablenotetext{1}{The offsets between the pointings of object exposures and sky exposures}
\end{table}

%% file: tab4.tex
\renewcommand{\thefootnote}{\alph{footnote}}
\scriptsize
\begin{center}
\renewcommand{\tabcolsep}{0.1cm}
\begin{longtable}{lcccccccccccccccccccccccccccccccccc}
\caption{Stellar and disk properties of the YSOs in L1641}\label{Tab:spt}\\
\hline\hline \multicolumn{1}{c}{} &\multicolumn{1}{c}{RA} &\multicolumn{1}{c}{DEC} &\multicolumn{1}{c}{}&\multicolumn{1}{c}{Lum} &\multicolumn{1}{c}{Av} &\multicolumn{1}{c}{$EW_{\rm H\alpha}$} &\multicolumn{1}{c}{$EW_{\rm H\beta}$} &\multicolumn{1}{c}{Log\,$\dot{M}_{\rm acc}^{\rm H\alpha}$} &\multicolumn{1}{c}{Log\,$\dot{M}_{\rm acc}^{\rm H\beta}$} &\multicolumn{1}{c}{} &\multicolumn{1}{c}{} &\multicolumn{1}{c}{} &\multicolumn{1}{c}{}\\
\multicolumn{1}{c}{} &\multicolumn{1}{c}{(J2000)} &\multicolumn{1}{c}{(J2000)} &\multicolumn{1}{c}{Spt}&\multicolumn{1}{c}{($L_{\odot}$)} &\multicolumn{1}{c}{(mag)} &\multicolumn{1}{c}{\AA} &\multicolumn{1}{c}{\AA} &\multicolumn{1}{c}{(M\accunit)} &\multicolumn{1}{c}{(M\accunit)} &\multicolumn{1}{c}{TTS\tablenotemark{a}} &\multicolumn{1}{c}{Disk\tablenotemark{b}}\\ \hline
\endfirsthead
\multicolumn{14}{c}{{\bfseries \tablename\ \thetable{} -- continued}}\\
\hline\hline \multicolumn{1}{c}{} &\multicolumn{1}{c}{RA} &\multicolumn{1}{c}{DEC} &\multicolumn{1}{c}{}&\multicolumn{1}{c}{Lum} &\multicolumn{1}{c}{Av} &\multicolumn{1}{c}{$EW_{\rm H\alpha}$} &\multicolumn{1}{c}{$EW_{\rm H\beta}$} &\multicolumn{1}{c}{Log\,$\dot{M}_{\rm acc}^{\rm H\alpha}$} &\multicolumn{1}{c}{Log\,$\dot{M}_{\rm acc}^{\rm H\beta}$} &\multicolumn{1}{c}{} &\multicolumn{1}{c}{} &\multicolumn{1}{c}{} &\multicolumn{1}{c}{}\\
\multicolumn{1}{c}{} &\multicolumn{1}{c}{(J2000)} &\multicolumn{1}{c}{(J2000)} &\multicolumn{1}{c}{Spt}&\multicolumn{1}{c}{($L_{\odot}$)} &\multicolumn{1}{c}{(mag)} &\multicolumn{1}{c}{\AA} &\multicolumn{1}{c}{\AA} &\multicolumn{1}{c}{(M\accunit)} &\multicolumn{1}{c}{(M\accunit)} &\multicolumn{1}{c}{TTS\tablenotemark{a}}  &\multicolumn{1}{c}{Disk\tablenotemark{b}}\\ \hline
\endhead
\hline\endfoot
\\[-1.8ex] \hline \hline
\endlastfoot
1  & 05 36 05.00  &$-$06 42 43.8  &M5.5  &     0.398&0.0 &$     -23.6$ & $     -14.9$& $     -9.1$ &$     -9.6$&C&Thick\\[0.3pt]
2  & 05 35 10.69  &$-$06 34 15.5  &M4.5  &    0.090&      0.0 &$     -36.3$ & $     -24.2$& $     -9.9$ &$     -10.3$&C&Thick\\[0.3pt]
3  & 05 34 52.29  &$-$06 35 08.5  &M4  &     0.128&0.0 &$     -8.4$ & $     -8.2$& ... &...&W&N\\[0.3pt]
4  & 05 35 50.56  &$-$06 33 17.3  &M3  &     0.254&0.0 &$     -3.8$ & $     -3.5$& ... &...&W&N\\[0.3pt]
5  & 05 35 10.75  &$-$06 34 44.5  &F5.5  &      4.774&     0.3 &$      1.7$ & $      2.5$& ... &...&W&N\\[0.3pt]
6  & 05 35 26.58  &$-$06 19 09.0  &M1.5  &      1.155&      1.3 &$     -3.3$ & $     -2.2$& ... &...&W&N\\[0.3pt]
7  & 05 34 39.20  &$-$06 25 24.2  &M4  &     0.140&0.0 &$     -4.9$ & $     -4.3$& ... &...&W&N\\[0.3pt]
8  & 05 35 37.32  &$-$06 23 26.3  &M1.5  &     0.278&     0.2 &$     -17.9$ & $     -9.4$& $     -9.2$ &$     -9.3$&C&TD\\
\end{longtable}
\end{center}
\normalsize

%% file: tab6.tex
\begin{center}
\renewcommand{\tabcolsep}{0.1cm}
\scriptsize
\begin{longtable}{lcccccccclccccccccccccccccccccccccc}
\caption{Accretion properties of the YSOs in L1641 from Hectochelle.}\label{Tab:EW}\\
\hline\hline  &\multicolumn{1}{c}{RA}&\multicolumn{1}{c}{DEC}&\multicolumn{1}{c}{$EW$(Li)}&\multicolumn{1}{c}{$EW$(H$\alpha$)}&\multicolumn{1}{c}{$FW$$_{H\alpha, 10\%}$}& &\multicolumn{1}{c}{Accretor} &\multicolumn{1}{c}{Log $\dot{M}_{acc}$} & &\\
\multicolumn{1}{c}{ID} &\multicolumn{1}{c}{(J2000)}&\multicolumn{1}{c}{(J2000)}&\multicolumn{1}{c}{(\AA)}&\multicolumn{1}{c}{(\AA)}&\multicolumn{1}{c}{(km/s)}& \multicolumn{1}{c}{TTS} &\multicolumn{1}{c}{Prob} &\multicolumn{1}{c}{(M$_{\odot}$/yr)} &\multicolumn{1}{c}{Epoch} \\  \hline 
 
\endfirsthead
 
\multicolumn{11}{c}{{\bfseries \tablename\ \thetable{} -- continued}}\\
\hline\hline  &\multicolumn{1}{c}{RA}&\multicolumn{1}{c}{DEC}&\multicolumn{1}{c}{$EW$(Li)}&\multicolumn{1}{c}{$EW$(H$\alpha$)}&\multicolumn{1}{c}{$FW$$_{H\alpha, 10\%}$}& &\multicolumn{1}{c}{Accretor} &\multicolumn{1}{c}{Log $\dot{M}_{acc}$} & &\\
\multicolumn{1}{c}{ID} &\multicolumn{1}{c}{(J2000)}&\multicolumn{1}{c}{(J2000)}&\multicolumn{1}{c}{(\AA)}&\multicolumn{1}{c}{(\AA)}&\multicolumn{1}{c}{(km/s)}& \multicolumn{1}{c}{TTS} &\multicolumn{1}{c}{Prob} &\multicolumn{1}{c}{(M$_{\odot}$/yr)} &\multicolumn{1}{c}{Epoch}\\  \hline 
 
\endhead
\hline\endfoot
\hline
1  & 05 36 05.00  &$-$06 42 43.8  &...&$     -25.2$  &      357&C&      100\%&$     -9.4$&2010 Feb 5\\[0.3pt]
  &  &  &     0.30&$     -21.1$  &      348&C&      100\%&$     -9.5$&2010 Mar 3\\[0.3pt]
2  & 05 35 10.69  &$-$06 34 15.5  &...&$     -41.1$  &      215&C&      83\%&$     -10.8$&2010 Nov 29\\[0.3pt]
3  & 05 34 52.29  &$-$06 35 08.5  &...&$     -10.8$  &      292&W&      96\%&...&2010 Feb 5\\[0.3pt]
  &  &  &...&$     -8.3$  &      229&W&      86\%&...&2010 Mar 3\\[0.3pt]
4  & 05 35 50.56  &$-$06 33 17.3  &...&$     -4.9$  &      104&W&      10\%&...&2010 Nov 29\\[0.3pt]
6  & 05 35 26.58  &$-$06 19 09.0  &...&$     -4.3$  &      192&W&      77\%&...&2010 Nov 29\\[0.3pt]
  &  &  &...&$     -5.3$  &      214&W&      83\%&...&2010 Mar 3\\
\hline
\end{longtable}
\normalsize
\end{center}

%% file: ms.bbl
\begin{thebibliography}{179}
\expandafter\ifx\csname natexlab\endcsname\relax\def\natexlab#1{#1}\fi

\bibitem[{{Alencar} \& {Batalha}(2002)}]{2002ApJ...571..378A}
{Alencar}, S.~H.~P., \& {Batalha}, C. 2002, \apj, 571, 378

\bibitem[{{Alencar} {et~al.}(2001){Alencar}, {Johns-Krull}, \&
  {Basri}}]{2001AJ....122.3335A}
{Alencar}, S.~H.~P., {Johns-Krull}, C.~M., \& {Basri}, G. 2001, \aj, 122, 3335

\bibitem[{{Alexander} \& {Armitage}(2009)}]{2009ApJ...704..989A}
{Alexander}, R.~D., \& {Armitage}, P.~J. 2009, \apj, 704, 989

\bibitem[{{Alexander} {et~al.}(2006){Alexander}, {Clarke}, \&
  {Pringle}}]{2006MNRAS.369..229A}
{Alexander}, R.~D., {Clarke}, C.~J., \& {Pringle}, J.~E. 2006, \mnras, 369, 229

\bibitem[{{Allen}(1995)}]{1995PhDT..........A}
{Allen}, L.~E. 1995, PhD thesis, AA(Department of Physics and Astronomy,
  University of Massachusetts, Amherst.)

\bibitem[{{Allen} {et~al.}(2012){Allen}, {Gutermuth}, {Kryukova}, {Megeath},
  {Pipher}, {Naylor}, {Jeffries}, {Wolk}, {Spitzbart}, \&
  {Muzerolle}}]{2012ApJ...750..125A}
{Allen}, T.~S., {Gutermuth}, R.~A., {Kryukova}, E., {et~al.} 2012, \apj, 750,
  125

\bibitem[{{Alves} \& {Bouy}(2012)}]{2012A&A...547A..97A}
{Alves}, J., \& {Bouy}, H. 2012, \aap, 547, A97

\bibitem[{{Andrews} \& {Williams}(2005)}]{2005ApJ...631.1134A}
{Andrews}, S.~M., \& {Williams}, J.~P. 2005, \apj, 631, 1134

\bibitem[{{Anthony-Twarog}(1982)}]{1982AJ.....87.1213A}
{Anthony-Twarog}, B.~J. 1982, \aj, 87, 1213

\bibitem[{{Armitage} {et~al.}(2003){Armitage}, {Clarke}, \&
  {Palla}}]{2003MNRAS.342.1139A}
{Armitage}, P.~J., {Clarke}, C.~J., \& {Palla}, F. 2003, \mnras, 342, 1139

\bibitem[{{Balbus} \& {Hawley}(1998)}]{1998RvMP...70....1B}
{Balbus}, S.~A., \& {Hawley}, J.~F. 1998, Reviews of Modern Physics, 70, 1

\bibitem[{{Bally} {et~al.}(1987){Bally}, {Stark}, {Wilson}, \&
  {Langer}}]{1987ApJ...312L..45B}
{Bally}, J., {Stark}, A.~A., {Wilson}, R.~W., \& {Langer}, W.~D. 1987, \apjl,
  312, L45

\bibitem[{{Baraffe} {et~al.}(1998){Baraffe}, {Chabrier}, {Allard}, \&
  {Hauschildt}}]{1998A&A...337..403B}
{Baraffe}, I., {Chabrier}, G., {Allard}, F., \& {Hauschildt}, P.~H. 1998, \aap,
  337, 403

\bibitem[{{Baraffe} {et~al.}(2002){Baraffe}, {Chabrier}, {Allard}, \&
  {Hauschildt}}]{2002A&A...382..563B}
---. 2002, \aap, 382, 563

\bibitem[{{Bertout} {et~al.}(2007){Bertout}, {Siess}, \&
  {Cabrit}}]{2007A&A...473L..21B}
{Bertout}, C., {Siess}, L., \& {Cabrit}, S. 2007, \aap, 473, L21

\bibitem[{{Bessell} \& {Brett}(1988)}]{1988PASP..100.1134B}
{Bessell}, M.~S., \& {Brett}, J.~M. 1988, \pasp, 100, 1134

\bibitem[{{Bouwman} {et~al.}(2006){Bouwman}, {Lawson}, {Dominik}, {Feigelson},
  {Henning}, {Tielens}, \& {Waters}}]{2006ApJ...653L..57B}
{Bouwman}, J., {Lawson}, W.~A., {Dominik}, C., {et~al.} 2006, \apjl, 653, L57

\bibitem[{{Brandner} {et~al.}(2000){Brandner}, {Grebel}, {Chu}, {Dottori},
  {Brandl}, {Richling}, {Yorke}, {Points}, \&
  {Zinnecker}}]{2000AJ....119..292B}
{Brandner}, W., {Grebel}, E.~K., {Chu}, Y.-H., {et~al.} 2000, \aj, 119, 292

\bibitem[{{Calvet} {et~al.}(2004){Calvet}, {Muzerolle}, {Brice{\~n}o},
  {Hern{\'a}ndez}, {Hartmann}, {Saucedo}, \& {Gordon}}]{2004AJ....128.1294C}
{Calvet}, N., {Muzerolle}, J., {Brice{\~n}o}, C., {et~al.} 2004, \aj, 128, 1294

\bibitem[{{Caratti O Garatti} {et~al.}(2012){Caratti O Garatti}, {Garcia
  Lopez}, {Antoniucci}, {Nisini}, {Giannini}, {Eisl{\"o}ffel}, {Ray},
  {Lorenzetti}, \& {Cabrit}}]{2012A&A...538A..64C}
{Caratti O Garatti}, A., {Garcia Lopez}, R., {Antoniucci}, S., {et~al.} 2012,
  \aap, 538, A64

\bibitem[{{Cardelli} {et~al.}(1989){Cardelli}, {Clayton}, \&
  {Mathis}}]{1989ApJ...345..245C}
{Cardelli}, J.~A., {Clayton}, G.~C., \& {Mathis}, J.~S. 1989, \apj, 345, 245

\bibitem[{{Carpenter} {et~al.}(2005){Carpenter}, {Wolf}, {Schreyer},
  {Launhardt}, \& {Henning}}]{2005AJ....129.1049C}
{Carpenter}, J.~M., {Wolf}, S., {Schreyer}, K., {Launhardt}, R., \& {Henning},
  T. 2005, \aj, 129, 1049

\bibitem[{{Carpenter} {et~al.}(2009){Carpenter}, {Bouwman}, {Mamajek}, {Meyer},
  {Hillenbrand}, {Backman}, {Henning}, {Hines}, {Hollenbach}, {Kim},
  {Moro-Martin}, {Pascucci}, {Silverstone}, {Stauffer}, \&
  {Wolf}}]{2009ApJS..181..197C}
{Carpenter}, J.~M., {Bouwman}, J., {Mamajek}, E.~E., {et~al.} 2009, \apjs, 181,
  197

\bibitem[{{Carrera} {et~al.}(2007){Carrera}, {Ebrero}, {Mateos}, {Ceballos},
  {Corral}, {Barcons}, {Page}, {Rosen}, {Watson}, {Tedds}, {Della Ceca},
  {Maccacaro}, {Brunner}, {Freyberg}, {Lamer}, {Bauer}, \&
  {Ueda}}]{2007A&A...469...27C}
{Carrera}, F.~J., {Ebrero}, J., {Mateos}, S., {et~al.} 2007, \aap, 469, 27

\bibitem[{{Chen} \& {Tokunaga}(1994)}]{1994ApJS...90..149C}
{Chen}, H., \& {Tokunaga}, A.~T. 1994, \apjs, 90, 149

\bibitem[{{Chiang} \& {Murray-Clay}(2007)}]{2007NatPh...3..604C}
{Chiang}, E., \& {Murray-Clay}, R. 2007, Nature Physics, 3, 604

\bibitem[{{Cieza} {et~al.}(2010){Cieza}, {Schreiber}, {Romero}, {Mora},
  {Merin}, {Swift}, {Orellana}, {Williams}, {Harvey}, \&
  {Evans}}]{2010ApJ...712..925C}
{Cieza}, L.~A., {Schreiber}, M.~R., {Romero}, G.~A., {et~al.} 2010, \apj, 712,
  925

\bibitem[{{Clarke} {et~al.}(2001){Clarke}, {Gendrin}, \&
  {Sotomayor}}]{2001MNRAS.328..485C}
{Clarke}, C.~J., {Gendrin}, A., \& {Sotomayor}, M. 2001, \mnras, 328, 485

\bibitem[{{Comer{\'o}n} {et~al.}(2003){Comer{\'o}n}, {Fern{\'a}ndez},
  {Baraffe}, {Neuh{\"a}user}, \& {Kaas}}]{2003A&A...406.1001C}
{Comer{\'o}n}, F., {Fern{\'a}ndez}, M., {Baraffe}, I., {Neuh{\"a}user}, R., \&
  {Kaas}, A.~A. 2003, \aap, 406, 1001

\bibitem[{{Connelley} \& {Greene}(2010)}]{2010AJ....140.1214C}
{Connelley}, M.~S., \& {Greene}, T.~P. 2010, \aj, 140, 1214

\bibitem[{{Currie}(2010)}]{2010arXiv1002.1715C}
{Currie}, T. 2010, ArXiv e-prints

\bibitem[{{Currie} \& {Kenyon}(2009)}]{2009AJ....138..703C}
{Currie}, T., \& {Kenyon}, S.~J. 2009, \aj, 138, 703

\bibitem[{{Currie} {et~al.}(2009){Currie}, {Lada}, {Plavchan}, {Robitaille},
  {Irwin}, \& {Kenyon}}]{2009ApJ...698....1C}
{Currie}, T., {Lada}, C.~J., {Plavchan}, P., {et~al.} 2009, \apj, 698, 1

\bibitem[{{Currie} \& {Sicilia-Aguilar}(2011)}]{2011ApJ...732...24C}
{Currie}, T., \& {Sicilia-Aguilar}, A. 2011, \apj, 732, 24

\bibitem[{{Dahm}(2008)}]{2008AJ....136..521D}
{Dahm}, S.~E. 2008, \aj, 136, 521

\bibitem[{{Dahm} \& {Hillenbrand}(2007)}]{2007AJ....133.2072D}
{Dahm}, S.~E., \& {Hillenbrand}, L.~A. 2007, \aj, 133, 2072

\bibitem[{{Dahm} \& {Simon}(2005)}]{2005AJ....129..829D}
{Dahm}, S.~E., \& {Simon}, T. 2005, \aj, 129, 829

\bibitem[{{D'Antona} \& {Mazzitelli}(1997)}]{1997MmSAI..68..807D}
{D'Antona}, F., \& {Mazzitelli}, I. 1997, Memorie della Societa Astronomica
  Italiana, 68, 807

\bibitem[{{Dotter} {et~al.}(2008){Dotter}, {Chaboyer}, {Jevremovi{\'c}},
  {Kostov}, {Baron}, \& {Ferguson}}]{2008ApJS..178...89D}
{Dotter}, A., {Chaboyer}, B., {Jevremovi{\'c}}, D., {et~al.} 2008, \apjs, 178,
  89

\bibitem[{{Dullemond} {et~al.}(2006){Dullemond}, {Natta}, \&
  {Testi}}]{2006ApJ...645L..69D}
{Dullemond}, C.~P., {Natta}, A., \& {Testi}, L. 2006, \apjl, 645, L69

\bibitem[{{Espaillat} {et~al.}(2012){Espaillat}, {Ingleby}, {Hern{\'a}ndez},
  {Furlan}, {D'Alessio}, {Calvet}, {Andrews}, {Muzerolle}, {Qi}, \&
  {Wilner}}]{2012ApJ...747..103E}
{Espaillat}, C., {Ingleby}, L., {Hern{\'a}ndez}, J., {et~al.} 2012, \apj, 747,
  103

\bibitem[{{Evans} {et~al.}(2009){Evans}, {Dunham}, {J{\o}rgensen}, {Enoch},
  {Mer{\'{\i}}n}, {van Dishoeck}, {Alcal{\'a}}, {Myers}, {Stapelfeldt},
  {Huard}, {Allen}, {Harvey}, {van Kempen}, {Blake}, {Koerner}, {Mundy},
  {Padgett}, \& {Sargent}}]{2009ApJS..181..321E}
{Evans}, N.~J., {Dunham}, M.~M., {J{\o}rgensen}, J.~K., {et~al.} 2009, \apjs,
  181, 321

\bibitem[{{Fabricant} {et~al.}(2005){Fabricant}, {Fata}, {Roll}, {Hertz},
  {Caldwell}, {Gauron}, {Geary}, {McLeod}, {Szentgyorgyi}, {Zajac}, {Kurtz},
  {Barberis}, {Bergner}, {Brown}, {Conroy}, {Eng}, {Geller}, {Goddard},
  {Honsa}, {Mueller}, {Mink}, {Ordway}, {Tokarz}, {Woods}, {Wyatt}, {Epps}, \&
  {Dell'Antonio}}]{2005PASP..117.1411F}
{Fabricant}, D., {Fata}, R., {Roll}, J., {et~al.} 2005, \pasp, 117, 1411

\bibitem[{{Fang} {et~al.}(2013){Fang}, {van Boekel}, {Bouwman}, {Henning},
  {Lawson}, \& {Sicilia-Aguilar}}]{2013A&A...549A..15F}
{Fang}, M., {van Boekel}, R., {Bouwman}, J., {et~al.} 2013, \aap, 549, A15

\bibitem[{{Fang} {et~al.}(2009){Fang}, {van Boekel}, {Wang}, {Carmona},
  {Sicilia-Aguilar}, \& {Henning}}]{2009A&A...504..461F}
{Fang}, M., {van Boekel}, R., {Wang}, W., {et~al.} 2009, \aap, 504, 461(Paper
  I)

\bibitem[{{Fang} {et~al.}(2012){Fang}, {van Boekel}, {King}, {Henning},
  {Bouwman}, {Doi}, {Okamoto}, {Roccatagliata}, \&
  {Sicilia-Aguilar}}]{2012A&A...539A.119F}
{Fang}, M., {van Boekel}, R., {King}, R.~R., {et~al.} 2012, \aap, 539, A119

\bibitem[{{Feigelson} {et~al.}(2003){Feigelson}, {Gaffney}, {Garmire},
  {Hillenbrand}, \& {Townsley}}]{2003ApJ...584..911F}
{Feigelson}, E.~D., {Gaffney}, III, J.~A., {Garmire}, G., {Hillenbrand}, L.~A.,
  \& {Townsley}, L. 2003, \apj, 584, 911

\bibitem[{{Feigelson} \& {Montmerle}(1999)}]{1999ARA&A..37..363F}
{Feigelson}, E.~D., \& {Montmerle}, T. 1999, \araa, 37, 363

\bibitem[{{F{\H u}r{\'e}sz} {et~al.}(2008){F{\H u}r{\'e}sz}, {Hartmann},
  {Megeath}, {Szentgyorgyi}, \& {Hamden}}]{2008ApJ...676.1109F}
{F{\H u}r{\'e}sz}, G., {Hartmann}, L.~W., {Megeath}, S.~T., {Szentgyorgyi},
  A.~H., \& {Hamden}, E.~T. 2008, \apj, 676, 1109

\bibitem[{{Fischer} \& {Marcy}(1992)}]{1992ApJ...396..178F}
{Fischer}, D.~A., \& {Marcy}, G.~W. 1992, \apj, 396, 178

\bibitem[{{Flaherty} \& {Muzerolle}(2008)}]{2008AJ....135..966F}
{Flaherty}, K.~M., \& {Muzerolle}, J. 2008, \aj, 135, 966

\bibitem[{{Flaherty} {et~al.}(2007){Flaherty}, {Pipher}, {Megeath}, {Winston},
  {Gutermuth}, {Muzerolle}, {Allen}, \& {Fazio}}]{2007ApJ...663.1069F}
{Flaherty}, K.~M., {Pipher}, J.~L., {Megeath}, S.~T., {et~al.} 2007, \apj, 663,
  1069

\bibitem[{{Furlan} {et~al.}(2006){Furlan}, {Hartmann}, {Calvet}, {D'Alessio},
  {Franco-Hern{\'a}ndez}, {Forrest}, {Watson}, {Uchida}, {Sargent}, {Green},
  {Keller}, \& {Herter}}]{2006ApJS..165..568F}
{Furlan}, E., {Hartmann}, L., {Calvet}, N., {et~al.} 2006, \apjs, 165, 568

\bibitem[{{G{\^a}lfalk} \& {Olofsson}(2008)}]{2008A&A...489.1409G}
{G{\^a}lfalk}, M., \& {Olofsson}, G. 2008, \aap, 489, 1409

\bibitem[{{Gammie}(1996)}]{1996ApJ...457..355G}
{Gammie}, C.~F. 1996, \apj, 457, 355

\bibitem[{{Garcia Lopez} {et~al.}(2006){Garcia Lopez}, {Natta}, {Testi}, \&
  {Habart}}]{2006A&A...459..837G}
{Garcia Lopez}, R., {Natta}, A., {Testi}, L., \& {Habart}, E. 2006, \aap, 459,
  837

\bibitem[{{Gatti} {et~al.}(2008){Gatti}, {Natta}, {Randich}, {Testi}, \&
  {Sacco}}]{2008A&A...481..423G}
{Gatti}, T., {Natta}, A., {Randich}, S., {Testi}, L., \& {Sacco}, G. 2008,
  \aap, 481, 423

\bibitem[{{Genzel} {et~al.}(1981){Genzel}, {Reid}, {Moran}, \&
  {Downes}}]{1981ApJ...244..884G}
{Genzel}, R., {Reid}, M.~J., {Moran}, J.~M., \& {Downes}, D. 1981, \apj, 244,
  884

\bibitem[{{Ghez} {et~al.}(1997){Ghez}, {McCarthy}, {Patience}, \&
  {Beck}}]{1997ApJ...481..378G}
{Ghez}, A.~M., {McCarthy}, D.~W., {Patience}, J.~L., \& {Beck}, T.~L. 1997,
  \apj, 481, 378

\bibitem[{{Ghez} {et~al.}(1993){Ghez}, {Neugebauer}, \&
  {Matthews}}]{1993AJ....106.2005G}
{Ghez}, A.~M., {Neugebauer}, G., \& {Matthews}, K. 1993, \aj, 106, 2005

\bibitem[{{Glassgold} {et~al.}(1997){Glassgold}, {Najita}, \&
  {Igea}}]{1997ApJ...480..344G}
{Glassgold}, A.~E., {Najita}, J., \& {Igea}, J. 1997, \apj, 480, 344

\bibitem[{{Guarcello} {et~al.}(2007){Guarcello}, {Prisinzano}, {Micela},
  {Damiani}, {Peres}, \& {Sciortino}}]{2007A&A...462..245G}
{Guarcello}, M.~G., {Prisinzano}, L., {Micela}, G., {et~al.} 2007, \aap, 462,
  245

\bibitem[{{G{\"u}del}(2004)}]{2004A&ARv..12...71G}
{G{\"u}del}, M. 2004, \aapr, 12, 71

\bibitem[{{G{\"u}del} {et~al.}(2007){G{\"u}del}, {Briggs}, {Arzner}, {Audard},
  {Bouvier}, {Feigelson}, {Franciosini}, {Glauser}, {Grosso}, {Micela},
  {Monin}, {Montmerle}, {Padgett}, {Palla}, {Pillitteri}, {Rebull}, {Scelsi},
  {Silva}, {Skinner}, {Stelzer}, \& {Telleschi}}]{2007A&A...468..353G}
{G{\"u}del}, M., {Briggs}, K.~R., {Arzner}, K., {et~al.} 2007, \aap, 468, 353

\bibitem[{{Gullbring} {et~al.}(1998){Gullbring}, {Hartmann}, {Briceno}, \&
  {Calvet}}]{1998ApJ...492..323G}
{Gullbring}, E., {Hartmann}, L., {Briceno}, C., \& {Calvet}, N. 1998, \apj,
  492, 323

\bibitem[{{Gullbring} {et~al.}(1996){Gullbring}, {Petrov}, {Ilyin}, {Tuominen},
  {Gahm}, \& {Loden}}]{1996A&A...314..835G}
{Gullbring}, E., {Petrov}, P.~P., {Ilyin}, I., {et~al.} 1996, \aap, 314, 835

\bibitem[{{Gustafsson} {et~al.}(2008){Gustafsson}, {Edvardsson}, {Eriksson},
  {J{\o}rgensen}, {Nordlund}, \& {Plez}}]{2008A&A...486..951G}
{Gustafsson}, B., {Edvardsson}, B., {Eriksson}, K., {et~al.} 2008, \aap, 486,
  951

\bibitem[{{Gutermuth} {et~al.}(2008){Gutermuth}, {Myers}, {Megeath}, {Allen},
  {Pipher}, {Muzerolle}, {Porras}, {Winston}, \& {Fazio}}]{2008ApJ...674..336G}
{Gutermuth}, R.~A., {Myers}, P.~C., {Megeath}, S.~T., {et~al.} 2008, \apj, 674,
  336

\bibitem[{{Haisch} {et~al.}(2001){Haisch}, {Lada}, \&
  {Lada}}]{2001ApJ...553L.153H}
{Haisch}, Jr., K.~E., {Lada}, E.~A., \& {Lada}, C.~J. 2001, \apjl, 553, L153

\bibitem[{{Hartigan} {et~al.}(1995){Hartigan}, {Edwards}, \&
  {Ghandour}}]{1995ApJ...452..736H}
{Hartigan}, P., {Edwards}, S., \& {Ghandour}, L. 1995, \apj, 452, 736

\bibitem[{{Hartigan} \& {Kenyon}(2003)}]{2003ApJ...583..334H}
{Hartigan}, P., \& {Kenyon}, S.~J. 2003, \apj, 583, 334

\bibitem[{{Hartmann}(1999)}]{1999NewAR..43....1H}
{Hartmann}, L. 1999, NewA Rev., 43, 1

\bibitem[{{Hartmann}(2001)}]{2001AJ....121.1030H}
---. 2001, \aj, 121, 1030

\bibitem[{{Hartmann} {et~al.}(1998){Hartmann}, {Calvet}, {Gullbring}, \&
  {D'Alessio}}]{1998ApJ...495..385H}
{Hartmann}, L., {Calvet}, N., {Gullbring}, E., \& {D'Alessio}, P. 1998, \apj,
  495, 385

\bibitem[{{Hartmann} {et~al.}(2006){Hartmann}, {D'Alessio}, {Calvet}, \&
  {Muzerolle}}]{2006ApJ...648..484H}
{Hartmann}, L., {D'Alessio}, P., {Calvet}, N., \& {Muzerolle}, J. 2006, \apj,
  648, 484

\bibitem[{{Hartmann} {et~al.}(1994){Hartmann}, {Hewett}, \&
  {Calvet}}]{1994ApJ...426..669H}
{Hartmann}, L., {Hewett}, R., \& {Calvet}, N. 1994, \apj, 426, 669

\bibitem[{{Hayashi} {et~al.}(1985){Hayashi}, {Nakazawa}, \&
  {Nakagawa}}]{1985prpl.conf.1100H}
{Hayashi}, C., {Nakazawa}, K., \& {Nakagawa}, Y. 1985, in Protostars and
  planets II (A86-12626 03-90). Tucson, AZ, University of Arizona Press, 1985,
  p. 1100-1153., ed. D.~C. {Black} \& M.~S. {Matthews}, 1100--1153

\bibitem[{{Herbig}(1998)}]{1998ApJ...497..736H}
{Herbig}, G.~H. 1998, \apj, 497, 736

\bibitem[{{Herbig} \& {Dahm}(2002)}]{2002AJ....123..304H}
{Herbig}, G.~H., \& {Dahm}, S.~E. 2002, \aj, 123, 304

\bibitem[{{Herczeg} \& {Hillenbrand}(2008)}]{2008ApJ...681..594H}
{Herczeg}, G.~J., \& {Hillenbrand}, L.~A. 2008, \apj, 681, 594

\bibitem[{{Hern{\'a}ndez} {et~al.}(2004){Hern{\'a}ndez}, {Calvet},
  {Brice{\~n}o}, {Hartmann}, \& {Berlind}}]{2004AJ....127.1682H}
{Hern{\'a}ndez}, J., {Calvet}, N., {Brice{\~n}o}, C., {Hartmann}, L., \&
  {Berlind}, P. 2004, \aj, 127, 1682

\bibitem[{{Hern{\'a}ndez} {et~al.}(2010){Hern{\'a}ndez}, {Morales-Calderon},
  {Calvet}, {Hartmann}, {Muzerolle}, {Gutermuth}, {Luhman}, \&
  {Stauffer}}]{2010ApJ...722.1226H}
{Hern{\'a}ndez}, J., {Morales-Calderon}, M., {Calvet}, N., {et~al.} 2010, \apj,
  722, 1226

\bibitem[{{Hern{\'a}ndez} {et~al.}(2007{\natexlab{a}}){Hern{\'a}ndez},
  {Hartmann}, {Megeath}, {Gutermuth}, {Muzerolle}, {Calvet}, {Vivas},
  {Brice{\~n}o}, {Allen}, {Stauffer}, {Young}, \&
  {Fazio}}]{2007ApJ...662.1067H}
{Hern{\'a}ndez}, J., {Hartmann}, L., {Megeath}, T., {et~al.}
  2007{\natexlab{a}}, \apj, 662, 1067

\bibitem[{{Hern{\'a}ndez} {et~al.}(2007{\natexlab{b}}){Hern{\'a}ndez},
  {Calvet}, {Brice{\~n}o}, {Hartmann}, {Vivas}, {Muzerolle}, {Downes}, {Allen},
  \& {Gutermuth}}]{2007ApJ...671.1784H}
{Hern{\'a}ndez}, J., {Calvet}, N., {Brice{\~n}o}, C., {et~al.}
  2007{\natexlab{b}}, \apj, 671, 1784

\bibitem[{{Hillenbrand}(1997)}]{1997AJ....113.1733H}
{Hillenbrand}, L.~A. 1997, \aj, 113, 1733

\bibitem[{{Hillenbrand}(2002)}]{2002astro.ph.10520H}
---. 2002, ArXiv Astrophysics e-prints

\bibitem[{{Hillenbrand} {et~al.}(2008){Hillenbrand}, {Bauermeister}, \&
  {White}}]{2008ASPC..384..200H}
{Hillenbrand}, L.~A., {Bauermeister}, A., \& {White}, R.~J. 2008, in
  Astronomical Society of the Pacific Conference Series, Vol. 384, 14th
  Cambridge Workshop on Cool Stars, Stellar Systems, and the Sun, ed. G.~{van
  Belle}, 200--+

\bibitem[{{Hillenbrand} \& {White}(2004)}]{2004ApJ...604..741H}
{Hillenbrand}, L.~A., \& {White}, R.~J. 2004, \apj, 604, 741

\bibitem[{{Hirota} {et~al.}(2007){Hirota}, {Bushimata}, {Choi}, {Honma},
  {Imai}, {Iwadate}, {Jike}, {Kameno}, {Kameya}, {Kamohara}, {Kan-Ya},
  {Kawaguchi}, {Kijima}, {Kim}, {Kobayashi}, {Kuji}, {Kurayama}, {Manabe},
  {Maruyama}, {Matsui}, {Matsumoto}, {Miyaji}, {Nagayama}, {Nakagawa},
  {Nakamura}, {Oh}, {Omodaka}, {Oyama}, {Sakai}, {Sasao}, {Sato}, {Sato},
  {Shibata}, {Shintani}, {Tamura}, {Tsushima}, \&
  {Yamashita}}]{2007PASJ...59..897H}
{Hirota}, T., {Bushimata}, T., {Choi}, Y.~K., {et~al.} 2007, \pasj, 59, 897

\bibitem[{{Hollenbach} {et~al.}(2000){Hollenbach}, {Yorke}, \&
  {Johnstone}}]{2000prpl.conf..401H}
{Hollenbach}, D.~J., {Yorke}, H.~W., \& {Johnstone}, D. 2000, Protostars and
  Planets IV, 401

\bibitem[{{Hsu} {et~al.}(2012){Hsu}, {Hartmann}, {Allen}, {Hern{\'a}ndez},
  {Megeath}, {Mosby}, {Tobin}, \& {Espaillat}}]{2012ApJ...752...59H}
{Hsu}, W.-H., {Hartmann}, L., {Allen}, L., {et~al.} 2012, \apj, 752, 59

\bibitem[{{Hsu} {et~al.}(2013){Hsu}, {Hartmann}, {Allen}, {Hern{\'a}ndez},
  {Megeath}, {Tobin}, \& {Ingleby}}]{2013ApJ...764..114H}
---. 2013, \apj, 764, 114

\bibitem[{{Jayawardhana} {et~al.}(2006){Jayawardhana}, {Coffey}, {Scholz},
  {Brandeker}, \& {van Kerkwijk}}]{2006ApJ...648.1206J}
{Jayawardhana}, R., {Coffey}, J., {Scholz}, A., {Brandeker}, A., \& {van
  Kerkwijk}, M.~H. 2006, \apj, 648, 1206

\bibitem[{{Jeffries} {et~al.}(2009){Jeffries}, {Jackson}, {James}, \&
  {Cargile}}]{2009MNRAS.400..317J}
{Jeffries}, R.~D., {Jackson}, R.~J., {James}, D.~J., \& {Cargile}, P.~A. 2009,
  \mnras, 400, 317

\bibitem[{{Johns} \& {Basri}(1995)}]{1995ApJ...449..341J}
{Johns}, C.~M., \& {Basri}, G. 1995, \apj, 449, 341

\bibitem[{{Johnstone} {et~al.}(1998){Johnstone}, {Hollenbach}, \&
  {Bally}}]{1998ApJ...499..758J}
{Johnstone}, D., {Hollenbach}, D., \& {Bally}, J. 1998, \apj, 499, 758

\bibitem[{{Kenyon} \& {Hartmann}(1995)}]{1995ApJS..101..117K}
{Kenyon}, S.~J., \& {Hartmann}, L. 1995, \apjs, 101, 117

\bibitem[{{Kurosawa} {et~al.}(2011){Kurosawa}, {Romanova}, \&
  {Harries}}]{2011MNRAS.416.2623K}
{Kurosawa}, R., {Romanova}, M.~M., \& {Harries}, T.~J. 2011, \mnras, 416, 2623

\bibitem[{{Kurucz}(1994)}]{1994KurCD..19.....K}
{Kurucz}, R. 1994, Solar abundance model atmospheres for 0,1,2,4,8 km/s.~Kurucz
  CD-ROM No.~19.~ Cambridge, Mass.: Smithsonian Astrophysical Observatory,
  1994., 19

\bibitem[{{Kurucz}(1979)}]{1979ApJS...40....1K}
{Kurucz}, R.~L. 1979, \apjs, 40, 1

\bibitem[{{Lada} {et~al.}(2006){Lada}, {Muench}, {Luhman}, {Allen}, {Hartmann},
  {Megeath}, {Myers}, {Fazio}, {Wood}, {Muzerolle}, {Rieke}, {Siegler}, \&
  {Young}}]{2006AJ....131.1574L}
{Lada}, C.~J., {Muench}, A.~A., {Luhman}, K.~L., {et~al.} 2006, \aj, 131, 1574

\bibitem[{{Lada} \& {Lada}(1995)}]{1995AJ....109.1682L}
{Lada}, E.~A., \& {Lada}, C.~J. 1995, \aj, 109, 1682

\bibitem[{{Lafreni{\`e}re} {et~al.}(2008){Lafreni{\`e}re}, {Jayawardhana},
  {Brandeker}, {Ahmic}, \& {van Kerkwijk}}]{2008ApJ...683..844L}
{Lafreni{\`e}re}, D., {Jayawardhana}, R., {Brandeker}, A., {Ahmic}, M., \& {van
  Kerkwijk}, M.~H. 2008, \apj, 683, 844

\bibitem[{{Lawrence} {et~al.}(2007){Lawrence}, {Warren}, {Almaini}, {Edge},
  {Hambly}, {Jameson}, {Lucas}, {Casali}, {Adamson}, {Dye}, {Emerson},
  {Foucaud}, {Hewett}, {Hirst}, {Hodgkin}, {Irwin}, {Lodieu}, {McMahon},
  {Simpson}, {Smail}, {Mortlock}, \& {Folger}}]{2007MNRAS.379.1599L}
{Lawrence}, A., {Warren}, S.~J., {Almaini}, O., {et~al.} 2007, \mnras, 379,
  1599

\bibitem[{{Leinert} {et~al.}(1993){Leinert}, {Zinnecker}, {Weitzel},
  {Christou}, {Ridgway}, {Jameson}, {Haas}, \& {Lenzen}}]{1993A&A...278..129L}
{Leinert}, C., {Zinnecker}, H., {Weitzel}, N., {et~al.} 1993, \aap, 278, 129

\bibitem[{{Lima} {et~al.}(2010){Lima}, {Alencar}, {Calvet}, {Hartmann}, \&
  {Muzerolle}}]{2010A&A...522A.104L}
{Lima}, G.~H.~R.~A., {Alencar}, S.~H.~P., {Calvet}, N., {Hartmann}, L., \&
  {Muzerolle}, J. 2010, \aap, 522, A104

\bibitem[{{Lin} \& {Papaloizou}(1993)}]{1993prpl.conf..749L}
{Lin}, D.~N.~C., \& {Papaloizou}, J.~C.~B. 1993, in Protostars and Planets III,
  ed. {E.~H.~Levy \& J.~I.~Lunine}, 749--835

\bibitem[{{Luhman} {et~al.}(2010){Luhman}, {Allen}, {Espaillat}, {Hartmann}, \&
  {Calvet}}]{2010ApJS..186..111L}
{Luhman}, K.~L., {Allen}, P.~R., {Espaillat}, C., {Hartmann}, L., \& {Calvet},
  N. 2010, \apjs, 186, 111

\bibitem[{{Luhman} {et~al.}(2003){Luhman}, {Stauffer}, {Muench}, {Rieke},
  {Lada}, {Bouvier}, \& {Lada}}]{2003ApJ...593.1093L}
{Luhman}, K.~L., {Stauffer}, J.~R., {Muench}, A.~A., {et~al.} 2003, \apj, 593,
  1093

\bibitem[{{Luhman} {et~al.}(2008){Luhman}, {Allen}, {Allen}, {Gutermuth},
  {Hartmann}, {Mamajek}, {Megeath}, {Myers}, \& {Fazio}}]{2008ApJ...675.1375L}
{Luhman}, K.~L., {Allen}, L.~E., {Allen}, P.~R., {et~al.} 2008, \apj, 675, 1375

\bibitem[{{Maddalena} {et~al.}(1986){Maddalena}, {Morris}, {Moscowitz}, \&
  {Thaddeus}}]{1986ApJ...303..375M}
{Maddalena}, R.~J., {Morris}, M., {Moscowitz}, J., \& {Thaddeus}, P. 1986,
  \apj, 303, 375

\bibitem[{{Mamajek} {et~al.}(2004){Mamajek}, {Meyer}, {Hinz}, {Hoffmann},
  {Cohen}, \& {Hora}}]{2004ApJ...612..496M}
{Mamajek}, E.~E., {Meyer}, M.~R., {Hinz}, P.~M., {et~al.} 2004, \apj, 612, 496

\bibitem[{{Mayne} {et~al.}(2012){Mayne}, {Harries}, {Rowe}, \&
  {Acreman}}]{2012MNRAS.423.1775M}
{Mayne}, N.~J., {Harries}, T.~J., {Rowe}, J., \& {Acreman}, D.~M. 2012, \mnras,
  423, 1775

\bibitem[{{Megeath} {et~al.}(2012){Megeath}, {Gutermuth}, {Muzerolle},
  {Kryukova}, {Flaherty}, {Hora}, {Allen}, {Hartmann}, {Myers}, {Pipher},
  {Stauffer}, {Young}, \& {Fazio}}]{2012AJ....144..192M}
{Megeath}, S.~T., {Gutermuth}, R., {Muzerolle}, J., {et~al.} 2012, \aj, 144,
  192

\bibitem[{{Mendigut{\'{\i}}a} {et~al.}(2011){Mendigut{\'{\i}}a}, {Eiroa},
  {Montesinos}, {Mora}, {Oudmaijer}, {Mer{\'{\i}}n}, \&
  {Meeus}}]{2011A&A...529A..34M}
{Mendigut{\'{\i}}a}, I., {Eiroa}, C., {Montesinos}, B., {et~al.} 2011, \aap,
  529, A34+

\bibitem[{{Menten} {et~al.}(2007){Menten}, {Reid}, {Forbrich}, \&
  {Brunthaler}}]{2007A&A...474..515M}
{Menten}, K.~M., {Reid}, M.~J., {Forbrich}, J., \& {Brunthaler}, A. 2007, \aap,
  474, 515

\bibitem[{{Mentuch} {et~al.}(2008){Mentuch}, {Brandeker}, {van Kerkwijk},
  {Jayawardhana}, \& {Hauschildt}}]{2008ApJ...689.1127M}
{Mentuch}, E., {Brandeker}, A., {van Kerkwijk}, M.~H., {Jayawardhana}, R., \&
  {Hauschildt}, P.~H. 2008, \apj, 689, 1127

\bibitem[{{Mer{\'{\i}}n} {et~al.}(2010){Mer{\'{\i}}n}, {Brown}, {Oliveira},
  {Herczeg}, {van Dishoeck}, {Bottinelli}, {Evans}, {Cieza}, {Spezzi},
  {Alcal{\'a}}, {Harvey}, {Blake}, {Bayo}, {Geers}, {Lahuis}, {Prusti},
  {Augereau}, {Olofsson}, {Walter}, \& {Chiu}}]{2010ApJ...718.1200M}
{Mer{\'{\i}}n}, B., {Brown}, J.~M., {Oliveira}, I., {et~al.} 2010, \apj, 718,
  1200

\bibitem[{{Meyer} \& {Beckwith}(2000)}]{2000LNP...548..341M}
{Meyer}, M.~R., \& {Beckwith}, S.~V.~W. 2000, in Lecture Notes in Physics,
  Berlin Springer Verlag, Vol. 548, ISO Survey of a Dusty Universe, ed.
  D.~{Lemke}, M.~{Stickel}, \& K.~{Wilke}, 341--+

\bibitem[{{Meyer} {et~al.}(1997){Meyer}, {Calvet}, \&
  {Hillenbrand}}]{1997AJ....114..288M}
{Meyer}, M.~R., {Calvet}, N., \& {Hillenbrand}, L.~A. 1997, \aj, 114, 288

\bibitem[{{Mohanty} {et~al.}(2005){Mohanty}, {Jayawardhana}, \&
  {Basri}}]{2005ApJ...626..498M}
{Mohanty}, S., {Jayawardhana}, R., \& {Basri}, G. 2005, \apj, 626, 498

\bibitem[{{Mottram} {et~al.}(2007){Mottram}, {Vink}, {Oudmaijer}, \&
  {Patel}}]{2007MNRAS.377.1363M}
{Mottram}, J.~C., {Vink}, J.~S., {Oudmaijer}, R.~D., \& {Patel}, M. 2007,
  \mnras, 377, 1363

\bibitem[{{Muench} {et~al.}(2007){Muench}, {Lada}, {Luhman}, {Muzerolle}, \&
  {Young}}]{2007AJ....134..411M}
{Muench}, A.~A., {Lada}, C.~J., {Luhman}, K.~L., {Muzerolle}, J., \& {Young},
  E. 2007, \aj, 134, 411

\bibitem[{{Muench} {et~al.}(2000){Muench}, {Lada}, \&
  {Lada}}]{2000ApJ...533..358M}
{Muench}, A.~A., {Lada}, E.~A., \& {Lada}, C.~J. 2000, \apj, 533, 358

\bibitem[{{Muench} {et~al.}(2002){Muench}, {Lada}, {Lada}, \&
  {Alves}}]{2002ApJ...573..366M}
{Muench}, A.~A., {Lada}, E.~A., {Lada}, C.~J., \& {Alves}, J. 2002, \apj, 573,
  366

\bibitem[{{Muench} {et~al.}(2003){Muench}, {Lada}, {Lada}, {Elston}, {Alves},
  {Horrobin}, {Huard}, {Levine}, {Raines}, \&
  {Rom{\'a}n-Z{\'u}{\~n}iga}}]{2003AJ....125.2029M}
{Muench}, A.~A., {Lada}, E.~A., {Lada}, C.~J., {et~al.} 2003, \aj, 125, 2029

\bibitem[{{Muzerolle} {et~al.}(2010){Muzerolle}, {Allen}, {Megeath},
  {Hern{\'a}ndez}, \& {Gutermuth}}]{2010ApJ...708.1107M}
{Muzerolle}, J., {Allen}, L.~E., {Megeath}, S.~T., {Hern{\'a}ndez}, J., \&
  {Gutermuth}, R.~A. 2010, \apj, 708, 1107

\bibitem[{{Muzerolle} {et~al.}(2001){Muzerolle}, {Calvet}, \&
  {Hartmann}}]{2001ApJ...550..944M}
{Muzerolle}, J., {Calvet}, N., \& {Hartmann}, L. 2001, \apj, 550, 944

\bibitem[{{Muzerolle} {et~al.}(1998){Muzerolle}, {Hartmann}, \&
  {Calvet}}]{1998AJ....116.2965M}
{Muzerolle}, J., {Hartmann}, L., \& {Calvet}, N. 1998, \aj, 116, 2965

\bibitem[{{Muzerolle} {et~al.}(2003){Muzerolle}, {Hillenbrand}, {Calvet},
  {Brice{\~n}o}, \& {Hartmann}}]{2003ApJ...592..266M}
{Muzerolle}, J., {Hillenbrand}, L., {Calvet}, N., {Brice{\~n}o}, C., \&
  {Hartmann}, L. 2003, \apj, 592, 266

\bibitem[{{Muzerolle} {et~al.}(2005){Muzerolle}, {Luhman}, {Brice{\~n}o},
  {Hartmann}, \& {Calvet}}]{2005ApJ...625..906M}
{Muzerolle}, J., {Luhman}, K.~L., {Brice{\~n}o}, C., {Hartmann}, L., \&
  {Calvet}, N. 2005, \apj, 625, 906

\bibitem[{{Najita} {et~al.}(2007){Najita}, {Strom}, \&
  {Muzerolle}}]{2007MNRAS.378..369N}
{Najita}, J.~R., {Strom}, S.~E., \& {Muzerolle}, J. 2007, \mnras, 378, 369

\bibitem[{{Natta} {et~al.}(2004){Natta}, {Testi}, {Muzerolle}, {Randich},
  {Comer{\'o}n}, \& {Persi}}]{2004A&A...424..603N}
{Natta}, A., {Testi}, L., {Muzerolle}, J., {et~al.} 2004, \aap, 424, 603

\bibitem[{{Natta} {et~al.}(2006){Natta}, {Testi}, \&
  {Randich}}]{2006A&A...452..245N}
{Natta}, A., {Testi}, L., \& {Randich}, S. 2006, \aap, 452, 245

\bibitem[{{Nguyen} {et~al.}(2009{\natexlab{a}}){Nguyen}, {Jayawardhana}, {van
  Kerkwijk}, {Brandeker}, {Scholz}, \& {Damjanov}}]{2009ApJ...695.1648N}
{Nguyen}, D.~C., {Jayawardhana}, R., {van Kerkwijk}, M.~H., {et~al.}
  2009{\natexlab{a}}, \apj, 695, 1648

\bibitem[{{Nguyen} {et~al.}(2009{\natexlab{b}}){Nguyen}, {Scholz}, {van
  Kerkwijk}, {Jayawardhana}, \& {Brandeker}}]{2009ApJ...694L.153N}
{Nguyen}, D.~C., {Scholz}, A., {van Kerkwijk}, M.~H., {Jayawardhana}, R., \&
  {Brandeker}, A. 2009{\natexlab{b}}, \apjl, 694, L153

\bibitem[{{O'dell} {et~al.}(1993){O'dell}, {Wen}, \&
  {Hu}}]{1993ApJ...410..696O}
{O'dell}, C.~R., {Wen}, Z., \& {Hu}, X. 1993, \apj, 410, 696

\bibitem[{{Osterloh} \& {Beckwith}(1995)}]{1995ApJ...439..288O}
{Osterloh}, M., \& {Beckwith}, S.~V.~W. 1995, \apj, 439, 288

\bibitem[{{Palla} \& {Stahler}(2000)}]{2000ApJ...540..255P}
{Palla}, F., \& {Stahler}, S.~W. 2000, \apj, 540, 255

\bibitem[{{Pontoppidan} {et~al.}(2005){Pontoppidan}, {Dullemond}, {van
  Dishoeck}, {Blake}, {Boogert}, {Evans}, {Kessler-Silacci}, \&
  {Lahuis}}]{2005ApJ...622..463P}
{Pontoppidan}, K.~M., {Dullemond}, C.~P., {van Dishoeck}, E.~F., {et~al.} 2005,
  \apj, 622, 463

\bibitem[{{Pott} {et~al.}(2010){Pott}, {Perrin}, {Furlan}, {Ghez}, {Herbst}, \&
  {Metchev}}]{2010ApJ...710..265P}
{Pott}, J.-U., {Perrin}, M.~D., {Furlan}, E., {et~al.} 2010, \apj, 710, 265

\bibitem[{{Quillen} {et~al.}(2004){Quillen}, {Blackman}, {Frank}, \&
  {Varni{\`e}re}}]{2004ApJ...612L.137Q}
{Quillen}, A.~C., {Blackman}, E.~G., {Frank}, A., \& {Varni{\`e}re}, P. 2004,
  \apjl, 612, L137

\bibitem[{{Reipurth} {et~al.}(1996){Reipurth}, {Pedrosa}, \&
  {Lago}}]{1996A&AS..120..229R}
{Reipurth}, B., {Pedrosa}, A., \& {Lago}, M.~T.~V.~T. 1996, \aaps, 120, 229

\bibitem[{{Rice} {et~al.}(2003){Rice}, {Wood}, {Armitage}, {Whitney}, \&
  {Bjorkman}}]{2003MNRAS.342...79R}
{Rice}, W.~K.~M., {Wood}, K., {Armitage}, P.~J., {Whitney}, B.~A., \&
  {Bjorkman}, J.~E. 2003, \mnras, 342, 79

\bibitem[{{Richling} \& {Yorke}(2000)}]{2000ApJ...539..258R}
{Richling}, S., \& {Yorke}, H.~W. 2000, \apj, 539, 258

\bibitem[{{Rieke} \& {Lebofsky}(1985)}]{1985ApJ...288..618R}
{Rieke}, G.~H., \& {Lebofsky}, M.~J. 1985, \apj, 288, 618

\bibitem[{{Rigliaco} {et~al.}(2011{\natexlab{a}}){Rigliaco}, {Natta},
  {Randich}, {Testi}, \& {Biazzo}}]{2011A&A...525A..47R}
{Rigliaco}, E., {Natta}, A., {Randich}, S., {Testi}, L., \& {Biazzo}, K.
  2011{\natexlab{a}}, \aap, 525, A47+

\bibitem[{{Rigliaco} {et~al.}(2011{\natexlab{b}}){Rigliaco}, {Natta},
  {Randich}, {Testi}, {Covino}, {Herczeg}, \&
  {Alcal{\'a}}}]{2011A&A...526L...6R}
{Rigliaco}, E., {Natta}, A., {Randich}, S., {et~al.} 2011{\natexlab{b}}, \aap,
  526, L6+

\bibitem[{{Rigliaco} {et~al.}(2012){Rigliaco}, {Natta}, {Testi}, {Randich},
  {Alcal{\`a}}, {Covino}, \& {Stelzer}}]{2012A&A...548A..56R}
{Rigliaco}, E., {Natta}, A., {Testi}, L., {et~al.} 2012, \aap, 548, A56

\bibitem[{{Robin} {et~al.}(2003){Robin}, {Reyl{\'e}}, {Derri{\`e}re}, \&
  {Picaud}}]{2003A&A...409..523R}
{Robin}, A.~C., {Reyl{\'e}}, C., {Derri{\`e}re}, S., \& {Picaud}, S. 2003,
  \aap, 409, 523

\bibitem[{{Robitaille} {et~al.}(2007){Robitaille}, {Whitney}, {Indebetouw}, \&
  {Wood}}]{2007ApJS..169..328R}
{Robitaille}, T.~P., {Whitney}, B.~A., {Indebetouw}, R., \& {Wood}, K. 2007,
  \apjs, 169, 328

\bibitem[{{Sestito} {et~al.}(2008){Sestito}, {Palla}, \&
  {Randich}}]{2008A&A...487..965S}
{Sestito}, P., {Palla}, F., \& {Randich}, S. 2008, \aap, 487, 965

\bibitem[{{Sicilia-Aguilar} {et~al.}(2006{\natexlab{a}}){Sicilia-Aguilar},
  {Hartmann}, {F{\"u}r{\'e}sz}, {Henning}, {Dullemond}, \&
  {Brandner}}]{2006AJ....132.2135S}
{Sicilia-Aguilar}, A., {Hartmann}, L.~W., {F{\"u}r{\'e}sz}, G., {et~al.}
  2006{\natexlab{a}}, \aj, 132, 2135

\bibitem[{{Sicilia-Aguilar} {et~al.}(2011){Sicilia-Aguilar}, {Henning},
  {Dullemond}, {Patel}, {Juh{\'a}sz}, {Bouwman}, \&
  {Sturm}}]{2011ApJ...742...39S}
{Sicilia-Aguilar}, A., {Henning}, T., {Dullemond}, C.~P., {et~al.} 2011, \apj,
  742, 39

\bibitem[{{Sicilia-Aguilar} {et~al.}(2010){Sicilia-Aguilar}, {Henning}, \&
  {Hartmann}}]{2010ApJ...710..597S}
{Sicilia-Aguilar}, A., {Henning}, T., \& {Hartmann}, L.~W. 2010, \apj, 710, 597

\bibitem[{{Sicilia-Aguilar} {et~al.}(2008){Sicilia-Aguilar}, {Henning},
  {Juh{\'a}sz}, {Bouwman}, {Garmire}, \& {Garmire}}]{2008ApJ...687.1145S}
{Sicilia-Aguilar}, A., {Henning}, T., {Juh{\'a}sz}, A., {et~al.} 2008, \apj,
  687, 1145

\bibitem[{{Sicilia-Aguilar} {et~al.}(2013){Sicilia-Aguilar}, {Henning}, {Linz},
  {Andr{\'e}}, {Stutz}, {Eiroa}, \& {White}}]{2013A&A...551A..34S}
{Sicilia-Aguilar}, A., {Henning}, T., {Linz}, H., {et~al.} 2013, \aap, 551, A34

\bibitem[{{Sicilia-Aguilar} {et~al.}(2006{\natexlab{b}}){Sicilia-Aguilar},
  {Hartmann}, {Calvet}, {Megeath}, {Muzerolle}, {Allen}, {D'Alessio},
  {Mer{\'{\i}}n}, {Stauffer}, {Young}, \& {Lada}}]{2006ApJ...638..897S}
{Sicilia-Aguilar}, A., {Hartmann}, L., {Calvet}, N., {et~al.}
  2006{\natexlab{b}}, \apj, 638, 897

\bibitem[{{Sicilia-Aguilar} {et~al.}(2009){Sicilia-Aguilar}, {Bouwman},
  {Juh{\'a}sz}, {Henning}, {Roccatagliata}, {Lawson}, {Acke}, {Feigelson},
  {Tielens}, {Decin}, \& {Meeus}}]{2009ApJ...701.1188S}
{Sicilia-Aguilar}, A., {Bouwman}, J., {Juh{\'a}sz}, A., {et~al.} 2009, \apj,
  701, 1188

\bibitem[{{Siess} {et~al.}(2000){Siess}, {Dufour}, \&
  {Forestini}}]{2000A&A...358..593S}
{Siess}, L., {Dufour}, E., \& {Forestini}, M. 2000, \aap, 358, 593

\bibitem[{{Skrutskie} {et~al.}(2006){Skrutskie}, {Cutri}, {Stiening},
  {Weinberg}, {Schneider}, {Carpenter}, {Beichman}, {Capps}, {Chester},
  {Elias}, {Huchra}, {Liebert}, {Lonsdale}, {Monet}, {Price}, {Seitzer},
  {Jarrett}, {Kirkpatrick}, {Gizis}, {Howard}, {Evans}, {Fowler}, {Fullmer},
  {Hurt}, {Light}, {Kopan}, {Marsh}, {McCallon}, {Tam}, {Van Dyk}, \&
  {Wheelock}}]{2006AJ....131.1163S}
{Skrutskie}, M.~F., {Cutri}, R.~M., {Stiening}, R., {et~al.} 2006, \aj, 131,
  1163

\bibitem[{{St{\"o}rzer} \& {Hollenbach}(1999)}]{1999ApJ...515..669S}
{St{\"o}rzer}, H., \& {Hollenbach}, D. 1999, \apj, 515, 669

\bibitem[{{Strom} {et~al.}(1989{\natexlab{a}}){Strom}, {Newton}, {Strom},
  {Seaman}, {Carrasco}, {Cruz-Gonzalez}, {Serrano}, \&
  {Grasdalen}}]{1989ApJS...71..183S}
{Strom}, K.~M., {Newton}, G., {Strom}, S.~E., {et~al.} 1989{\natexlab{a}},
  \apjs, 71, 183

\bibitem[{{Strom} {et~al.}(1989{\natexlab{b}}){Strom}, {Strom}, {Edwards},
  {Cabrit}, \& {Skrutskie}}]{1989AJ.....97.1451S}
{Strom}, K.~M., {Strom}, S.~E., {Edwards}, S., {Cabrit}, S., \& {Skrutskie},
  M.~F. 1989{\natexlab{b}}, \aj, 97, 1451

\bibitem[{{Strom} {et~al.}(1993){Strom}, {Strom}, \&
  {Merrill}}]{1993ApJ...412..233S}
{Strom}, K.~M., {Strom}, S.~E., \& {Merrill}, K.~M. 1993, \apj, 412, 233

\bibitem[{{Szentgyorgyi} {et~al.}(2011){Szentgyorgyi}, {Furesz}, {Cheimets},
  {Conroy}, {Eng}, {Fabricant}, {Fata}, {Gauron}, {Geary}, {McLeod}, {Zajac},
  {Amato}, {Bergner}, {Caldwell}, {Dupree}, {Goddard}, {Johnston}, {Meibom},
  {Mink}, {Pieri}, {Roll}, {Tokarz}, {Wyatt}, {Epps}, {Hartmann}, \&
  {Meszaros}}]{2011PASP..123.1188S}
{Szentgyorgyi}, A., {Furesz}, G., {Cheimets}, P., {et~al.} 2011, \pasp, 123,
  1188

\bibitem[{{Tobin} {et~al.}(2009){Tobin}, {Hartmann}, {Furesz}, {Mateo}, \&
  {Megeath}}]{2009ApJ...697.1103T}
{Tobin}, J.~J., {Hartmann}, L., {Furesz}, G., {Mateo}, M., \& {Megeath}, S.~T.
  2009, \apj, 697, 1103

\bibitem[{{Tognelli} {et~al.}(2011){Tognelli}, {Prada Moroni}, \&
  {Degl'Innocenti}}]{2011A&A...533A.109T}
{Tognelli}, E., {Prada Moroni}, P.~G., \& {Degl'Innocenti}, S. 2011, \aap, 533,
  A109

\bibitem[{{Vink} {et~al.}(2002){Vink}, {Drew}, {Harries}, \&
  {Oudmaijer}}]{2002MNRAS.337..356V}
{Vink}, J.~S., {Drew}, J.~E., {Harries}, T.~J., \& {Oudmaijer}, R.~D. 2002,
  \mnras, 337, 356

\bibitem[{{Watson} {et~al.}(2009){Watson}, {Schr{\"o}der}, {Fyfe}, {Page},
  {Lamer}, {Mateos}, {Pye}, {Sakano}, {Rosen}, {Ballet}, {Barcons}, {Barret},
  {Boller}, {Brunner}, {Brusa}, {Caccianiga}, {Carrera}, {Ceballos}, {Della
  Ceca}, {Denby}, {Denkinson}, {Dupuy}, {Farrell}, {Fraschetti}, {Freyberg},
  {Guillout}, {Hambaryan}, {Maccacaro}, {Mathiesen}, {McMahon}, {Michel},
  {Motch}, {Osborne}, {Page}, {Pakull}, {Pietsch}, {Saxton}, {Schwope},
  {Severgnini}, {Simpson}, {Sironi}, {Stewart}, {Stewart}, {Stobbart}, {Tedds},
  {Warwick}, {Webb}, {West}, {Worrall}, \& {Yuan}}]{2009A&A...493..339W}
{Watson}, M.~G., {Schr{\"o}der}, A.~C., {Fyfe}, D., {et~al.} 2009, \aap, 493,
  339

\bibitem[{{White} \& {Basri}(2003)}]{2003ApJ...582.1109W}
{White}, R.~J., \& {Basri}, G. 2003, \apj, 582, 1109

\bibitem[{{White} \& {Hillenbrand}(2004)}]{2004ApJ...616..998W}
{White}, R.~J., \& {Hillenbrand}, L.~A. 2004, \apj, 616, 998

\bibitem[{{Wilson} {et~al.}(2005){Wilson}, {Dame}, {Masheder}, \&
  {Thaddeus}}]{2005A&A...430..523W}
{Wilson}, B.~A., {Dame}, T.~M., {Masheder}, M.~R.~W., \& {Thaddeus}, P. 2005,
  \aap, 430, 523

\bibitem[{{Winston} {et~al.}(2009){Winston}, {Megeath}, {Wolk}, {Hernandez},
  {Gutermuth}, {Muzerolle}, {Hora}, {Covey}, {Allen}, {Spitzbart}, {Peterson},
  {Myers}, \& {Fazio}}]{2009AJ....137.4777W}
{Winston}, E., {Megeath}, S.~T., {Wolk}, S.~J., {et~al.} 2009, \aj, 137, 4777

\bibitem[{{Winston} {et~al.}(2010){Winston}, {Megeath}, {Wolk}, {Spitzbart},
  {Gutermuth}, {Allen}, {Hernandez}, {Covey}, {Muzerolle}, {Hora}, {Myers}, \&
  {Fazio}}]{2010AJ....140..266W}
---. 2010, \aj, 140, 266

\bibitem[{{Wolk}(2009)}]{2009AIPC.1094..959W}
{Wolk}, S.~J. 2009, in American Institute of Physics Conference Series, Vol.
  1094, American Institute of Physics Conference Series, ed. {E.~Stempels},
  959--962

\bibitem[{{Wright} {et~al.}(2010){Wright}, {Eisenhardt}, {Mainzer}, {Ressler},
  {Cutri}, {Jarrett}, {Kirkpatrick}, {Padgett}, {McMillan}, {Skrutskie},
  {Stanford}, {Cohen}, {Walker}, {Mather}, {Leisawitz}, {Gautier}, {McLean},
  {Benford}, {Lonsdale}, {Blain}, {Mendez}, {Irace}, {Duval}, {Liu}, {Royer},
  {Heinrichsen}, {Howard}, {Shannon}, {Kendall}, {Walsh}, {Larsen}, {Cardon},
  {Schick}, {Schwalm}, {Abid}, {Fabinsky}, {Naes}, \&
  {Tsai}}]{2010AJ....140.1868W}
{Wright}, E.~L., {Eisenhardt}, P.~R.~M., {Mainzer}, A.~K., {et~al.} 2010, \aj,
  140, 1868

\bibitem[{{York} {et~al.}(2000){York}, {Adelman}, {Anderson}, {Anderson},
  {Annis}, {Bahcall}, {Bakken}, {Barkhouser}, {Bastian}, {Berman}, {Boroski},
  {Bracker}, {Briegel}, {Briggs}, {Brinkmann}, {Brunner}, {Burles}, {Carey},
  {Carr}, {Castander}, {Chen}, {Colestock}, {Connolly}, {Crocker}, {Csabai},
  {Czarapata}, {Davis}, {Doi}, {Dombeck}, {Eisenstein}, {Ellman}, {Elms},
  {Evans}, {Fan}, {Federwitz}, {Fiscelli}, {Friedman}, {Frieman}, {Fukugita},
  {Gillespie}, {Gunn}, {Gurbani}, {de Haas}, {Haldeman}, {Harris}, {Hayes},
  {Heckman}, {Hennessy}, {Hindsley}, {Holm}, {Holmgren}, {Huang}, {Hull},
  {Husby}, {Ichikawa}, {Ichikawa}, {Ivezi{\'c}}, {Kent}, {Kim}, {Kinney},
  {Klaene}, {Kleinman}, {Kleinman}, {Knapp}, {Korienek}, {Kron}, {Kunszt},
  {Lamb}, {Lee}, {Leger}, {Limmongkol}, {Lindenmeyer}, {Long}, {Loomis},
  {Loveday}, {Lucinio}, {Lupton}, {MacKinnon}, {Mannery}, {Mantsch}, {Margon},
  {McGehee}, {McKay}, {Meiksin}, {Merelli}, {Monet}, {Munn}, {Narayanan},
  {Nash}, {Neilsen}, {Neswold}, {Newberg}, {Nichol}, {Nicinski}, {Nonino},
  {Okada}, {Okamura}, {Ostriker}, {Owen}, {Pauls}, {Peoples}, {Peterson},
  {Petravick}, {Pier}, {Pope}, {Pordes}, {Prosapio}, {Rechenmacher}, {Quinn},
  {Richards}, {Richmond}, {Rivetta}, {Rockosi}, {Ruthmansdorfer}, {Sandford},
  {Schlegel}, {Schneider}, {Sekiguchi}, {Sergey}, {Shimasaku}, {Siegmund},
  {Smee}, {Smith}, {Snedden}, {Stone}, {Stoughton}, {Strauss}, {Stubbs},
  {SubbaRao}, {Szalay}, {Szapudi}, {Szokoly}, {Thakar}, {Tremonti}, {Tucker},
  {Uomoto}, {Vanden Berk}, {Vogeley}, {Waddell}, {Wang}, {Watanabe},
  {Weinberg}, {Yanny}, \& {Yasuda}}]{2000AJ....120.1579Y}
{York}, D.~G., {Adelman}, J., {Anderson}, Jr., J.~E., {et~al.} 2000, \aj, 120,
  1579

\bibitem[{{Zhang} \& {Wang}(2009)}]{2009AJ....138.1830Z}
{Zhang}, M., \& {Wang}, H. 2009, \aj, 138, 1830

\end{thebibliography}
